\documentclass[aps,pre,twocolumn,balancelastpage,showpacs,reprint]{revtex4-1}

\pdfoutput=1
\usepackage{blindtext}
\usepackage{centernot}
\usepackage{graphicx}
\usepackage{amsmath}
\usepackage{times}
\usepackage{amssymb}
\usepackage{mathrsfs}
\usepackage{chemarr}
\usepackage{color}
\usepackage{url}
\usepackage{version}
\usepackage[hidelinks]{hyperref}

\usepackage{mwe,tikz}\usepackage[percent]{overpic}

\usepackage{bm}
\usepackage[export]{adjustbox}
\definecolor{linkcolor}{rgb}{0,0,0.6}

\newcommand{\p}{\partial}

\newcommand{\eps}{\varepsilon}

\begin{document}


\title{Spatial Organization of Active Particles with Field Mediated Interactions}

\author{Ruben Zakine, Jean-Baptiste Fournier and Fr\'ed\'eric van Wijland}

\affiliation{Université de Paris, Laboratoire Mati\`ere et Syst\`emes Complexes (MSC), UMR 7057 CNRS, F-75205 Paris, France}

\date{\today}

\pacs{...}

\begin{abstract}
We consider a system of independent point-like particles performing a Brownian motion while interacting with a Gaussian fluctuating background. These particles are in addition endowed with a discrete two-state internal degree of freedom that is subjected to a nonequilibrium source of noise, which affects their coupling with the background field. We explore the phase diagram of the system and pinpoint the role of the nonequilibrium drive in producing a nontrivial patterned spatial organization. We are able, by means of a weakly nonlinear analysis, to account for the parameter-dependence of the boundaries of the phase and pattern diagram in the stationary state.

\end{abstract}

\maketitle 

\section{Introduction}

Interactions between objects mediated by an elastic medium are ubiquitous in the soft-matter realm. 
Examples include  surfaces, colloids or proteins in soft-matter media such as critical binary mixtures~\cite{Fisher78,Hertlein08}, liquid crystals~\cite{Ajdari91,poulin_1997}, capillary interfaces~\cite{Nicolson49,Hu05Nat} and bio-membranes~\cite{Goulian93EPL,Dan93Lang,Dommersnes99EPJB,BitbolPlos12,vanderWel16}.
Effective interactions, already at the level of equilibrium systems, may lead to a  rich phase behavior~\cite{poulin_1997,brown_2000,noguchi_2017}. Yet relaxing the constraint of detailed balance is expected to allow for an even richer phenomenology. Such nonequilibrium examples include biophysical agents such as  bacteria or cells. These are good examples of active particles whose chemical action on the medium~\cite{miller_2001}, or whose hydrodynamic interactions with the medium~\cite{stenhammar_2017}, give birth to mediated (non local) interactions between particles that cannot be encapsulated in an effective energy picture. Down to even smaller scales, recent experiments on reactive proteins suggest interesting collective behavior made possible by the induced deformation of the lipid bilayer in which proteins are embedded in~\cite{fribourg_2014,sumino_2014}. This class of systems also encompasses synthetic objects that actively shape the medium in which they are confined. The latter include liquid crystal colloids endowed with a tunable degree of freedom~\cite{evans_2013}, whose interactions have been experimentally explored~\cite{criante_2013} with a view to controling self-assembly at the micrometer scale. 

The goal of the present work is to build a generic model belonging to the same family of systems as those just described, in particular by incorporating an externally switchable (active) degree of freedom. To do so, we shall retain the physical ingredients shared by these systems.

First, we resolve particles at the individual level and we assume their individual motion to be diffusive. We further endow our particles with a two-state, spin-like, internal degree of freedom that can be switched by an external drive (this is where activity will come into play). Second, we choose to describe the embedding medium by a coarse-grained field. Third, we consider a coupling between the medium and the degrees of freedom --both spin and position-- of the particles, that will lead to mediated interactions. In order to focus on the latter, we omit from our model any direct interaction (hard-core, attractive or else) between particles. The dynamics of the medium itself is assumed to be local and purely relaxational (model A-like).
 The out-of-equilibrium nature of the system comes from the active conformation switch of the particle that breaks detailed balance. Beyond proteins in biomembranes, such externally driven conformational changes can also be found in synthetic soft-matter systems (see \cite{fribourg_2014,grawitter_2018,chen_2004} for recent references).

Even with the simplifying assumptions that have led to our model,  our particles do evolve far from equilibrium, and no  free-energy based method is available. Predicting collective phenomena thus requires to implement a variety of approaches, both numerical (Monte Carlo) and analytical (mean-field equations, noiseless reaction-diffusion equations). We present the details of the model and in particular its key parameters in Sec.~\ref{sec:Model}. 
Its stationary phase diagram is explored in Sec.~\ref{sec:Simu}, by means of Monte Carlo dynamical simulations, both for our active system and its equilibrium counterpart (that we properly define). A variety of patterned phases emerge in some regions of our parameter space. The subsequent mean-field analysis of Sec.~\ref{sec:Mean-field} allows us to understand the phase boundaries of the phase diagram given by our simulations. This very good qualitative (and good quantitative) agreement between the solution of the mean-field partial differential equations (PDE) and dynamical simulations suggests that the mean-field approach might also prove powerful to describe the physical nature of our patterned phases. Thus, in Sec.~\ref{sec:Pattern_analysis}, we embark into a linear and a nonlinear analysis of these equations, which allows us to describe the pattern content of our physical problem. Particular emphasis is placed on the extra mathematical difficulty of dealing with a conserved mode in a pattern forming system, a question that was hitherto sidelined in the existing literature of active inclusions in membranes~\cite{reigada_2005_1,reigada_2005_2}.
 As it will turn out, this is the existence of the conserved mode (expressing that particles are conserved, regardless of their internal spin state) that gives birth to a rich phenomenology of patterns. Our analysis of such patterns will draw from recent theoretical work~\cite{riecke1992ginzburg_landau,matthews_2000, cox_2003, winterbottom_2005}.
In our final two sections, we discuss the role of the nonequilibrium drive in producing patterned phases. To this end we introduce, in Sec.\ref{sec:Mixed_active_eq}, a model that interpolates between the active system and its equilibrium counterpart. This allows us to probe the robustness of patterns with respect to a partial restoration of reversibility (via tunable coupling to the same thermal bath the particles and the field are in contact with). To further pinpoint at the microscopic level the processes by which entropy is created in our active system, we establish a spatial map of entropy production that we superimpose to the patterns we obtain.  The various regimes observed throughout our simulations can be reasonably rationalized using this versatile entropy production as a quantitative indicator. This last piece of our analysis can be found in Sec.~\ref{sec:Entropy_creation}.

\section{Model}
\label{sec:Model}
The first ingredient of our model is a fluctuating field $\phi$ standing for the surrounding medium in which our particles are embedded. Our analysis is confined to a two-dimensional medium.
We choose to use a free field with a Gaussian Hamiltonian endowed with the following features: the value of the field at rest and without particles is $0$, and the field has a finite correlation length $\xi$. The Hamiltonian of the field then reads
\begin{align}
H_0=\int\! d^2 x \left[\frac r 2 \phi^2+ \frac c 2 (\bm \nabla\phi)^2\right],
\label{eq:hamiltonian}
\end{align}
with $\xi=\sqrt{c/r}$. We assume the medium is in contact with a thermostat at temperature $T$.  We further assume a separation of scales between the medium constituents and the particles, so that the medium can be described by a continuous field on a continuous space. However we retain the individual localized nature of the particle which we describe by their position $\bm r_k$. The value $\phi(\bm r_k)$ of the field at the position $\bm r_k$ of particle $k$ is elastically constrained to the value $\pm \phi_0$ by the internal degree of freedom $S_k=\pm1$ of the particle. This leads us to use the following interaction energy between $N$ particles and the field 
\begin{align}
H_\mathrm{int}=\sum_{k=1}^N \frac B 2 \left(\phi(\bm r_k)- S_k\phi_0\right)^2,
\label{eq:hamiltonian_interaction}
\end{align}
where $B$ is the strength of the particle-field coupling. Note that, as discussed in the introduction, particles experience no direct interactions (not even hard-core repulsion).  The Ising spin variable $S_k$ refers to the two internal states the particle is assumed to be found in. More realistic models will of course be system-dependent. For instance, to describe conically shaped proteins that locally constrain lipid membrane curvature, a Helfrich Hamiltonian~\cite{helfrich_1973} should be used instead of Eq.~\eqref{eq:hamiltonian}. A description of the membrane thickness with the Landau-Ginzburg Hamiltonian~\eqref{eq:hamiltonian} is perhaps better adapted to the description of protein-protein interactions experiencing hydrophobic mismatch interactions and coarse-grained packing interactions, already existing in pure one-component lipid bilayer~\cite{west_2009}. Once energy functions are specified, we turn to the question of how to implement dynamical evolution. Regarding the background field itself, discarding possible conservation laws or hydrodynamic interactions (either or both could prove relevant in a variety of physical systems), we  resort to a purely relaxational dynamics consistent with the contact to a thermal bath at temperature $T$:
\begin{align}
&\partial_t \phi(\bm x,t) =-\Gamma\frac{\delta H}{\delta \phi(\bm x,t)}+\sqrt{2\Gamma T}\,\zeta(\bm x,t),\label{eq:field_evol}\\
&\langle\zeta(\bm x,t)\zeta(\bm x',t')\rangle =\delta(\bm x-\bm x')\delta(t-t'),
\end{align}
where $H=H_0+H_\mathrm{int}$, $\Gamma$ is a mobility coefficient, $T$ is the temperature in energy units and $\zeta(\bm x,t)$ a Gaussian white noise with zero average. As far as particles are concerned, their (low Reynolds) motion is described by an overdamped Langevin equation:
\begin{align}
&\frac{d\bm r_k}{d t}=-\mu \frac{\partial H}{\partial \bm r_k}+\sqrt{2\mu T}\bm\xi_k(t),\label{eq:langevin}\\
&\langle \xi_k^\alpha(t) \xi_\ell^\beta(t')\rangle=\delta_{\alpha\beta}\delta_{k\ell}\delta(t-t'),
\end{align}
where $\mu$ is a mobility coefficient, and the $\xi_k^\alpha(t)$ are the  components of independent Gaussian white noises with zero average. We use the simplifying assumption that $\bm \xi_k$ and $\zeta$ are independent. So far, at fixed spin variables, our dynamics is consistent with detailed balance. The nonequilibrium drive will arise from the dynamics the spins are endowed with. With an external source of energy (such as photons or ATP in biological systems) in mind, we introduce temperature and state independent flipping rates $\alpha$ and $\gamma$:
\begin{align}\label{eq:spin-flips}
S_k=\mathrm{-1}\, \xrightleftharpoons[~\gamma~]{~\alpha~} \,S_k=\mathrm{+1}.
\end{align}
For the purpose of benchmarking genuinely nonequilibrium effects, we shall later introduce a detailed-balance preserving spin-flip dynamics. 
The final simplifying step is to work in terms of dimensionless parameters. {We introduce a characteristic size $a$ which will be used to spatially discretize the field $\phi$}, we normalize energies by $T$, times by $a^2/(\Gamma c)$ and we absorb $c$ in a redefinition of the field $\phi$. We thus carry out the replacements $\bm x/a\to \bm x$, $\Gamma ct/a^2\to t$, $c\phi^2/T\to\phi^2$, $c\phi_0^2/T\to\phi_0^2$, $a^2r/c\to r$, $B/c\to B$, $T\mu/(\Gamma c)\to \mu$, $a^2\alpha/(\Gamma c)\to\alpha$ and $a^2\gamma/(\Gamma c)\to\gamma$. In a nutshell, rescaling time, space, fields and constants boils down to $a=c=T=\Gamma=1$. Our model being now defined, we present the results of our numerical Monte Carlo-based exploration of its properties.

\section{Numerical simulations}
\label{sec:Simu}

We perform Monte Carlo dynamical simulations on a two dimensional square lattice of size $L_x\times L_y$ with periodic boundary conditions. The Gaussian field is defined on each site $(i,j)$  and takes continuous real values $\phi_{ij}$. The field $\phi$ evolves according to the explicit stochastic Euler scheme corresponding to the dimensionless form of Eq.~\eqref{eq:field_evol}:
\begin{align}
\begin{split}
\phi_{ij}(t+\Delta t)=&\phi_{ij}(t)-\Delta t\big[r\phi_{ij}(t)-\nabla^2\phi_{ij}(t)\\
 &+ B\sum_{k=1}^N (\phi_{\bm r_k}(t)-S_k\phi_0) \big]+ G(0,2\Delta t),
 \end{split}
\end{align}
where the discrete Laplacian is defined as
\begin{align}
&\nabla^2\phi_{ij}\equiv \phi_{i+1,j+1}+\phi_{i-1,j}+\phi_{i,j+1}+\phi_{i,j-1}-4\phi_{ij},
\label{eq:laplacian}
\end{align}
 $\bm r_k\equiv(i_k,j_k)$ being the position of the particle $k$ on the lattice, and $G(0,2\Delta t)$ being a Gaussian variable of mean $0$ and variance $2\Delta t$.

At each time step, we update the field, and then the particles' positions.
The particles lie on the lattice sites. Since our model involves non-interacting particles, we a priori allow for several particles to occupy the same lattice site. 
We implement a tower sampling algorithm~\cite{krauth_2006} to choose what action a particle should do, namely, either jump on a neighboring site, or stay on the same site or flip its spin. 
The total energy variation when the particle $k$ moves from site $(i_k,j_k)$ to $(i'_k,j'_k)$ is given by 
\begin{align}
\Delta H_{\bm r_k\to \bm r'_k}=\frac{B}{2}(\phi_{i'_k,j'_k}-\phi_{i_k,j_k})(\phi_{i'_k,j'_k}+\phi_{i_k,j_k}-2\phi_0 S_k).
\end{align}
The following jump probability $P_{(i_k,j_k)\to(i'_k,j'_k)}$ between times $t$ and $t+\Delta t$ implements a discrete version of the Langevin equation~\eqref{eq:langevin}:
\begin{align}
P_{(i_k,j_k)\to(i'_k,j'_k)} =\mu\Delta t \exp\left(-\frac{\Delta H_{\bm r_k\to \bm r'_k}}{2}\right).
\end{align}
According to our model, the spin flip probability of particle $k$, $P^f_k$, is fixed by the rates $\alpha$ and $\gamma$, except when we consider detailed-balance preserving flipping rates for the purpose of comparison to the out-of-equilibrium case. Each case will be specified below.
We take $\Delta t$ small enough to ensure that the probabilities verify
\begin{align}
\sum_{(i'_k,j'_k)} P_{(i_k,j_k)\to(i'_k,j'_k)}+P^f_k<1,
\end{align}
then  the probability $P^n_k$ that particle $k$ neither jumps nor flips is given by $P^n_k=1-[\sum_{(i'_k,j'_k)} P_{(i_k,j_k)\to(i'_k,j'_k)}+P^f_k]$.
We take $\Delta t=2\times 10^{-5}$.

\begin{figure}
\includegraphics[width=.98\columnwidth]{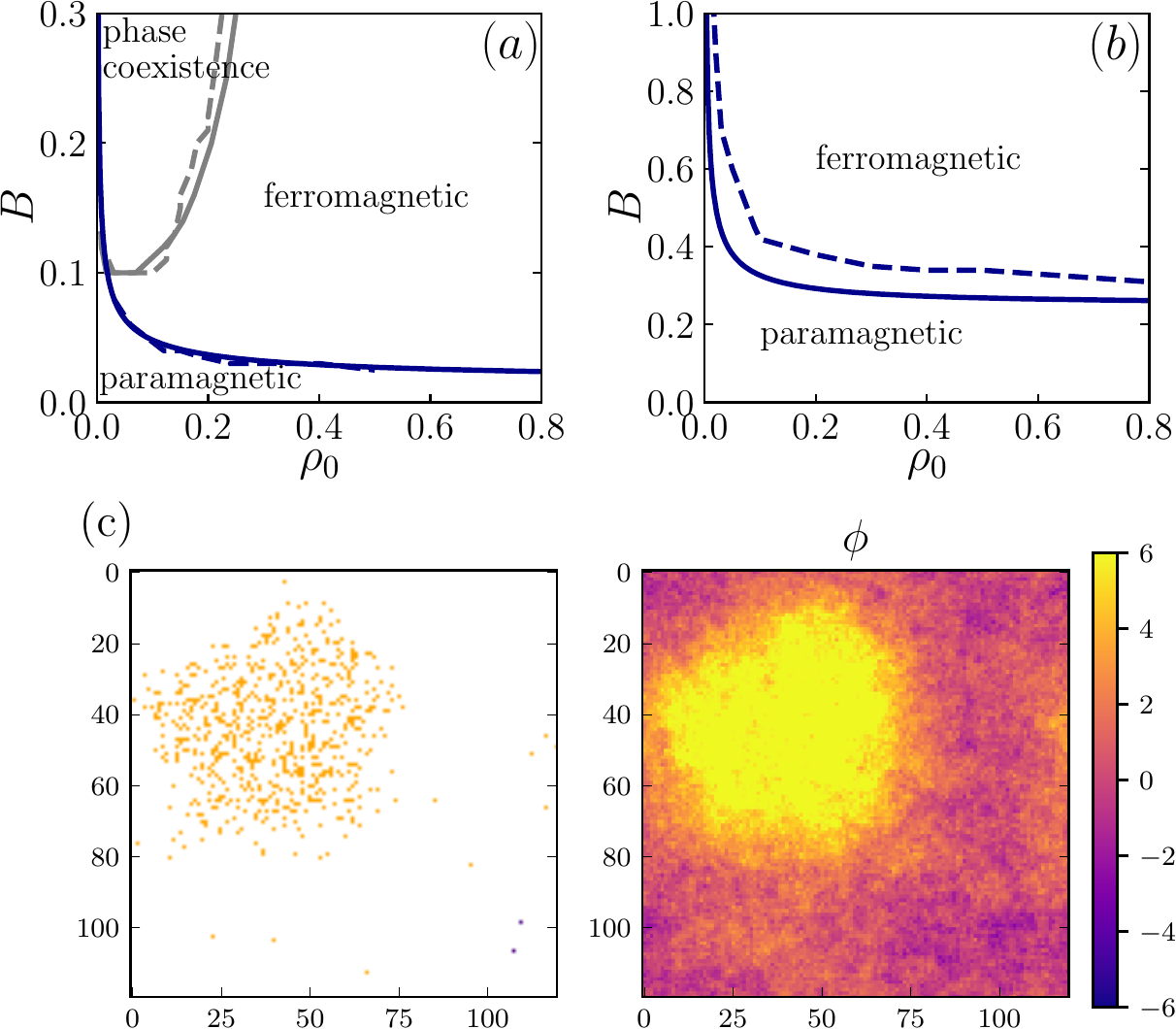}
\caption{Phase diagram (a) and (b), and snapshot (c) of the equilibrium system. (a) $r=0.01$, $\phi_0=8$. (b) $r=0.01$, $\phi_0=2$. Solid lines: mean-field predictions for the paramagnetic--ferromagnetic transition (blue) and for the binodal curve of the phase separation (gray). The corresponding dashed lines are the frontiers given by the Monte Carlo simulations. Phase coexistence is achieved for larger values of $B$ out of the range of the plot.
(c) Snapshot of the particle positions (left) and the corresponding underlying field $\phi$ (right) in the phase coexistence region for $r=0.01$, $B=0.26$, $\phi_0=8$, $\rho_0=0.05$. In the left snapshot, spin up particles are in yellow, spin down particles are in purple. }
\label{fig:phase_diag_eq}
\end{figure}

\subsection{Equilibrium benchmark}
\label{sec:MC_equilibrium_benchmark}
Though we are mostly interested in the active system, for the purpose of discussion and comparison we first study the system with spin flips that preserve the detailed balance condition. This is useful to sort out generic collective phenomena already present in our equilibrium Hamiltonian from those induced by activity. In this equilibrium benchmark system, the probability of a spin flip, say from spin up to spin down, between $t$ and $t+\Delta t$ is
\begin{equation}
P^f_k(\uparrow\to\downarrow)=\eta \Delta t \exp\left(-\frac{H_{k\downarrow}-H_{k\uparrow}}{2}\right),
\end{equation}
where $\eta$ is an inverse time-scale and where $H_{k\uparrow}$ (resp. $H_{k\downarrow}$) refers to the energy of the system when spin $k$ is up (resp. down). 

When endowed with this equilibrium reversible dynamics  the particles+field system already displays a nontrivial phase diagram explored with dynamical Monte Carlo simulations (see Fig.~\ref{fig:phase_diag_eq}).
When the coupling $B$ with the field  is small, a paramagnetic fluid (the average value of the spins is zero) is observed. When the coupling is increased, the system displays a paramagnetic-ferromagnetic transition. When the density is small, further increasing the coupling yields a phase separation between a paramagnetic gas and a ferromagnetic liquid~(Fig.\ref{fig:phase_diag_eq}c).
As a consistency check, we verified  that the equilibrium phase diagrams do not depend on dynamical parameters.
The specific simulations shown in Fig.~\ref{fig:phase_diag_eq} were performed with $\mu=5$, $\eta=1$.

\subsection{Active system}
\begin{figure}
\includegraphics[width=1\columnwidth]{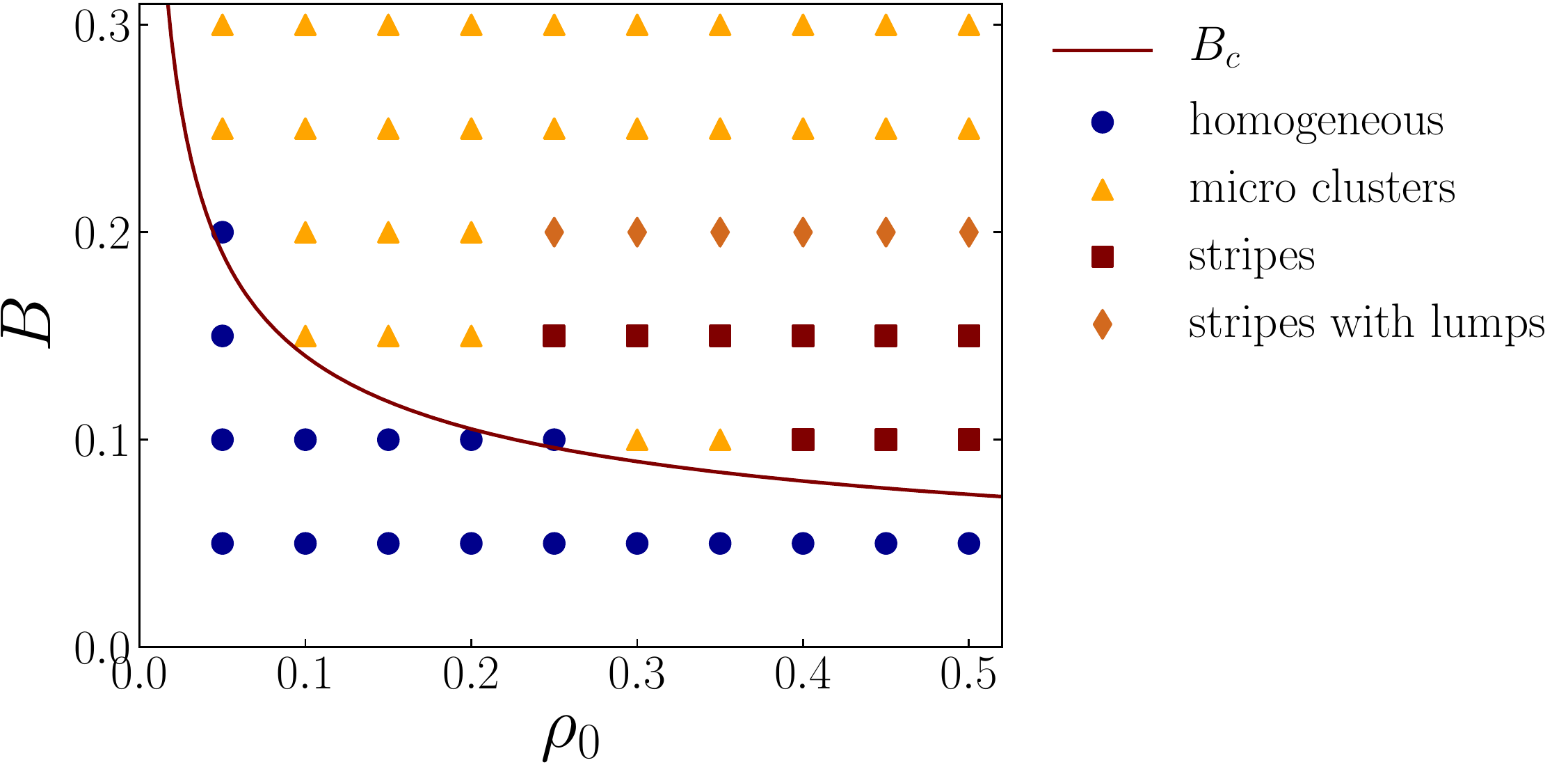}\\
\includegraphics[width=1\columnwidth]{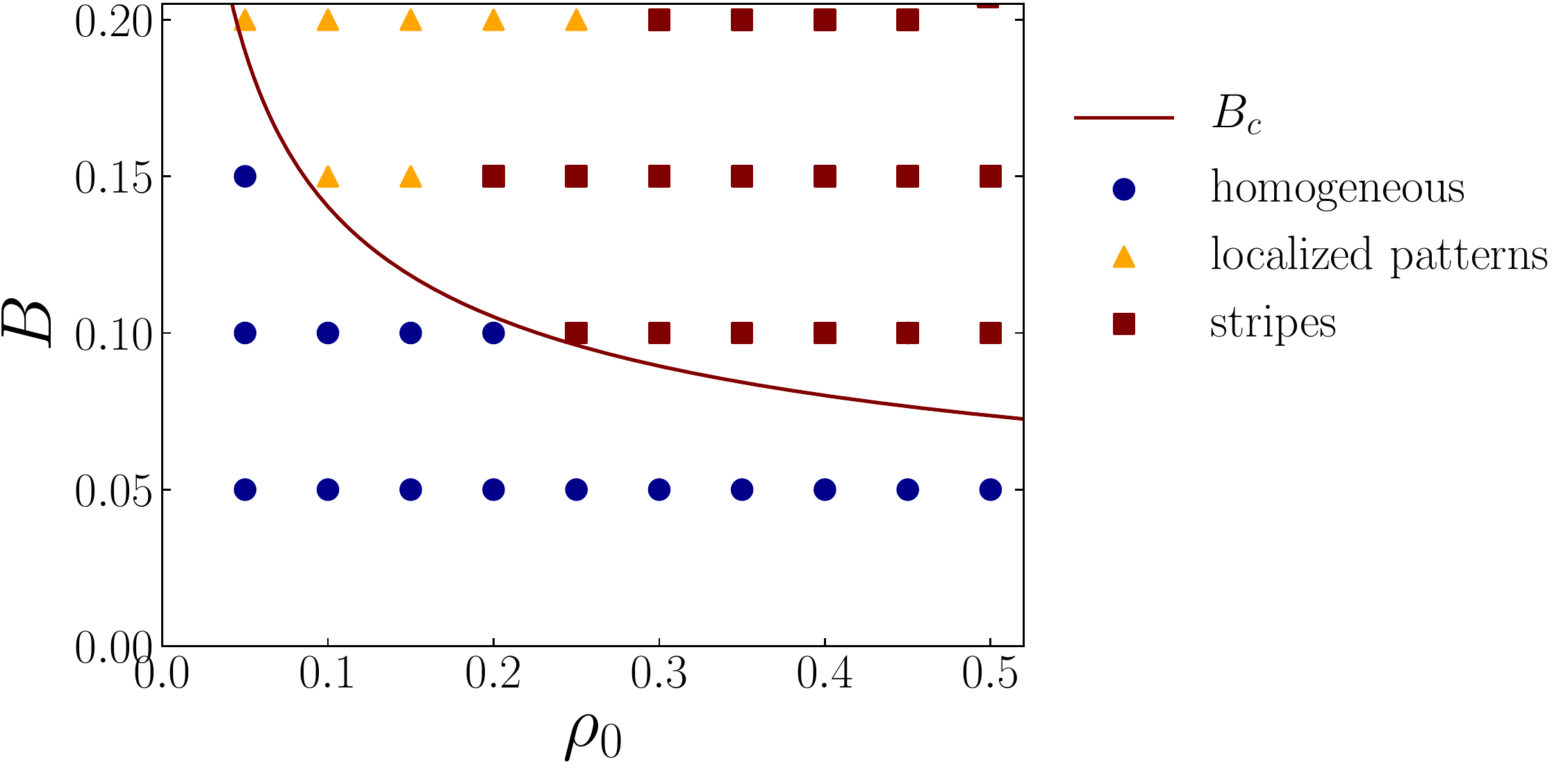}
\caption{\textbf{Top:} Monte Carlo phase diagram in space $(\rho_0,B)$ in the nonequilibrium steady-state. 
\textbf{Bottom:} Solved-PDE (Sec.~\ref{sec:Numerical_solution}) phase diagram in space $(\rho_0,B)$.
Parameters: $r=0.01$, $\phi_0=8$, $\mu=5$, $s=1/2$ and $\omega=0.2$). 
Solid burgundy line: pattern apparition threshold determined from a linear stability analysis (see Eq.~\eqref{eq:condition_kc}).}
\label{fig:phaseDiag_B_rho}
\end{figure}
In the active system, the flipping probabilities corresponding to Eq.~\eqref{eq:spin-flips} are given by
\begin{align}
P_k^f=\begin{cases}\gamma\Delta t &\mbox{if } S_k=+1\\
\alpha\Delta t &\mbox{if } S_k=-1.
\end{cases}
\end{align}
If flipping rates are symmetric ($\alpha=\gamma$) the average magnetization is zero and the system cannot develop a homogeneous ferromagnetic state. In the asymmetric case, it is convenient to define the total flip rate $\omega=\alpha+\gamma$ and the mean fraction $s=\alpha/\omega$ of spin-up particles in steady state. The flipping rates are thus given by $\alpha=s\omega$ and $\gamma=(1-s)\omega$.

\begin{figure}
\includegraphics[width=0.8\columnwidth]{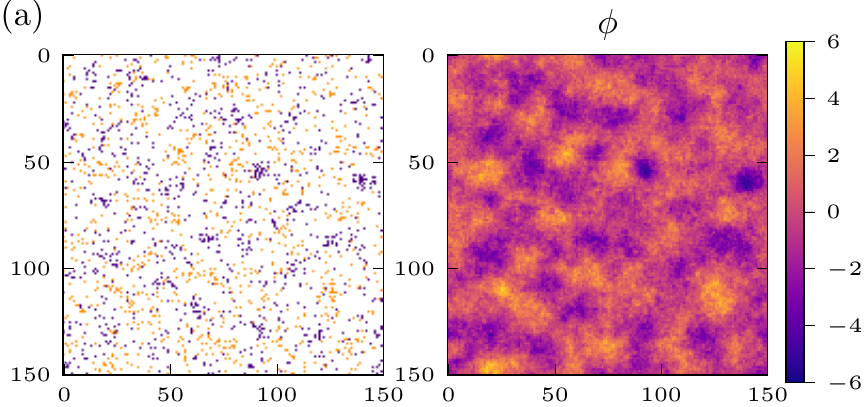}\\
\includegraphics[width=0.8\columnwidth]{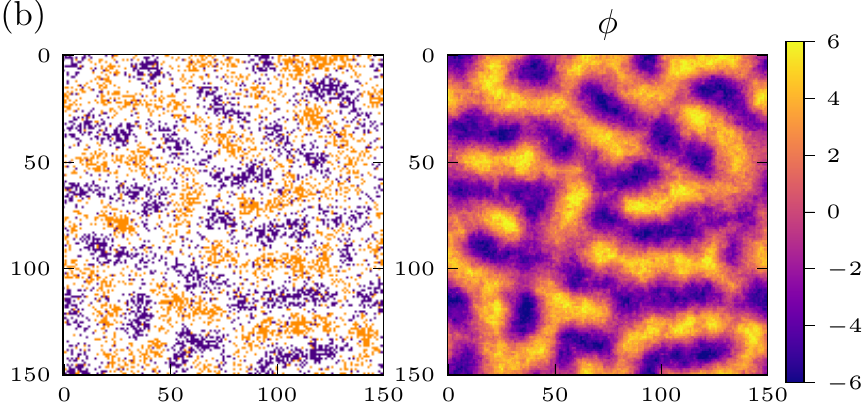}\\
\includegraphics[width=0.8\columnwidth]{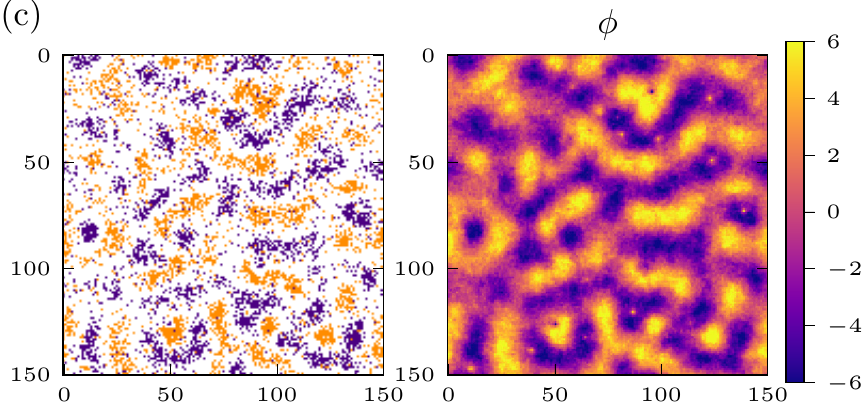}\\
\includegraphics[width=0.8\columnwidth]{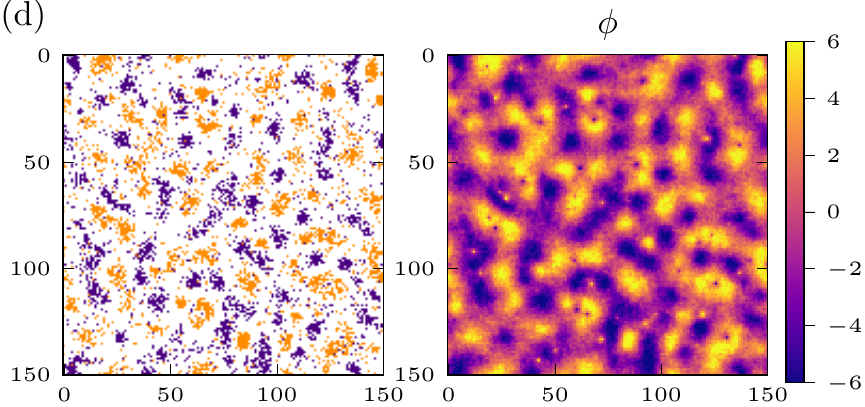}
\caption{Snapshots of the system corresponding to points in the phase diagram $(\rho_0,B)$ shown in Fig.~\ref{fig:phaseDiag_B_rho}. 
In the left snapshots, spin up particles are in yellow, spin down particles are in purple.
Parameters: same as in Fig.~\ref{fig:phaseDiag_B_rho}.
(a) $\rho_0=0.1 $, $B=0.2$: micro clusters. (b) $\rho_0=0.4 $, $B=0.15$: stripes. (c) $\rho_0=0.4 $, $B=0.20$: stripes with lumps. (d) $\rho_0=0.4 $, $B=0.25$: unstructured stripes, micro clusters.
}
\label{fig:snapshots_B_rho}
\end{figure}

\begin{figure}
\includegraphics[width=1\columnwidth]{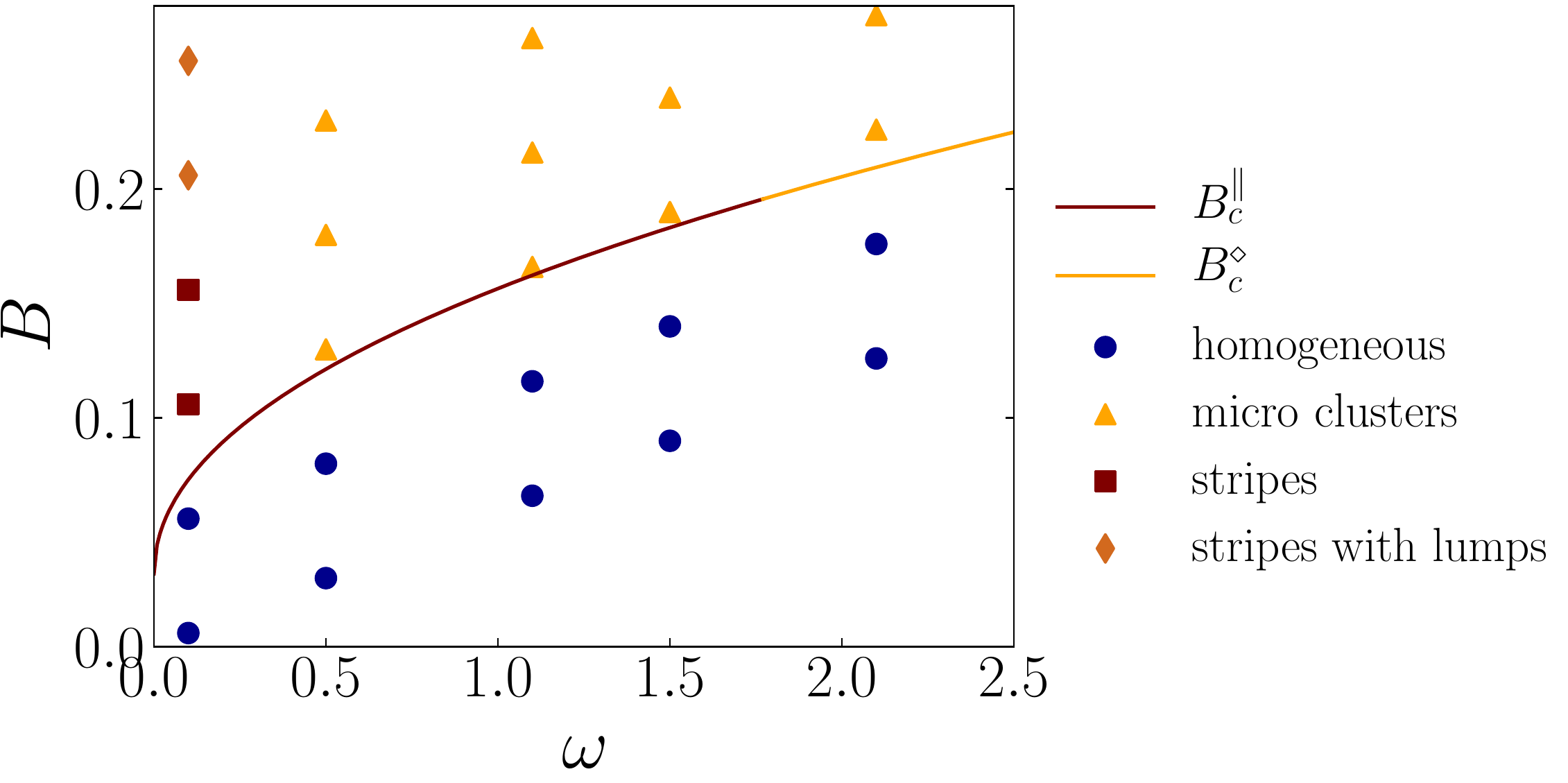}\\
\includegraphics[width=1\columnwidth]{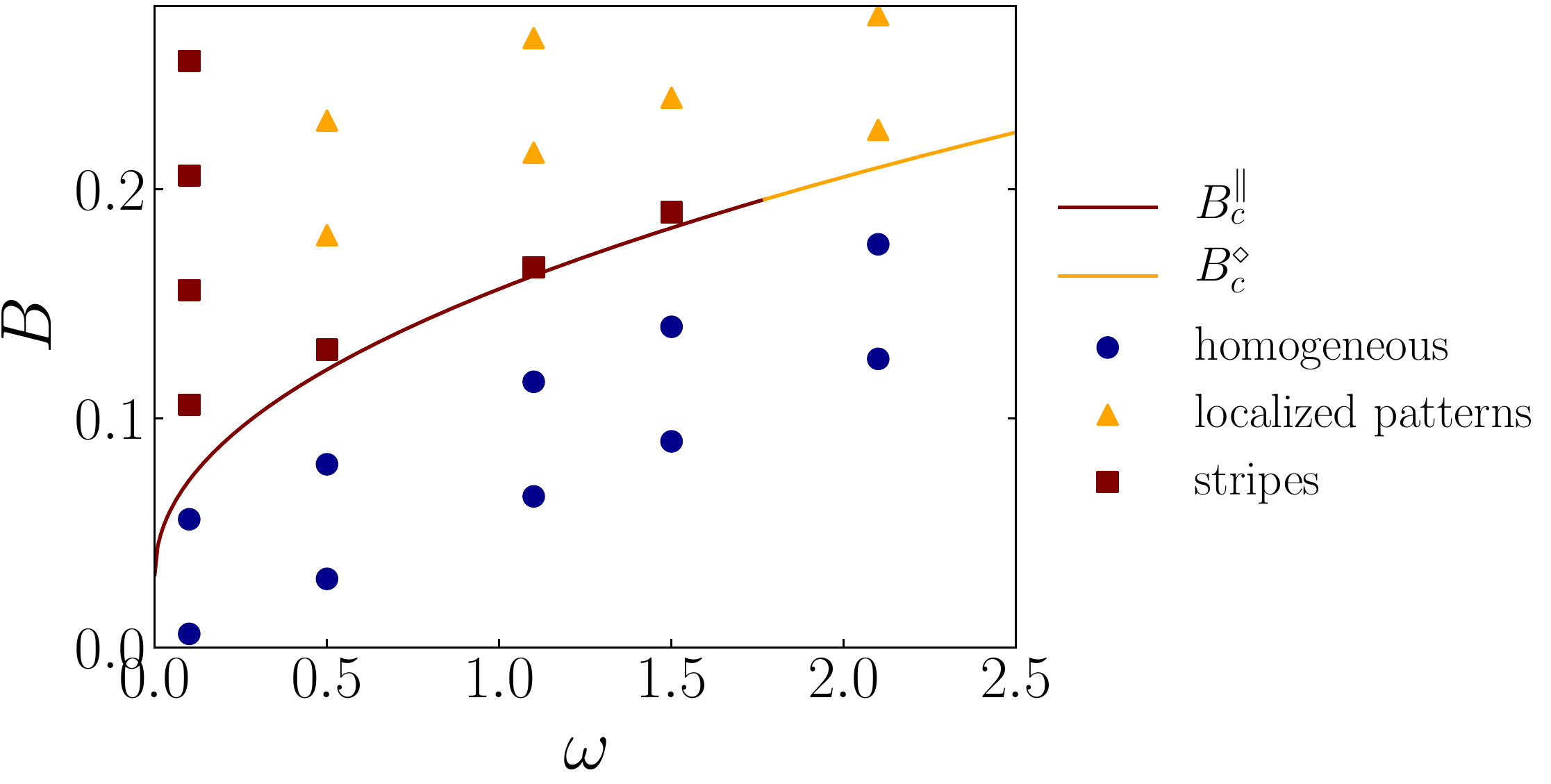}
\caption{\textbf{Top:} Monte Carlo phase diagram in space $(\omega,B)$ in the nonequilibrium steady-state. 
\textbf{Bottom:} Solved-PDE (Sec.~\ref{sec:Numerical_solution}) phase diagram in space $(\omega,B)$. 
Parameters: $r=0.01$, $s=1/2$, $\rho_0=0.3$, $\phi_0=8$, $\mu=5$. Solid burgundy line: pattern apparition threshold computed from linear stability analysis (see Eq.~\eqref{eq:condition_kc} in the section on the patterns analysis). Weakly nonlinear analysis predicts that rolls are stable to squares close to $B_c^\parallel$. Solid yellow line: pattern apparition threshold computed from linear stability analysis (see Eq.~\eqref{eq:condition_kc}). Weakly nonlinear analysis predicts that squares are stable to rolls close to $B_c^\diamond$. Rolls become unstable to squares at $(\omega,B)=(1.77, 0.1956)$ for $\rho_0=0.3$ (see Sec.~\ref{sec:rollsVSsquares}).}
\label{fig:phaseDiag_B_omega}
\end{figure}

In the following, we explore the phase diagram of the system for $s=1/2$ (Fig.~\ref{fig:phaseDiag_B_rho}, top).
When the coupling $B$ to the field is weak, the system remains homogeneous (and paramagnetic). At low densities, when increasing $B$, finite size clusters of both magnetization appear (see Fig.~\ref{fig:snapshots_B_rho}a). At higher densities, the phenomenology becomes richer. Increasing $B$ from the homogeneous phase, macroscopic stripes of both magnetization (Fig.~\ref{fig:snapshots_B_rho}b) are observed. 
As $B$ is further increased, the stripes harbor the continuous nucleation of small lumps of particles of opposite magnetization (Fig.~\ref{fig:snapshots_B_rho}c). These lumps grow, drift, then merge with adjacent bands of same magnetization. Increasing $B$ again, the proliferation of lumps leads to a system of micro-clusters (Fig.~\ref{fig:snapshots_B_rho}d).
In the patterned phase (stripes or clusters), increasing the flipping frequency $\omega$ yields local mixing of the spins, which results in the homogeneization of the whole system (Fig.~\ref{fig:phaseDiag_B_omega}, top).

\begin{figure}
\centering
\includegraphics[width=1.05\columnwidth]{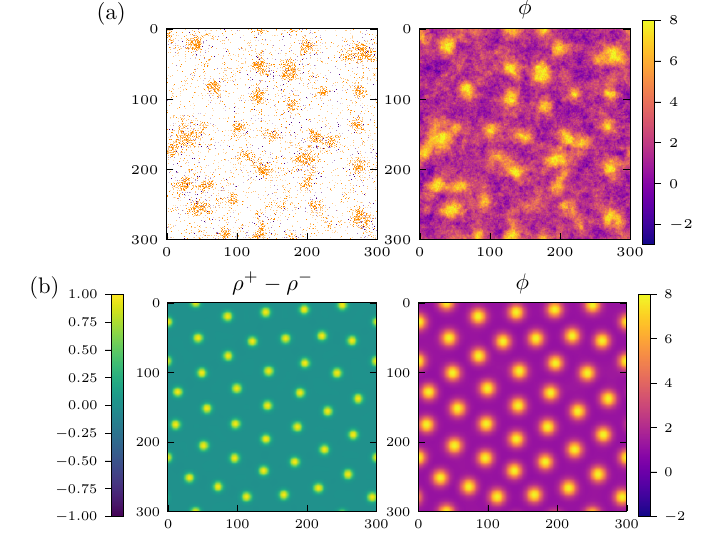}
\caption{Snapshots of (a) Monte Carlo simulation, (b) PDEs solution, 
in the strongly asymmetric case ($s=0.9$) in the nonequilibrium steady-state. 
In the top-left snapshot, spin up particles are in yellow, spin down particles are in purple. 
The PDEs solution exhibit stable hexagonal pattern which is partly destroyed in the Monte Carlo simulation.
See Sec.~\ref{sec:Numerical_solution} for details on the mean field solution. 
Parameters: $r=0.02$, $B=0.16$, $\rho_0=0.1$, $\phi_0=10$, $\mu=5$, $\omega=0.25$, $s=0.9$, and $N=9000$ in the Monte Carlo simulation.
}
\label{fig:snapshots_hexagons}
\end{figure}

If now asymmetric flipping rates ($s\ne 1/2$) are considered, the phase diagram features similar transitions. The homogeneous phase is however ferromagnetic since the mean number of spins up and spins down is different. {In addition, because of the breaking of the up-down symmetry, hexagonal patterns can be observed (see Fig.~\ref{fig:snapshots_hexagons})}.
In the following section, we work out a mean-field analysis which predicts the transition between different regimes. The appearance of lumps lies at the transition between two different pattern forming regimes.


\section{Mean-field behavior}
\label{sec:Mean-field}

\subsection{Equilibrium free energy}
\label{sec:mean-field_equilibrium}

We briefly treat the equilibrium case where particles are allowed to flip while respecting detailed balance (equilibrium benchmark). Our goal is to determine the phase diagram of the system. We first identify a conserved field, $\rho=\rho^++ \rho ^-$, and a non conserved field $\psi=\rho^+-\rho^-$. From the Hamiltonian $H$, we write down a mean-field free energy density:
\begin{align}
\begin{split}
f_\mathrm{MF}=\frac{r}{2}\phi^2&+B\frac{\rho+\psi}{4}(\phi-\phi_0)^2+B\frac{\rho-\psi}{4}(\phi+\phi_0)^2\\
&\quad+\frac{\rho+\psi}{2}\ln\frac{\rho+\psi}{2} +\frac{\rho-\psi}{2}\ln\frac{\rho-\psi}{2},
\end{split}
\end{align}
where the first three terms are directly inferred from the energy functional $H$, while the last two ones reflect the particles' entropy.
Since $\rho$ is the only conserved quantity, we minimize $f_\mathrm{MF}$ with respect to $\phi$ and $\psi$. We obtain $\phi=B\phi_0\psi/(r+B\rho)$ and $\psi=\rho \tanh(B\phi_0\phi)$. This imposes a self-consistent equation on $\phi=B\phi_ 0\rho \tanh(B\phi_0\phi)/(r+B\rho)$ similar to what is obtained for the magnetization in the Ising model. Searching for homogeneous phases, yields either $\phi=0$ (paramagnetic phase), or $\phi\ne 0$ (ferromagnetic phase). At low values of $B$, the system is uniform and there is a continuous paramagnetic--ferromagnetic transition at $
B_c^{(\mathrm{MF})}=(1+\sqrt{1+4r\,\phi_0^2/\rho})/(2\phi_0^2)$.
At higher values of $B$, we numerically solve the double tangent construction on $f_\mathrm{MF}(\rho)$ (already minimized with respect to $\psi$ and $\phi$). The system undergoes a phase separation between a low density paramagnetic phase and a high density ferromagnetic phase. These mean-field predictions correspond to the continuous lines of Fig.~\ref{fig:phase_diag_eq} while the results of the Monte Carlo simulations are indicated by the dashed lines. We have checked that the agreement is all the better as we are working at large $\phi_0$.

\subsection{Mean field dynamics}
We consider now the original model of interest where flips are fixed by an external and independent source of energy. Out of equilibrium, we can no longer rely on the free energy to construct the phase diagram.
Since particles execute Brownian motions, we consider the noiseless limit of the Dean--Kawasaki equations~\cite{dean1996langevin} for the up and down particle densities. Taking spin exchange into account (and neglecting the corresponding Poisson noise as well), we arrive at the deterministic evolution equations for $\rho^\pm$:
\begin{align}
\partial_t\rho^+&=
\mu\bm\nabla\cdot[\rho^+\bm\nabla \frac{\partial f_\mathrm{MF} }{\partial \rho^+}]+\alpha\rho^--\gamma\rho^+,
\label{eq:evolution_rho+}
\\
\partial_t\rho^-&=\mu\bm\nabla\cdot[\rho^-\bm\nabla \frac{\partial f_\mathrm{MF} }{\partial \rho^-}]
-\alpha\rho^-+\gamma\rho^+
\label{eq:evolution_rho-}\\
\partial_t\phi&= \nabla^2 \phi - r\phi- B \rho^+(\phi-\phi_0)- B \rho^-(\phi+\phi_0).
\label{eq:evolution_phi_naive}
\end{align}
It will prove convenient to write these equations in terms of the conserved field $\rho$, and of the non-conserved field $\psi$. We also parametrize the rates $\alpha$ and $\gamma$ by means of $\omega=\alpha+\gamma$ and $s=\alpha/\omega$ (the latter being the steady-state fraction of spin up particles). The dynamical evolutions of the fields then read
\begin{align}
\partial_t\rho&=
\mu \nabla^2\rho+\mu B\,\bm\nabla\cdot\left[(\rho\,\phi-\psi\phi_0)\bm\nabla \phi\right],\label{eq:evolution_rho}
\\
\begin{split}
\partial_t\psi&=
\mu \nabla^2\psi+\mu B\,\bm\nabla\cdot\left[(\psi\phi-\rho\phi_0)\bm\nabla \phi\right]\\
&\quad\quad\quad\quad-\omega \,\psi + (2s-1)\omega \rho,
\end{split}\label{eq:evolution_psi}\\
\partial_t\phi&= \nabla^2 \phi - r\phi- B \rho\phi+ B\phi_0\psi.\label{eq:evolution_phi}
\end{align}
These three equations are the starting point of our analysis of the patterns that form in the steady-state of our system. It is important to note that $s=\frac 12$ will play a special role because then these equations are invariant upon the up-down symmetry $(\rho,\psi,\phi)\to(\rho,-\psi,-\phi)$. Before we embark in a detailed analytical study of their pattern content, we begin with a numerical solution of these nonlinear coupled PDEs.

\subsection{Numerical solution of the coupled partial differential equations}
\label{sec:Numerical_solution}

We shall show that the numerical solution of the nonlinear coupled PDEs (which are noiseless) is relevant to analyze the stochastic simulations to the extent that phases and phase boundaries are quite faithfully captured. 
The coupled PDEs are solved on a lattice of size $L_x\times L_y=150\times 150$. The three fields $\rho^+$, $\rho^-$ and $\phi$ are discretized in time and space; an explicit Euler scheme to update the three fields is implemented.
The explicit Euler scheme is easy to implement and it converges in the domains of the phase diagram we are interested in. Our discretized equations take the following form:
\begin{align}
\begin{split}
\rho_{ij}^+(t+\Delta t)=\rho_{ij}^+(t)+&\Delta t\bigg[ 
\mu \nabla^2\rho_{ij}^+\\
&+\mu B\,\nabla_x( \rho^+_{ij}(\phi_{ij}-\phi_0)\nabla_x \phi_{ij})\\
&+\mu B\,\nabla_y( \rho^+_{ij}(\phi_{ij}-\phi_0)\nabla_y \phi_{ij})\\
&-\gamma\rho^+_{ij}+\alpha\rho^-_{ij}\bigg](t),
\end{split}
\end{align}
\begin{align}
\begin{split}
\phi_{ij}(t+\Delta t)=\phi_{ij}(t)+&\Delta t\big[\nabla^2\phi_{ij}-r\phi_{ij}\\
&-B\rho^+_{ij}(\phi_{ij}-\phi_0)\\
&-B\rho_{ij}^-(\phi_{ij}+\phi_0)\big](t),
\end{split}
\end{align}
and the discretized equation on $\rho^-$ is formally identical to the discretized equation on $\rho^+$ up to the exchange $\rho^+\leftrightarrow\rho^-$, $\phi_0\to-\phi_0$, $\alpha\leftrightarrow\gamma$.
The discrete spatial derivatives of any field $g_{ij}$ are defined as
\begin{align}
&\nabla_x g_{ij}\equiv \frac{1}{2}(g_{i+1,j}-g_{i-1,j}),\\
&\nabla_yg_{ij}\equiv \frac{1}{2}(g_{i,j+1}-g_{i,j-1}),
\end{align}
and the Laplacian has already been defined in Eq.~\eqref{eq:laplacian}.
 We confirm that different initial conditions lead to same stationary density profiles. We check the conservation of total density, namely $(L_x L_y)^{-1}\sum_{ij}\rho_{ij}=\rho_0 $, along with the positivity of $\rho^+$ and $\rho^-$ on each site.\\
 
\begin{figure}
\includegraphics[width=.99\columnwidth]{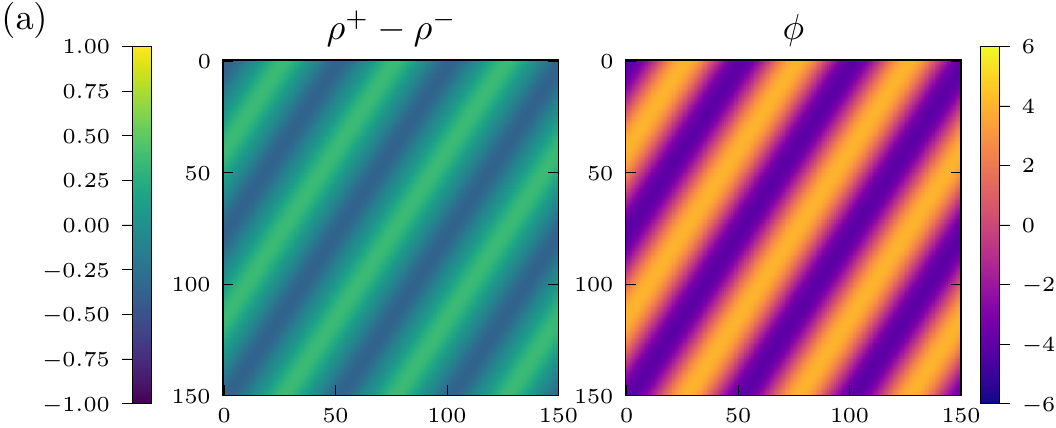}\\
\includegraphics[width=.99\columnwidth]{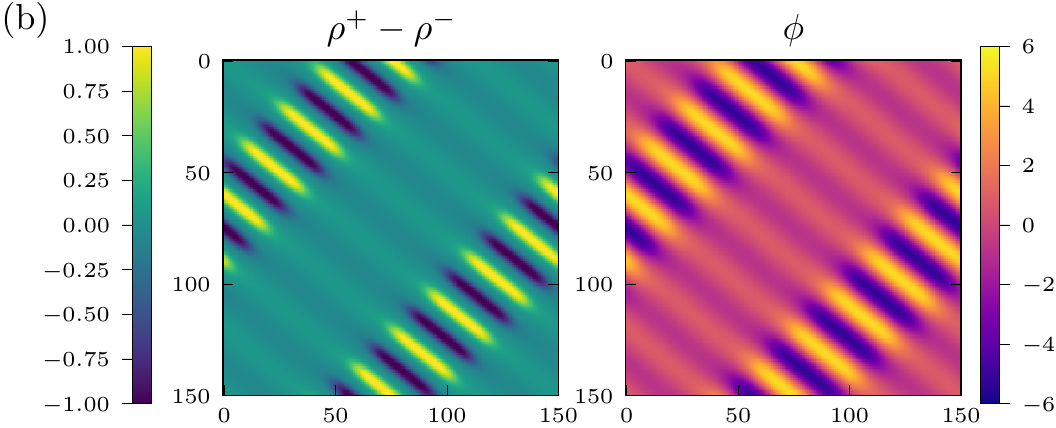}
\caption{(a) Snapshot of PDEs solution with $\omega=0.1$ and $B=0.106$. (b) Snapshot of PDEs solution with $\omega=0.5$ and $B=0.18$. In (a) we observe extended stripe-like patterns, while in (b) pattern localization occurs in the direction transverse to the stripes. Parameters: $r=0.01$, $\rho_0=0.3$, $\phi_0=8$, $\mu=5$, $s=1/2$. }
\label{fig:snapshot_B_omega_resolution}
\end{figure}

To ease comparison of the PDE solution with the Monte Carlo simulation, the PDE phase diagram is plotted Fig.~\ref{fig:phaseDiag_B_omega} (bottom) for the same physical parameters as those of the Monte Carlo results of Fig.~\ref{fig:phaseDiag_B_omega} (top). The results of the PDEs numerical solution match the results of the Monte Carlo simulations. 
The solution of the PDEs is also in good agreement with the predictions obtained from a weakly nonlinear analysis in Sec.~\ref{sec:weakly_nonlinear_analysis}: we can observe either homogeneous patterns (Fig.~\ref{fig:snapshot_B_omega_resolution}, top), or spatially localized patterns (Fig.~\ref{fig:snapshot_B_omega_resolution}, bottom), depending on the parameters. The coming section is devoted to an analysis of these patterns.

\section{Pattern analysis}
\label{sec:Pattern_analysis}

\subsection{Linear stability analysis}
By resorting to a linear stability analysis (LSA), the range of parameters for which a uniform stationary state is destabilized can be found. While LSA tells us about the first unstable mode, the question of which are the selected modes that eventually build up into patterns requires a full analysis of the nonlinear equations. The homogeneous and stationary solution to Eqs.~\eqref{eq:evolution_rho}, \eqref{eq:evolution_psi}, \eqref{eq:evolution_phi} is characterized by the following values of the fields
\begin{align}
\rho_h&=\rho_0,\\
\psi_h&=(2s-1)\rho_0,\\
\phi_h&=(2s-1)\frac{B\rho_0}{\tilde{r}}\phi_0,
\end{align}
with 
\begin{equation}
\tilde{r}=r+B\rho_0=\tilde{\xi}^{-2}
\end{equation}
where $\tilde{\xi}$ is the renormalized correlation length of the field $\phi$. We set $\rho_1=\rho-\rho_h$, $\psi_1=\psi-\psi_h$ and  $\phi_1=\phi-\phi_h$ and we expand Eqs.~\eqref{eq:evolution_rho}, \eqref{eq:evolution_psi}, \eqref{eq:evolution_phi} to linear order in the  $\rho_1$, $\phi_1$, $\psi_1$ fields. We expand the fields in Fourier modes $\sim e^{i\bm k\cdot\bm x}$ and we arrive at a linear system for the Fourier components $\partial_t(\rho_1,\psi_1,\phi_1)^T=M(\rho_1,\psi_1,\phi_1)^T$
\begin{widetext}
with
\begin{align}
M=\begin{pmatrix}
-\mu k^2 & 0 & -\mu B\rho_0\phi_0(2s-1)(B\rho_0/\tilde{r}-1)k^2\cr
(2s-1)\omega & -\mu k^2-\omega & -\mu B\rho_0\phi_0 [(2s-1)^2B\rho_0/\tilde{r}-1]k^2\cr
-B^2\rho_0\phi_0(2s-1)/\tilde{r} & B\phi_0 & -k^2-\tilde r
\end{pmatrix},
\end{align}
\end{widetext}
with $k=\|\bm k\| $.
The eigenvalues of $M$ can be shown to be always real which excludes oscillating patterns close to the threshold. We denote them  by $\sigma_i$, with $\sigma_1<\sigma_2<\sigma_3$. Solving $\det M(k)=0$ yields the modes for which temporal growth is marginal. In practice, we have $\det M=-\mu^2 k^2 Q(k^2)$, where $Q(X)=X^2+q_1 X+ q_2$ is degree 2 polynomial, with 
\begin{align}
q_1&=\frac{\omega}{\mu}+\tilde r-B^2 \rho_0 \phi_0^2+\frac{B^3 \rho_0^2\phi_0^2(2\tilde r-B\rho_0)(1-2s)^2}{\tilde r^2},\\
q_2&=\frac{\omega}{\mu\,\tilde r^2}\left(\tilde r^3-B^2\rho_0 r^2\phi_0^2(1-2s)^2\right).
\end{align}
Three different physical cases must be distinguished, depending on the roots $X_-$, $X_+$ of $Q$. 
\begin{figure}
\includegraphics[width=0.99\columnwidth]{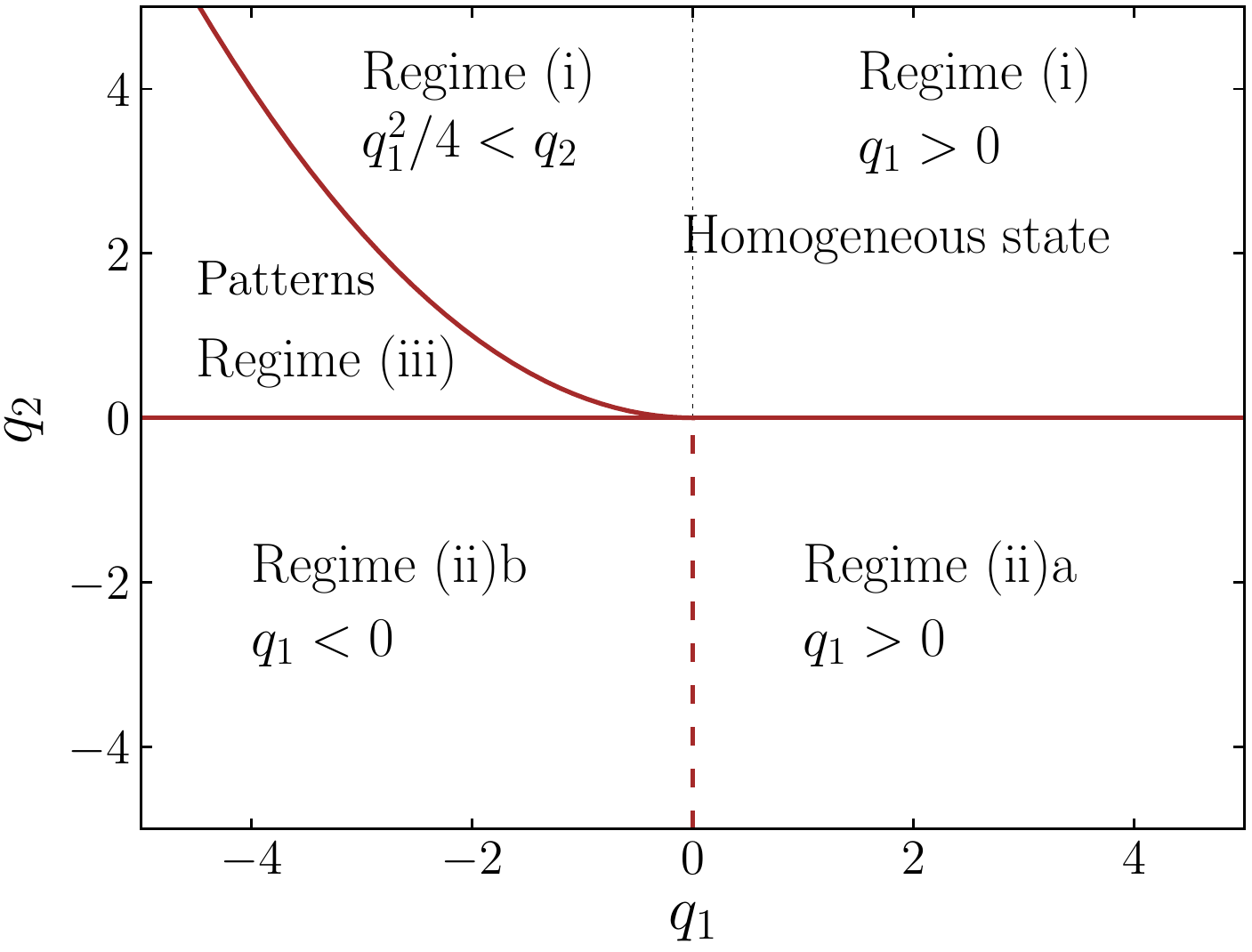}
\caption{Phase diagram from the linear stability analysis in space $(q_1,q_2)$. Solid brown line: boundaries between domains given by the LSA. Dashed brown: boundary from numerical solution of the PDEs. When $\omega=0$ we observe coarsening if $q_1<0$ and a stable homogeneous state if $q_1>0$. Regime (i): the homogeneous state is stable. Regime (ii)a: we numerically observe coarsening. Regime (ii)b: we numerically observe pattern formation. In both regimes (ii)a and (ii)b, the LSA predicted unstable modes down to $k\to0$, yet the system behavior can be very different from (ii)a to (ii)b. 
Regime (iii): finite wavelength patterns at stability threshold.}
\label{fig:stabilityRegime}
\end{figure}
\begin{figure}
  \begin{tikzpicture}
    \path (-4,0) node {\includegraphics[width=0.99\columnwidth]{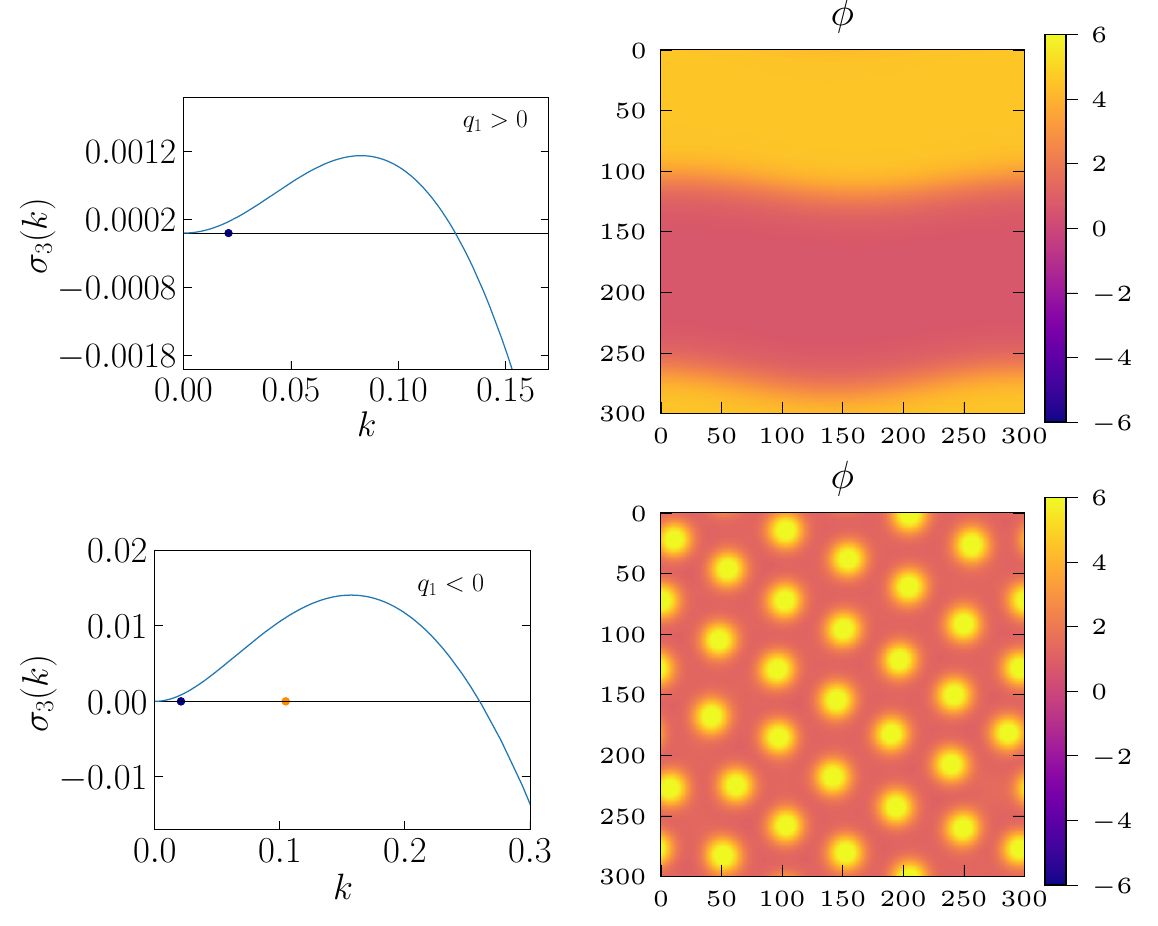}};
    \draw (-8,2.8) node[anchor=south west] {\bf (a)};
    \draw (-8,-0.5) node[anchor=south west] {\bf (b)};
  \end{tikzpicture}
\caption{ Larger eigenvalue $\sigma_3(k)$ in regime (ii)a (top left) and regime ii(b) (bottom left), and the corresponding solution of the PDEs (right).
In the left part, the other eigenvalues are strictly negative and are out of range of the plot.
Blue dot: minimal wavenumber $k_\mathrm{min}={2 \pi}/{L}$ sampled in the simulation. The instability is correctly captured in both cases.
While the linear stability analysis gives similar results in both cases, the solution  of PDEs (right column) shows  coarsening in one case, and pattern formation in the other case.  
Parameters: $r=0.02$, $B=0.16$, $\rho_0=0.1$, $\phi_0=9$, $\mu=5$, $s=0.9$. 
\textbf{\bf (a)} $\omega=10\omega_c$.
\textbf{\bf (b)} $\omega=0.5\omega_c$. Critical value $\omega_c$ defined in eq.~\eqref{eq:omega_c_definition}.
}
\label{fig:plotRegime2a_2b}
\end{figure}
\begin{itemize}
\item case (i): $q_1\geq0$ and $q_2\geq0$, or $q_2<q_1^2/4$, then $Q$ has no real positive roots. One can show that the three eigenvalues of $M$ are negative: the homogeneous state is stable.
\item case (ii): $q_2<0$ then $Q$ has only one positive root $X_+$. We set $k_+\equiv X_+^{1/2}$. In this regime, unstable modes go from $k=0$ to $k=k_+$ and we numerically observe either coarsening, or pattern formation depending on the sign of $q_1$, as shown in Fig.~\ref{fig:plotRegime2a_2b}.  When $\omega$ is non zero, we sit in regime (ii) where $q_2<0$ is equivalent to 
\begin{align}
(2s-1)^2>\frac{1}{B\phi_0^2}\left(1+\frac{B\rho_0}{r}\right)^2\left(1+\frac{r}{B\rho_0}\right),
\label{eq:condition_on_s}
\end{align}
which surprisingly does not depend upon the dynamical parameters. Physically, the instability comes from the frustrated field $\phi$ whose value at rest and without particles is $0$, different from $\phi_h$ in the presence of particles with non-symmetric flipping rates. This regime is referred to as type $\mathrm{II}_s$ in~\cite{cross_93}.
\item case (iii): $q_1^2/4\geq q_2>0$; patterns appear at finite wavelength (referred to as type $\mathrm{I}_s$ in \cite{cross_93}). We have $X_-,X_+>0$ and we set $k_\pm\equiv X_\pm^{1/2}$. The eigenvalue $\sigma_3(k)$ is positive for $k\in[k_-,k_+]$. At the onset of instability, the only growing mode is indexed by $k_c$ with, at the threshold, $k_c=k_-=k_+$. 
\end{itemize}
Note also that when $\omega=0$ we observe an equilibrium coarsening of the two populations of particles, under the condition $q_1<0$ (which is equivalent to sitting in the equilibrium ordered phase).

In summary, (ii) and (iii) are the two regimes where the homogeneous state is destabilized.
For nonzero flipping rates, the only way to transition from regime (i) to regime (ii), or from regime (iii) to (ii), is by changing the equilibrium parameters, namely $r$, $B$, $\rho_0$, $\phi_0$ and $s$. By contrast, at fixed equilibrium parameters,  we transition from regime (i) to regime (iii) by changing $\omega$ or $\mu$. In the following, we will focus on the transition caused by a change in the dynamics, and consequently, on instabilities starting at finite wavelength.

We begin our analysis with the simpler $s=1/2$ symmetric case, where the number of particles of each spin is identical in the steady-state. This ensures, after Eq.~\eqref{eq:condition_on_s}, that we are always in the pattern forming regime (iii). The matrix $M$ is now block diagonal and eigenvalues can be cast in a compact form:
\begin{align}
\sigma_1&=-\mu k^2,\\
\sigma_2&=\frac{1}{2}\left(-\tilde{r}-\omega-(1+\mu)k^2-\sqrt{\Lambda}\right),\\
\sigma_3&=\frac{1}{2}\left(-\tilde{r}-\omega-(1+\mu)k^2+\sqrt{\Lambda}\right),
\end{align}
with 
\begin{align}
\Lambda=[k^2(\mu-1)+\omega-\tilde{r}]^2+4\mu k^2 \rho_0 B^2\phi_0^2>0.
\end{align}
For the purpose of discussion we use $\omega=\alpha+\gamma$ as the control parameter. Physically, we recall that for high flipping rates, the system remains homogeneous since particles locally efficiently mix, whereas for $\omega= 0$, the system undergoes an equilibrium coarsening (see phase diagram Fig.~\ref{fig:phaseDiag_B_omega}).
Solving $q_1^2-4q_2=0$ yields a critical value of $\omega$:
\begin{align}
\omega_c=\mu(B\phi_0\sqrt{\rho_0}-\sqrt{\tilde r})^2,
\label{eq:omega_c_definition}
\end{align}
below which the homogeneous system is no longer stable.
To study the system close to this transition, we write $\omega=\omega_c-\eps^2$, where the distance to the threshold $\eps^2>0$ becomes our control parameter.
Since we sit in regime (iii), destabilization occurs at a mode $k_c=k_\pm>0$ when $\eps=0$. Thus, when $\omega<\omega_c$, $\sigma_3(k)\geqslant0$ for $k\in[k_-,k_+]$, with
\begin{align}
k_\pm=\frac{\sqrt{\rho_0 B^2 \phi_0^2-\tilde{r}-\Omega\pm\sqrt{(\rho_0 B^2\phi_0^2-\tilde{r}-\Omega)^2-4\tilde{r}\Omega}}}{\sqrt{2}},
\end{align}
and where $\Omega=\omega/\mu$.
The condition of existence of the $k_\pm$ modes (namely that $X_\pm$ are real) is given by
\begin{align}
B\phi_0\sqrt{\rho_0}\geqslant\sqrt{r+B\rho_0}+\sqrt{\omega/\mu}.
\label{eq:condition_kc}
\end{align}
At $\omega=\omega_c$ equality is achieved in Eq.~\eqref{eq:condition_kc} and this allows us to infer the critical wavelength of patterns  $\lambda_c\equiv 2\pi/k_c= 2\pi(\tilde{r}\omega_c/\mu)^{-1/4}$. This suggests that close to the threshold the patterns spatial periodicity $\lambda_{ p}$ could be the combination $\lambda_{ p}= 2\pi/k_{ p}=2\pi(\tilde{r} \omega/\mu)^{-1/4}$.  

\begin{figure}
\includegraphics[width=.9\columnwidth]{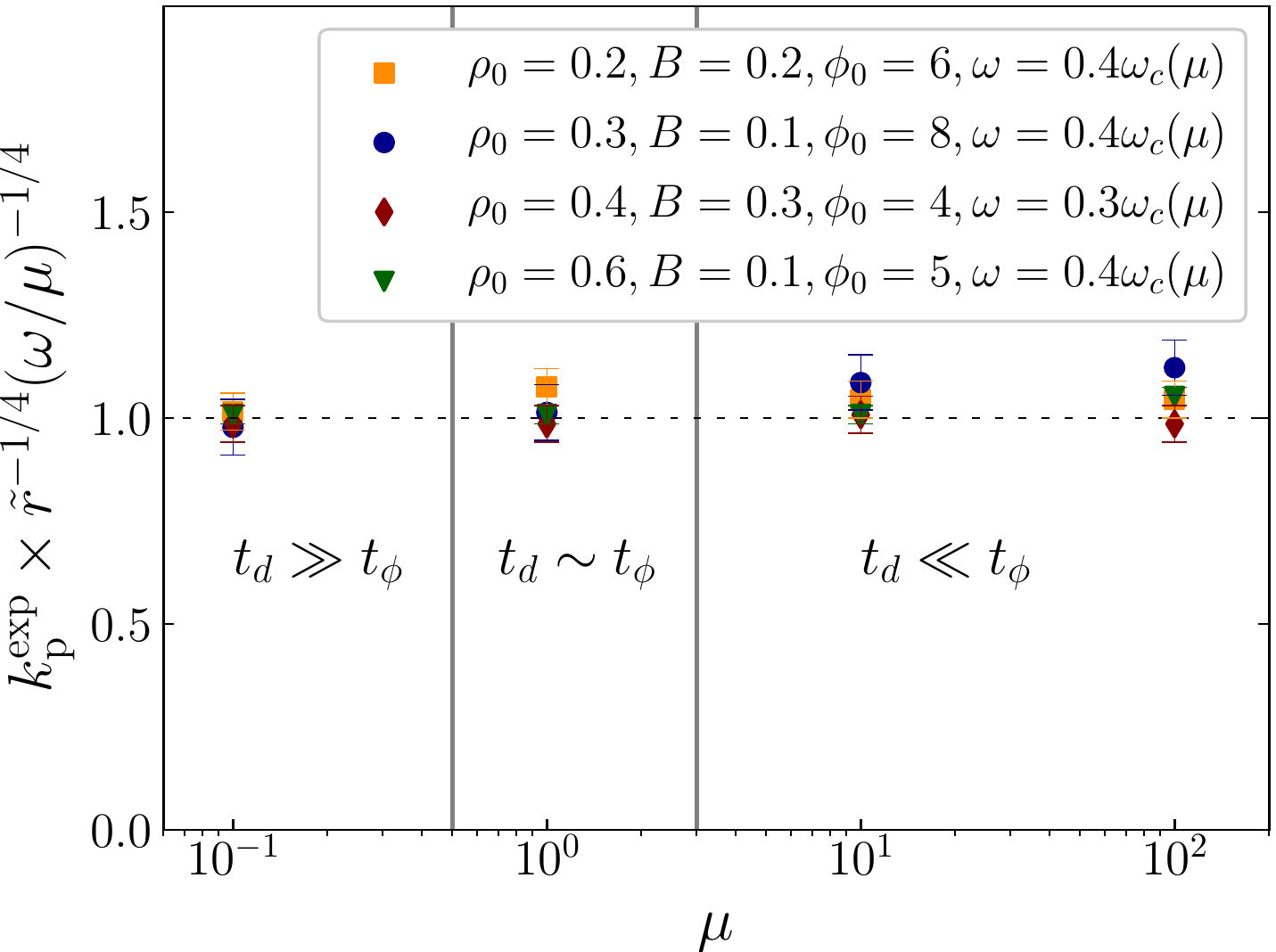}
\caption{Normalized wavenumber $k_p^\mathrm{exp}(\tilde r\omega/\mu)^{-1/4}$ measured from the simulation as a function of the mobility $\mu$ of the particles.  The wavenumber is given by the pattern wavelength and reads $k_p^\mathrm{exp}=2\pi/\lambda_p^\mathrm{exp}$.
Fixed parameters: $r=0.01$, $s=0.5$. 
We vary $B$, $\phi_0$ and $\rho_0\equiv N/(L_xL_y)$ for each simulation. 
We vary also $\mu$ and $\omega$ such that we always sit in the pattern forming regime. 
In particular, we set $\omega$ as a fraction of $\omega_c=\mu(B\phi_0\sqrt{\rho_0}-\sqrt{\tilde r})^2$. For $\omega=0.3\omega_c$ the relation $k_p^\mathrm{exp}=(\tilde r\omega/\mu)^{1/4}$ is still valid. This equality is no longer true when $\omega\lesssim 0.1\omega_c$. As $\mu$ changes from $10^ {-1}$ to $10^{2}$, we sit in different regimes where the particles are slow or fast with respect to the field dynamics.}
\label{fig:k_vs_mu}
\end{figure}

This prediction has been checked in simulations of a quasi 1D system of size $1000\times 10$ to force pattern formation along one direction, hence allowing us to achieve a good precision on the wavelength. We note on Fig.~\ref{fig:k_vs_mu} that the prediction on the pattern periodicity $\lambda_{ p}$ applies beyond the pattern formation threshold.
Interestingly enough, $\lambda_{p}$  can be expressed as the geometric mean of the renormalized correlation length $\tilde{\xi}$ of the Gaussian field $\phi$ in presence of inclusions and of the typical diffusion length $\ell_d\sim \sqrt{\mu/\omega}$ of a particle between two flips. The formula  $\lambda_{p}\sim(\tilde{\xi}\ell_d)^{1/2}$, expresses, at the level of a cluster, the balance between accretion via interactions vs. loss by diffusion. It would certainly be interesting to see $\lambda_{ p }$ emerge from a handwaving argument. Finally, it is worth noticing that close to threshold the selected wavelength does not depend upon the field mobility: it is only the particles' mobility with respect to the spins' flipping rate that matters. An estimate of the diffusion time of a particle over a characteristic correlation length $\tilde\xi$ of the field is $t_d\approx\tilde{\xi}^2/(2\mu)$. On the other hand, the correlation time $t_\phi$ of the field over a scale $\tilde \xi$ is given by $t_\phi\approx 1/(2 \tilde r)$. Hence particles are fast with respect to the field when $t_d\ll t_\phi$, or $1\ll\mu$. In Fig.~\ref{fig:k_vs_mu}, one can indeed see  that the selected wavelength does not change, whether particles are slow or fast with respect to the field.
We now turn to an analysis of the patterns that form beyond threshold.

\subsection{Weakly nonlinear analysis}
\label{sec:weakly_nonlinear_analysis}
\begin{figure}
\includegraphics[width=.49\columnwidth]{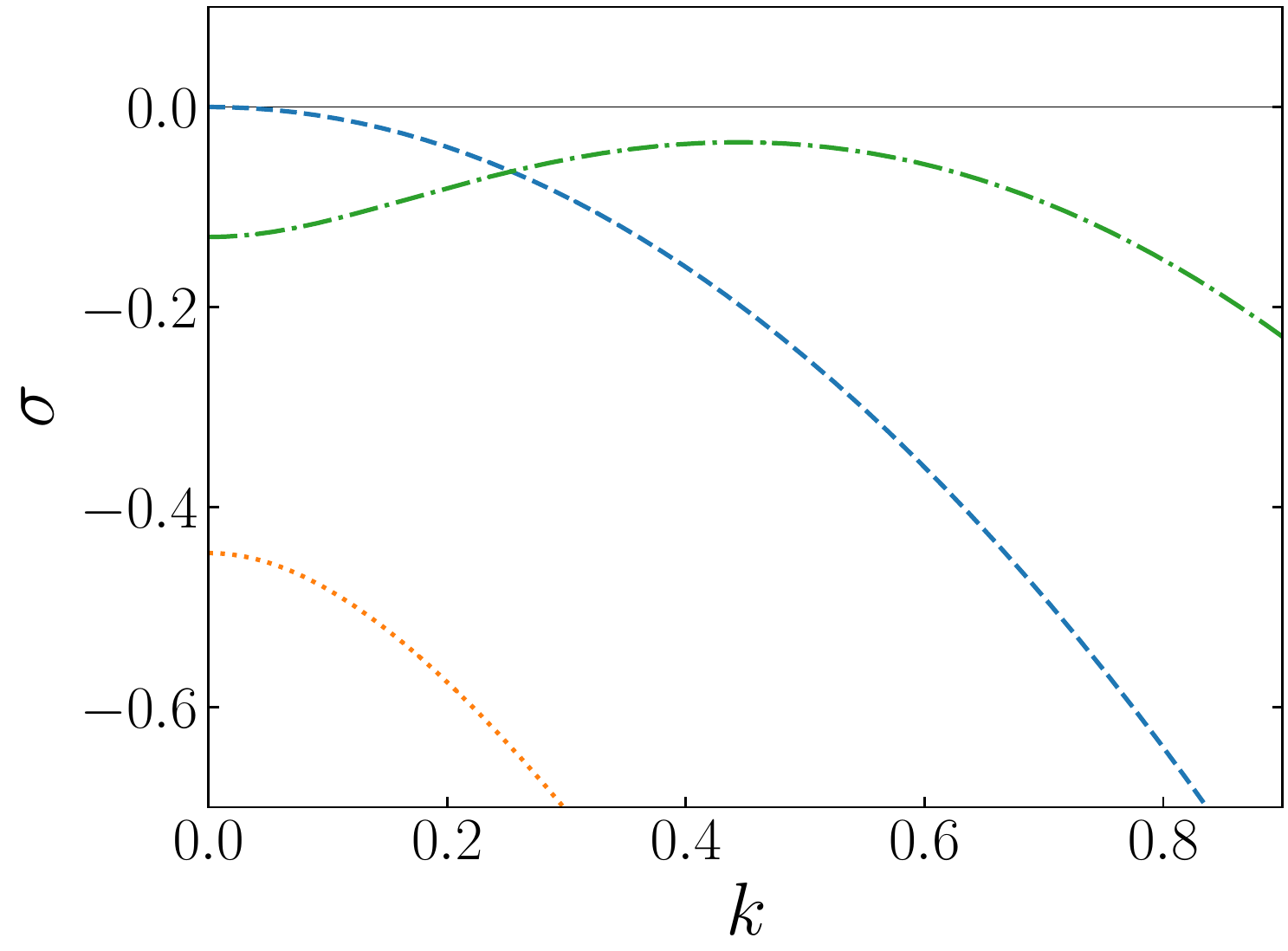}
\includegraphics[width=.49\columnwidth]{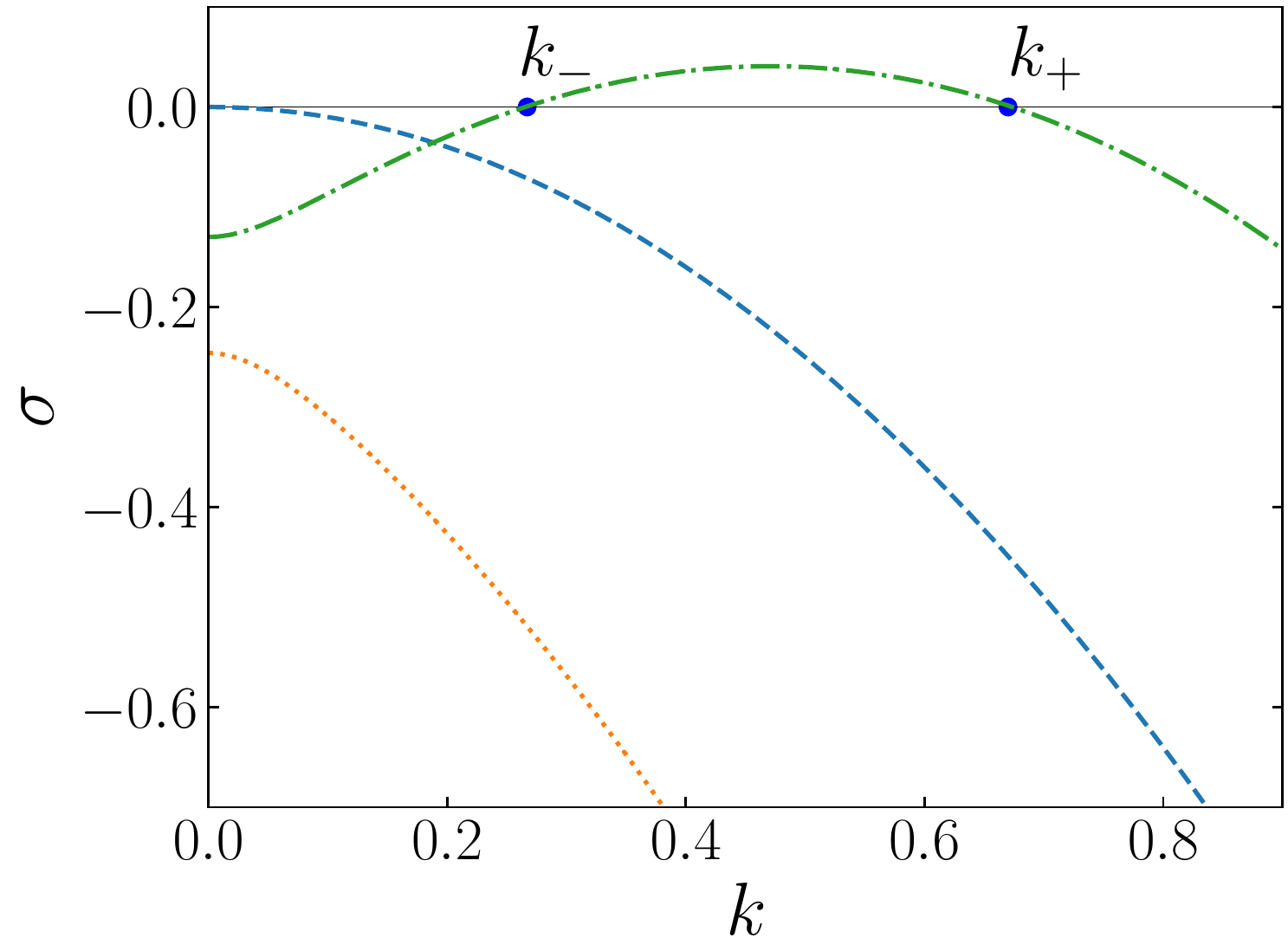}
\caption{Eigenvalues of $M$ as a function of $k$ in the homogeneous stable  regime (left), and in the pattern forming regime (iii) (right). $\sigma_3$ is the highest eigenvalue and is positive in the pattern formation regime.
\textbf{Left:} $\omega>\omega_c$. \textbf{Right:} $\omega<\omega_c$. Parameters: $r=0.01$, $B=0.3$, $\rho_0=0.4$, $\phi_0=5$, $\mu=1$, $s=0.5$. }
\label{fig:eigenvalues}
\end{figure}

In order to gain insight into nonlinear effects at $s=1/2$, we can derive an amplitude equation for the fields by extending the approach of Swift and Hohenberg~\cite{swift_hohenberg_77}. A direct, though naive, way of proceeding would be to extract the equations for the relevant fields for which we can find modes with exponential growth. In particular, in the basis where $M$ is diagonal, there is only one direction (corresponding to eigenvalue $\sigma_3$) along which we observe the temporal growth of the Fourier modes (see Fig.~\ref{fig:eigenvalues}).
The eigen-fields are given by the LSA. By denoting $D=P^{-1}MP$ where the transformation matrix $P$  writes
\begin{align}
&P=\begin{pmatrix}
1 & 0 & 0\cr
0 & a(k) & b(k)\cr
0 & 1 & 1
\end{pmatrix},
\end{align}
with 
\begin{align}
a(k)=\frac{(1-\mu)k^2+\tilde{r}-\omega-\sqrt{\Lambda}}{2B\phi_0},\\
b(k)=\frac{(1-\mu)k^2+\tilde{r}-\omega+\sqrt{\Lambda}}{2B\phi_0},
\end{align}
we define $(U,V,W)^T=P^{-1}(\rho_1,\psi_1,\phi_1)^T$, with $\rho_1$, $\psi_1$ and $\phi_1$ which are no longer infinitesimally small perturbations.
The new fields $(U,V,W)$ now verify $U\sim e^{\sigma_1 t}$, $V\sim e^{\sigma_2 t}$ and $W\sim e^{\sigma_3 t}$. 
However, we have to be careful: the existence of the conserved quantity $\rho$ implies that the mode $k=0$ is marginal for the field $U$ and has to be taken into account~\cite{coullet_1985,riecke_self-trapping_1992, matthews_2000}. 
Indeed nonlinear terms couple modes $k\to 0$ with modes $k\sim k_c$ in products $O(UW)$.
In the absence of a marginal growth, it would be correct to focus only on modes with exponential growth around $k_c$ and we could expand $\sigma_1$, $\sigma_2$ and $\sigma_3$ around $\eps=0$ and $k\sim k_c$. 
This would lead to neglect terms of the form $V^qW^p$ and $U^qW^p$ (for $q\geqslant1$ and $p\geqslant0$) since they exponentially go to $0$ when $k\sim k_c$. Interestingly, one of the erroneous conclusions we would arrive at is that square patterns could never be stable, in conflict with observations of PDEs' solution. 
Now keeping both relevant fields $U$ and $W$, the evolution equations read
\begin{align}
\partial_t U&=\sigma_1 U+\mathcal{N}_1(U,W),\label{eq:nonlinear_U}\\
\partial_t W&=\sigma_3 W+\mathcal{N}_2(U,W),\label{eq:nonlinear_W}
\end{align}
where $\mathcal{N}_1$ and $\mathcal{N}_2$ are nonlinear operators that couple $U$ and $W$. To lowest order in $\eps$, we find that $\mathcal{N}_1$ contains terms $\sim O(W^2)$ and that $\mathcal{N}_2$ contains terms $\sim O(UW)$. Thus, $U$ will saturate to $O(W^2)$, which renormalizes $O(W^3)$ terms in $\mathcal{N}_2$. Of course, this previous analysis holds for the case $s=1/2$, where equations are invariant upon the $(\rho,\psi,\phi)\to(\rho,-\psi,-\phi)$ symmetry. If $s\neq 1/2$, new terms appear in nonlinear equations \eqref{eq:nonlinear_U} and \eqref{eq:nonlinear_W}. To lowest order, new terms in $\mathcal{N}_2$ will take the form $O((2s-1)W^2)$, $O((2s-1)^2 W)$ and $O((2s-1)^2 W^3)$ such that the resulting equations remain consistent with the symmetry $(\rho,\psi,\phi,s)\to(\rho,-\psi,-\phi,\frac 12-s)$. These terms are directly responsible for the stability of hexagonal patterns~\cite{cox_2003} {as confirmed in the numerical simulations (see Fig.~\ref{fig:snapshots_hexagons})}. We are now going to derive, in a pragmatic fashion, the amplitude equations for the fields when $s=1/2$. Our derivation is inspired by the methods presented in~\cite{winterbottom_2005}.

We sit in the regime where patterns appear and we ask what the selected patterns beyond threshold are?
Weakly nonlinear analysis begins by noticing that $\sigma_3\sim \eps^2-a(k_c^2-k^2)^2$, above the pattern threshold. We work in units of the the slow time scale by defining $T=\eps^2 t$; similarly in units of the large wavelength scale, we set $X=\eps x$ which governs the evolution of the envelope of the fast growing patterns that develop at wavenumber $k_c$. 
The stationary homogeneous solution is perturbed when $\omega<\omega_c$. We expand the fields in a power series of the parameter $\eps$. 
In the symmetric case $s=1/2$, the stable patterns are usually rolls and squares~\cite{cox_2003}.
To study their relative stability in two dimensions, using $\psi_h=\phi_h=0$, the expansion for the fields reads
\begin{align}
\rho&=\rho_0+\eps^2R(X,Y)+\sum_{n=1}^\infty	\eps^n \rho_n(x,y,X,Y),\\
\psi&=\sum_{n=1}^\infty\eps^n \psi_n(x,y,X,Y),\\
\phi&=\sum_{n=1}^\infty\eps^n \phi_n(x,y,X,Y),
\end{align}
with $Y=\eps y$, and where $R(X,Y)$ is the large scale envelope of the marginal mode $k=0$ that has to be added in the expansion of the conserved field with the appropriate scaling to obtain a closure relation (see \cite{coullet_1985,matthews_2000}). 
The functions $\rho_n$, $\psi_n$ and $\phi_n$ are expected to be products of slow dynamics envelopes and fast growing patterns. These considerations allow us to write differential operators with the chain rule, namely, $\partial_x\to \partial_x+\eps\partial_X$, $\partial_y\to \partial_y+\eps\partial_Y$ and $\partial_t\to \eps^2\partial_T$. Next, we expand Eq.~\eqref{eq:evolution_rho}, \eqref{eq:evolution_psi}, \eqref{eq:evolution_phi} to successive orders to get a closed set of equations. In the canonical case of the Swift-Hohenberg equation \cite{swift_hohenberg_77,cross_93}, the closed relation for the lowest order amplitude is obtained to order $O(\eps^3)$. In our case of existence of a conserved quantity, we have to extract field evolution up to order $O(\eps^4)$ to get a closed system of equations. We are going to proceed recursively to extract the evolution of the fields. To order $O(\eps)$, we find the following system:
\begin{align}
\mathcal{L}
\begin{pmatrix}
   \rho_1  \\
   \psi_1\\
   \phi_1 
\end{pmatrix}=0,
\end{align}
with 
\begin{align}
\mathcal{L}=
\begin{pmatrix}
   \mu \nabla^2 & 0 & 0  \\
   0 &  \mu \nabla^2 -\omega_c & -\mu B\rho_0\phi_0\nabla^2 \\
   0 & B\phi_0  & \nabla^2-\tilde r   
\end{pmatrix},
\end{align}
and where $\nabla^2=\partial_x^2+\partial_y^2$. The solution of this system reads
\begin{align}
\rho_1(x,y,X,Y)=&0,\\
\psi_1(x,y,X,Y)=&P_1(X,Y)e^{i k_c x}+Q_1(X,Y)e^{i k_c y}+c.c.,\\
\phi_1(x,y,X,Y)=&\lambda_1 \psi_1,
\end{align}
with $\lambda_1=B\phi_0/(\tilde r+k_c^2)$ a simple scalar coefficient, $P_1$ and $Q_1$ are scalar functions of $X$ and $Y$, and where $c.c.$ stands for complex conjugate.
To order $O(\eps^2)$, the system we arrive at is
\begin{align}
\mathcal{L}
\begin{pmatrix}
   \rho_2  \\
   \psi_2\\
   \phi_2 
\end{pmatrix}=\zeta^{(2)}(\rho_1,\psi_1,\phi_1)
\label{eq:amplitude_order2}
\end{align}
with $\zeta^{(2)}=(\zeta^{(2)}_1,\zeta^{(2)}_2,\zeta^{(2)}_3)^T$ a vector which only depends on first order fields. The components of $\zeta^{(2)}$ read
\begin{align}
\begin{split}
\zeta^{(2)}_1=&\mu B [\phi_0(\psi_1\partial_x^2\phi_1+\partial_x\psi_1\partial_x\phi_1)\\
&+\phi_0(\psi_1\partial_y^2\phi_1+\partial_y\psi_1\partial_y\phi_1)\\
&-\rho_0(\partial_x\phi_1)^2-\rho_0\phi_1\partial_x^2\phi_1\\
&-\rho_0(\partial_y\phi_1)^2-\rho_0\phi_1\partial_y^2\phi_1],\end{split}\\
\zeta^{(2)}_2=&2\mu[B\rho_0\phi_0(\partial_{xX}+\partial_{yY})\phi_1-(\partial_{xX}+\partial_{yY}) \psi_1],\\
\zeta^{(2)}_3=&-2 (\partial_{xX}+\partial_{yY}) \phi_1 .
\end{align}
The solution of Eq.~\eqref{eq:amplitude_order2} reads:
\begin{align}
\begin{split}
\rho_2=&\lambda_2 P_1^2(X,Y)e^{2i k_c x} 
+ \lambda_2 Q_1^2(X,Y)e^{2i k_c y}\\
 &+ 2\lambda_2 P_1(X,Y)Q_1(X,Y)e^{i k_c x+ik_c y}\\
 &+2\lambda_2 P_1(X,Y)Q_1^*(X,Y)e^{i k_c x-ik_c y}+c.c.,
\end{split}\\
\psi_2=& P_2(X,Y)e^{i k_c x}+Q_2(X,Y)e^{i k_c y}+c.c.,\\
\begin{split}
\phi_2=& \lambda_1[P_2(X,Y)+2ik_c\frac{\lambda_1}{\tilde r+k_c^2}\partial_X P_1(X,Y)]e^{i k_c x}\\
&+\lambda_1[Q_2(X,Y)+2ik_c\frac{\lambda_1}{\tilde r+k_c^2}\partial_Y Q_1(X,Y)]e^{i k_c y}\\
&+c.c.,
\end{split}
\end{align}
with $\lambda_2\equiv-B\lambda_1(\rho_0\lambda_1-\phi_0)/2=\lambda_1^2(k_c^2+r)/2$ and where $Q_1^*$ is the complex conjugate of $Q_1$. At $O(\eps^3)$, we find the equation on $P_1$ (resp. $Q_1$) by collecting the terms proportional to $e^{ik_c x}$ (resp. $e^{ik_c y}$) in the two equations involving $\psi_3$ and $\phi_3$. A linear combination of these equations allows us to eliminate the second order amplitudes $P_2$ and $Q_2$ and to extract the time evolution on $P_1$ and $Q_1$. 
To order $O(\eps^3)$ we arrive at the following equations:
\begin{align}
\begin{split}
\partial_T P_1=& a_1 P_1 + a_2 \partial_{XX}P_1\\
 &- a_3 |P_1|^2 P_1-a_4 |Q_1|^2 P_1-a_5 R P_1 
\end{split}\\
\begin{split}
\partial_T Q_1=& a_1 Q_1 + a_2 \partial_{YY}Q_1\\
&- a_3 |Q_1|^2 Q_1-a_4 |P_1|^2 Q_1-a_5 R Q_1
\end{split}
\end{align}
with 
\begin{align}
a_1&=\frac{\tilde r}{\mu k_c^2+\tilde r},\\
a_2&=\frac{4\mu k_c^2 \sqrt{\tilde r} }{B \phi_0 (\mu k_c^2 +\tilde r)\sqrt{\rho_0}},\\
a_3&=\frac{B^3 \mu  k_c^2 \phi_0^2 \left(B \phi_0^2 \left(\left(k_c^2+\tilde r\right)^2-B^2 \rho_0^2\right)+2 \left(k_c^2+\tilde r\right)^2\right)}{2 \left(k_c^2+\tilde r\right)^2 \left(\mu  k_c^2 \rho_0B^2  \phi_0^2+\left(k_c^2+\tilde r\right)^2\right)}\\
a_4&=\frac{2  \mu  k_c^2 B^3\phi_0^2 \left(B^2 \rho_0 \phi_0^2 \left(k_c^2+ r\right)+\left(k_c^2+\tilde r\right)^2\right)}{(k_c^2+\tilde r)^2 \left(\mu  k_c^2 \rho_0B^2  \phi_0^2 +\left(k_c^2+\tilde r\right)^2 \right)},\\
a_5&=-\frac{ \mu  k_c^2 B^2\phi_0^2 \left(k_c^2+ r\right)}{\mu  k_c^2 \rho_0 B^2  \phi_0^2+\left(k_c^2+\tilde r\right)^2}.
\end{align}
To order $O(\eps^4)$, we close the system with the time evolution of $R(X,Y)$, which is obtained by extracting coefficients of the mode $k=0$ in the $\rho_4$ equation. We obtain
\begin{align}
\begin{split}
\partial_T R=& \mu \tilde\nabla R+ \kappa_1(\partial_X^2|P_1|^2+\partial_Y^2|Q_1|^2)\\
&+\kappa_2(\partial_Y^2|P_1|^2+\partial_X^2|Q_1|^2),
\end{split}
\end{align}
with 
\begin{align}
\kappa_1&=\frac{\mu B^2 \phi_0^2(k_c^2-r)}{(k_c^2+\tilde r)^2},\\
\kappa_2&=-\frac{\mu B^2 \phi_0^2(k_c^2+r)}{(k_c^2+\tilde r)^2}.
\end{align}
We then perform a change of scale to fall back onto the canonical system found in \cite{cox_2003,matthews_2000}. Setting
\begin{align}
 &T\to T/a_1;\\
 &X \to X\sqrt{a_2/a_1}\quad ;\quad Y \to Y\sqrt{a_2/a_1};\\
 &P_1\to P_1 \sqrt{a_1/a_3}\quad;\quad Q_1\to Q_1 \sqrt{a_1/a_3};\\
 & R\to R\, a_1/a_5,
\end{align}
we define
\begin{align}
g&=\frac{a_4}{a_3}>0,\\
b_1&=\frac{\mu}{a_2}>0,\\
b_2&=\frac{(\kappa_1+\kappa_2)a_5}{2a_2 a_3}>0,\\
b_3&=\frac{(\kappa_1-\kappa_2)a_5}{2a_2 a_3}<0,
\end{align}
and we finally obtain 
\begin{align}
\begin{split}
\partial_T P_1=& P_1 + \partial_{XX}P_1\\
 &-  |P_1|^2 P_1-g |Q_1|^2 P_1- R P_1 
\end{split}\label{eq:amplitude_1}\\
\begin{split}
\partial_T Q_1=&  Q_1 + \partial_{YY}Q_1\\
&-  |Q_1|^2 Q_1 - g|P_1|^2 Q_1- R Q_1
\end{split}\label{eq:amplitude_2}\\
\begin{split}
\partial_T R=& b_1 \tilde \nabla^2 R +b_2 \tilde \nabla(|P_1|^2+|Q_1|^2)\\
&+b_3(\partial_X^2-\partial_Y^2)(|P_1|^2-|Q_1|^2),
\end{split}
\label{eq:amplitude_3}
\end{align}
with $\tilde \nabla\equiv \partial_X^2+\partial_Y^2$.

\subsection{Roll and square stability}
\label{sec:rollsVSsquares}
\begin{figure}
\includegraphics[width=0.9\columnwidth]{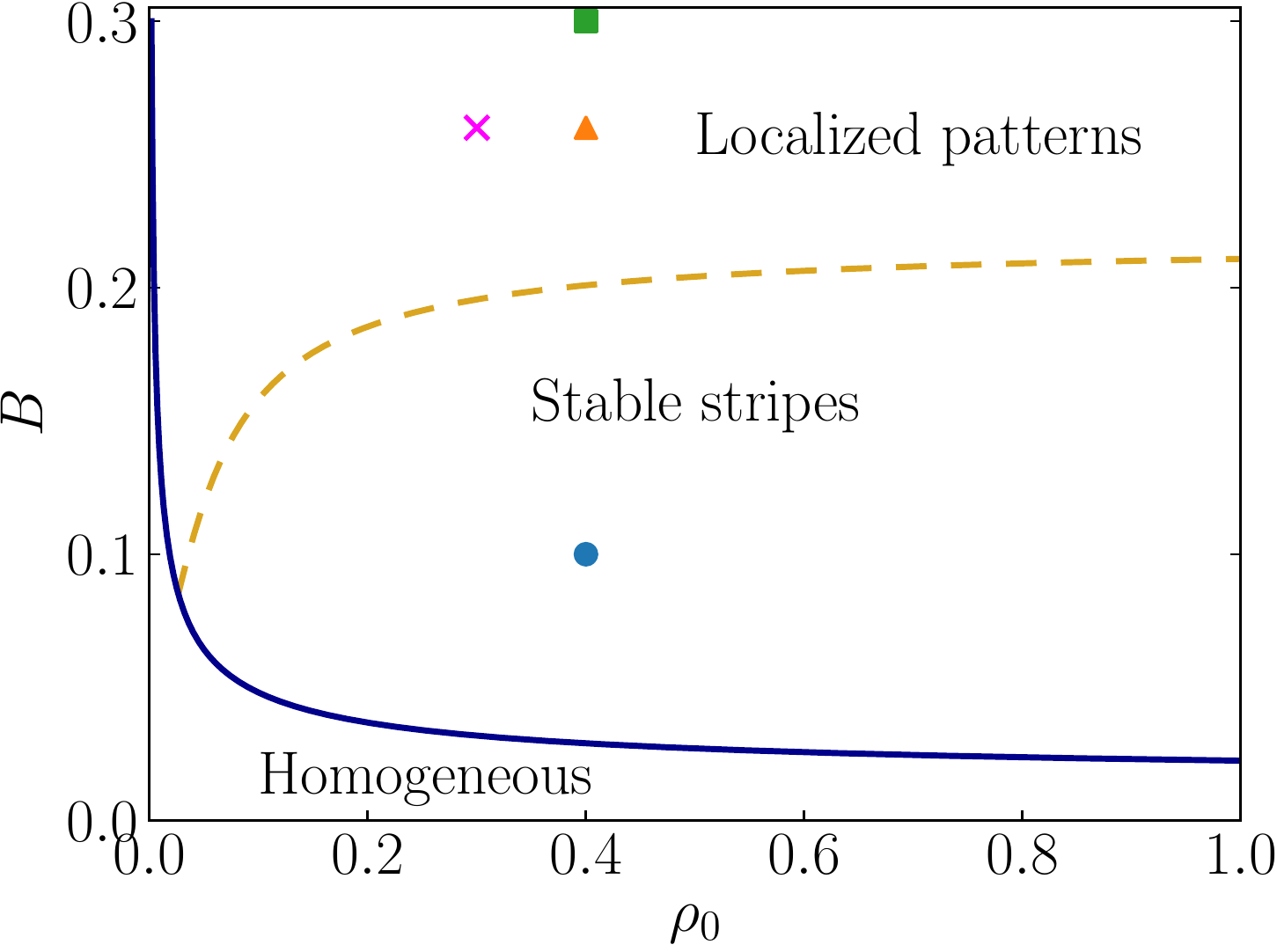}
\caption{Pattern phase diagram predicted from the weakly nonlinear analysis. Blue line: existence of critical frequency $\omega_c$. Yellow dashed line: $g=1$ is the stability boundary of roll with respect to squares when $\omega\lesssim \omega_c$. 
Blue bullet, orange triangle and green square correspond to different simulations performed with same density $\rho_0=0.4$ but different $B$ and $\omega$ to keep $\omega=0.9\omega_c$. 
Blue bullet: $B=0.1$, we observe stripes in PDE solution and large clusters in Monte Carlo simulations. 
Orange triangle: $B=0.26$, we observe pattern localization in PDE solution and small clusters in the simulation. 
Green square: $B=0.3$, we observe square patterns in PDE solution and small clusters in the simulation. The blue bullet, the orange triangle and the green square correspond to the patterns shown in Figs.~\ref{fig:snapshots_rolls_localized_and_squares}a, d and c, respectively.
Magenta cross: $L_x=L_y=175$ yields square patterns similar to Fig.~\ref{fig:snapshots_rolls_localized_and_squares}c, whereas $L_x=L_y=300$ yields localized stripes similar to Fig.~\ref{fig:snapshots_rolls_localized_and_squares}d.
}
\label{fig:phaseDiagram_instabilities}
\end{figure}
\begin{figure}
\includegraphics[width=.48\columnwidth]{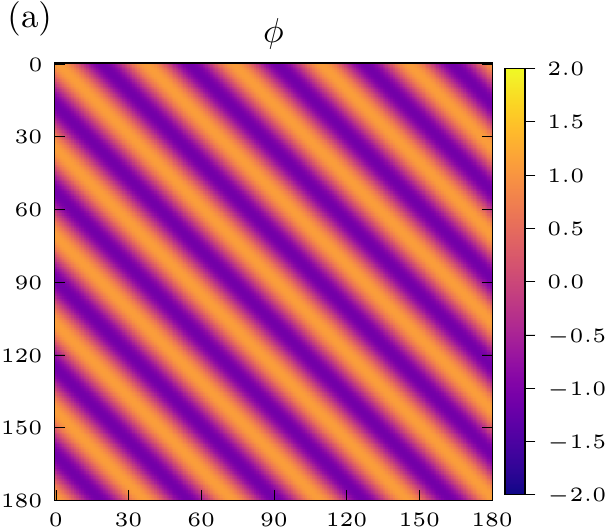}
\includegraphics[width=.48\columnwidth]{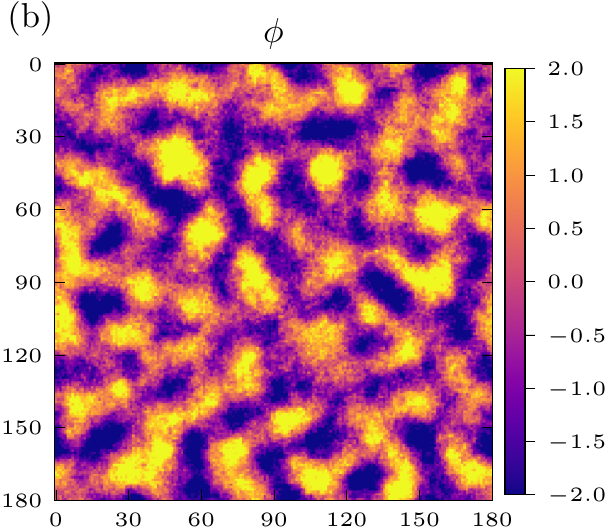}
\\
\includegraphics[width=.32\columnwidth]{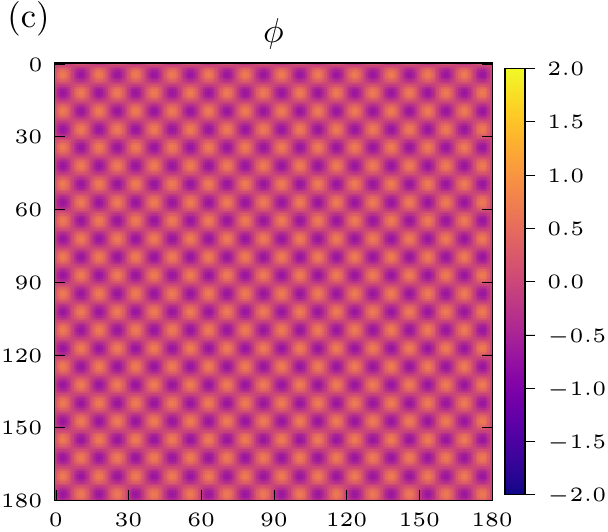}
\includegraphics[width=.32\columnwidth]{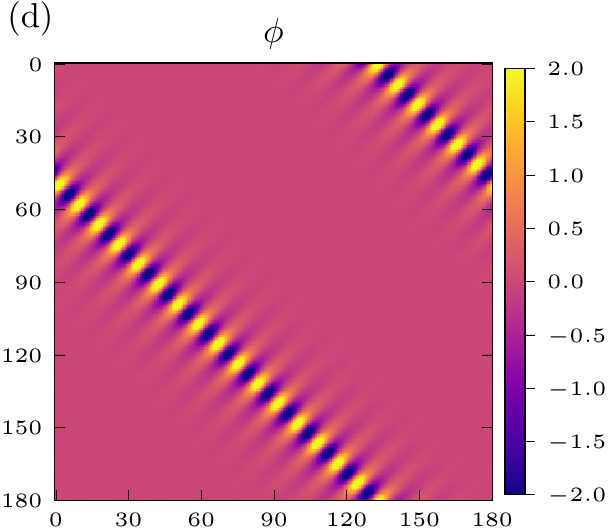}
\includegraphics[width=.32\columnwidth]{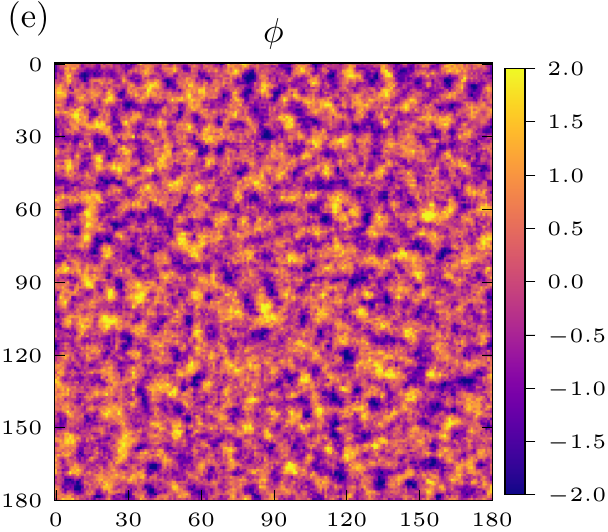}
\\
\includegraphics[width=.48\columnwidth]{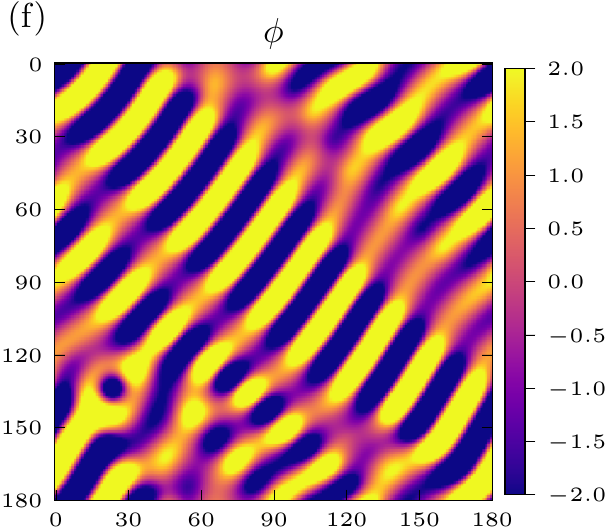}
\includegraphics[width=.48\columnwidth]{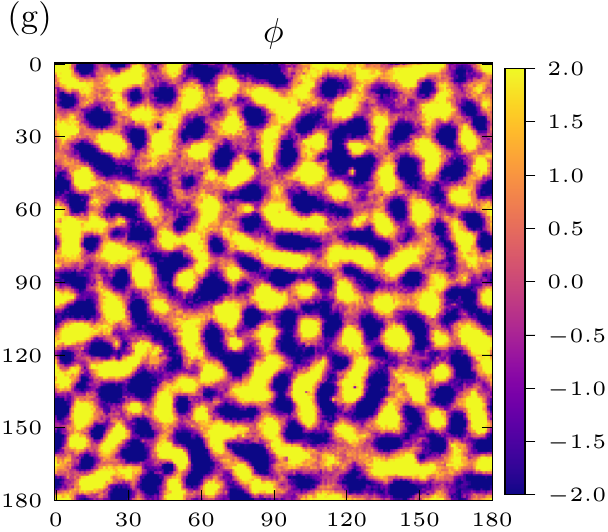}
\caption{ Simulation and solution of the PDEs close to pattern apparition threshold (a,b,c,d,e), and for $\omega$ far below $\omega_c$ (f,g). Shared parameters: $r=0.01$, $\rho_0=0.4$, $\phi_0=8$, $\mu=5$. 
For (a), (b), (c), (d) and (e) we have $\omega=0.9\omega_c$.  
For (f) and (g) we have $\omega=0.2\omega_c$.
(a) $B=0.1$, PDEs solution shows stripes in agreement with Fig.~\ref{fig:phaseDiagram_instabilities}. 
(b) $B=0.1$, Monte Carlo (MC) simulation shows structures of same size.
(c) $B=0.3$, square pattern (PDE).
(d) $B=0.26$, pattern localization (PDE).
(e) $B=0.26$, micro clusters (MC).
(f) $B=0.22$, stripes and localized clusters (PDE).
(g) $B=0.22$, stripes, clusters and lumps (MC).}
\label{fig:snapshots_rolls_localized_and_squares}
\end{figure}

As we now deal with amplitude equations \eqref{eq:amplitude_1}, \eqref{eq:amplitude_2} and \eqref{eq:amplitude_3} in a canonical form, the results obtained by \cite{cox_2003} are now directly transposable to our analysis. In particular, we can extract the stability boundaries of roll and square patterns, and predict  modulational instabilities. The outcome of this analysis is that
\begin{itemize}
\item[--] when $b_2+b_3>b_1$, rolls are unstable to one-dimensional disturbances (phase or amplitude modulation along the wave vector of patterns);
\item[--] if $b_2-b_3>b_1$ rolls undergo a two-dimensional instability, which is expressed through a transverse modulation of the rolls; see the dashed line in the phase diagram of Fig~\ref{fig:phaseDiagram_instabilities};
\item[--] if $g>1$, squares are unstable to rolls. Squares also undergo a modulational instability when $b_2>(1+g)b_1/2$ (we do not observe such patterns in the PDE solution); it turns out that in our model the condition for rolls to be unstable to squares is the same as the two-dimensional instability for rolls, and is thus described by the same dashed line in Fig~\ref{fig:phaseDiagram_instabilities}.
\end{itemize}
Since we have $b_3<0$, the condition $b_2-b_3>b_1$ preempts $b_2+b_3>b_1$; it is shown in \cite{cox_2003} that the former then controls pattern formation. This explains why we observe the two-dimensional instability for rolls in Fig~\ref{fig:snapshots_rolls_localized_and_squares}d. In our model, squares and transverse modulated rolls may exist separately at the same point of parameter space but they are ultimately selected by the geometry, the size, and the aspect ratio of the system (see magenta cross, Fig~\ref{fig:phaseDiagram_instabilities}).

\newpage
\section{From equilibrium system to active system}
\label{sec:Mixed_active_eq}

 So far, we have focused on the active system where spin flips are driven by a noise independent of temperature. Our analysis of the corresponding reaction-diffusion equations has shown the existence of a wealth of stationary patterns controlled by the values of the parameters of our model. These patterns simply do not exist in equilibrium when flips are controlled by temperature. To what extent does restoring a fraction of equilibrium spin flips within active flips suppresses the patterns we have obtained? Conversely, is adding a bit of activity over otherwise equilibrium flips sufficient to drive the system to a patterned stationary state? This section is about exploring the model system obtained by interpolating between fully active spin flips and equilibrium ones.
 
To implement both active and temperature controlled flips, the flipping rate $w(S_k, \phi)$ for spin $S_k$ is now the sum of the active rate and the equilibrium rate which depends on the field value at the particle's location $\phi_k\equiv\phi(\bm r_k)$:
\begin{align}
w(S_k,\phi) = \begin{cases}  s\omega +\eta e^{B \phi_k \phi_0}\equiv w_k^- &\mbox{if } S_k=-1 \\
 (1-s)\omega+\eta e^{-B \phi_k \phi_0}\equiv w_k^+ & \mbox{if } S_k=+1 \end{cases},
\end{align}
and where $\eta$ is the equilibrium flipping rate if spins did not interact with the Gaussian field.

\subsection{Mean-field analysis }
 
\begin{figure}
\includegraphics[width=.99\columnwidth]{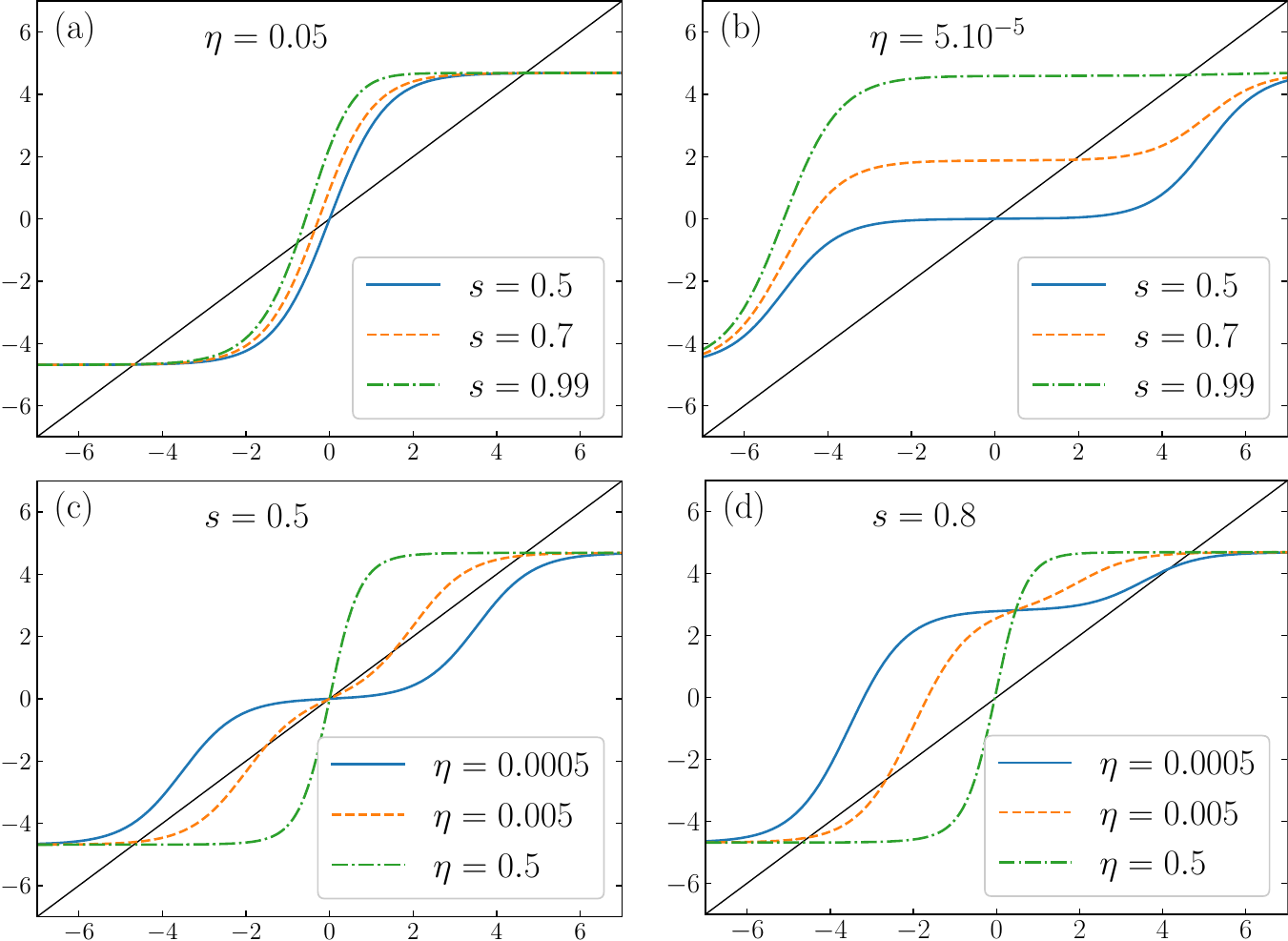}
\caption{Graphic solution of equation $G(\phi_h)=\phi_h$. Parameters: $r=0.01$, $B=0.3$, $\rho_0=0.5$, $\phi_0=5$, $\mu=1$ and $\omega=0.1$. When $\eta\sim\omega$ the solution is close to the equilibrium one and $s$ does not play important role.}
\label{fig:mixed_active_eq}
\end{figure}

The mean-field equations are the same as Eq.~\eqref{eq:evolution_rho+}~--~\eqref{eq:evolution_phi_naive} with $\alpha$ (resp. $\gamma$) changed into $w^-$ (resp. $w^+$):
\begin{align}
\partial_t\rho^+&=
\mu\bm\nabla\cdot[\rho^+\bm\nabla \frac{\partial f_\mathrm{MF} }{\partial \rho^+}]+w^-\rho^--w^+\rho^+,
\label{eq:evolution_rho+_mixed}
\\
\partial_t\rho^-&=\mu\bm\nabla\cdot[\rho^-\bm\nabla \frac{\partial f_\mathrm{MF} }{\partial \rho^-}]
-w^-\rho^-+w^+\rho^+,
\label{eq:evolution_rho-_mixed}\\
\partial_t\phi&= \nabla^2 \phi - r\phi- B \rho^+(\phi-\phi_0)- B \rho^-(\phi+\phi_0).
\label{eq:evolution_phi_naive_mixed}
\end{align}
 First, we search for a homogeneous stationary solution of this system. The self-consistent equation now verified by $\phi_h$ is more involved. We find that a homogeneous solution has
\begin{align}
\rho_h&=\rho_0,\\
\phi_h&=G(\phi_h)\\
\psi_h&=\frac{\tilde r}{B\phi_0}\phi_h,
\end{align}
where the function $G(\phi_h)$ is given by
\begin{align}
G(\phi_h)=\frac{B\rho_0}{ \tilde r}\frac{(2s-1)\omega+2\eta\sinh(B\phi_h\phi_0)}{\omega+2\eta\cosh(B\phi_h\phi_0)}\phi_0.
\end{align}
We show a graphical solution of $G(\phi_h)=\phi_h$ in Fig.~\ref{fig:mixed_active_eq}.
We remark on Fig.~\ref{fig:mixed_active_eq}(a) that the active fraction $s$ does not play a significant role when equilibrium flips are of the same order of magnitude as the active flip $\omega$. 
By contrast, when equilibrium flips are negligible, the homogeneous state is completely controlled by the fraction $s$ (see Fig.~\ref{fig:mixed_active_eq}b). 
The most interesting regime is for $2\eta/\omega\sim 10^{-1}$ where the self-consistent equation has up to five solutions for some parameters (see Fig.~\ref{fig:mixed_active_eq}c), unlike what we can observe in equilibrium (one or three solutions) or for the full active regime (one solution only). Finding out about the relative stability of these solutions comes first. This is the purpose of the following subsection.
\begin{figure*}
\includegraphics[width=0.71\columnwidth]{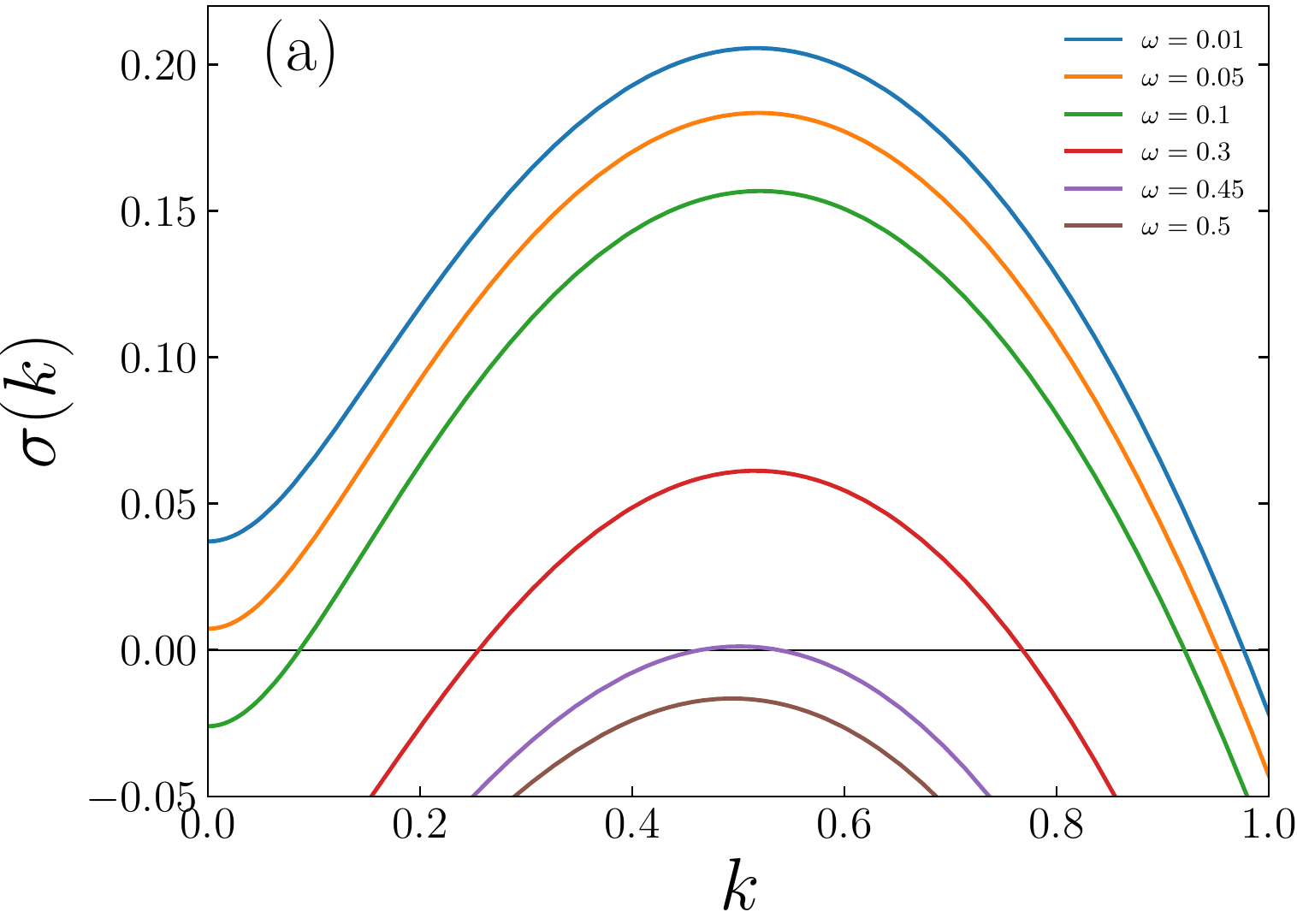}
\includegraphics[width=0.65\columnwidth]{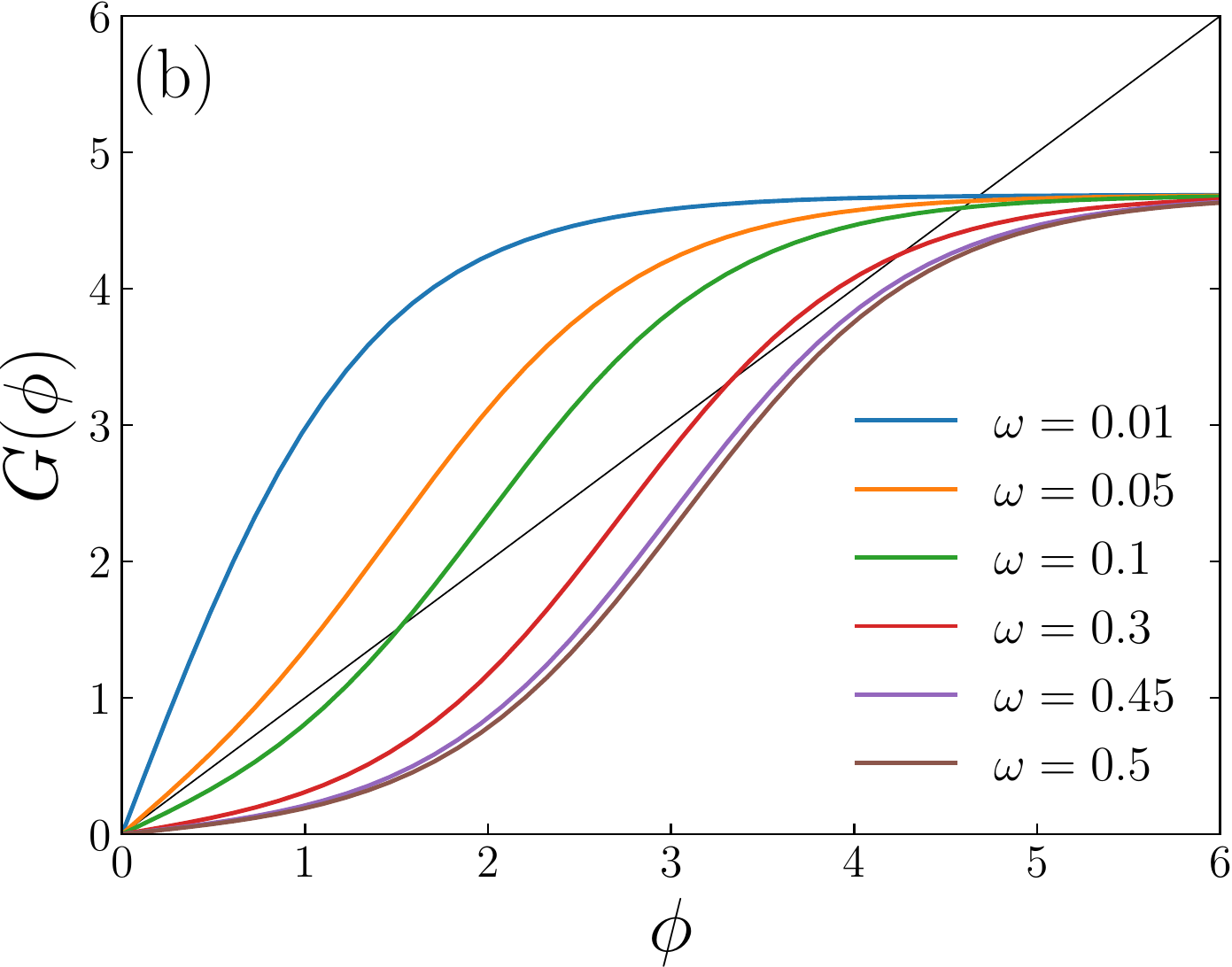}
\includegraphics[width=0.67\columnwidth]{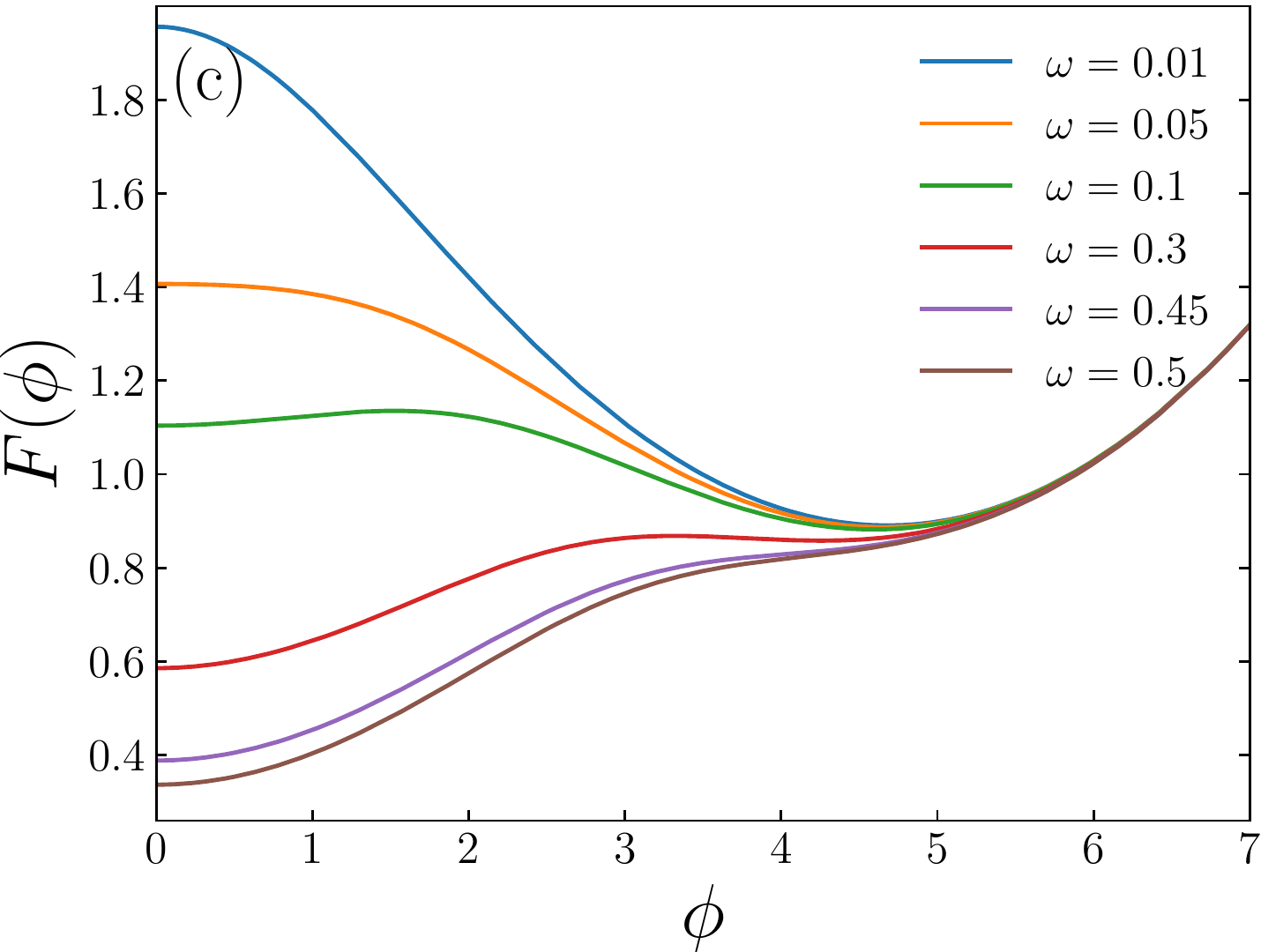}
\caption{For different values of $\omega$: (a) Most unstable eigenvalues from linear instability analysis around $\phi_h=0$. (b) Graphical solving of the self-consistent equation for the `magnetization' field $\phi$. (c) Mean field free energy $F(\phi)$. Starting from the value $\omega=0.5$ and decreasing $\omega$ yields different regimes. We observe successively a homogeneous state with zero magnetization, a patterned phase, and then a homogeneous ferromagnetic phase when such a state has the minimum energy. Other parameters: $r=0.01$, $B=0.3$, $\rho_0=0.5$, $\phi_0=5$, $\mu=1$, $\eta=0.005$ and $s=0.5$.  }
\label{fig:3figures_mixedActiveEq}
\end{figure*}
\begin{figure}
\includegraphics[width=0.7\columnwidth]{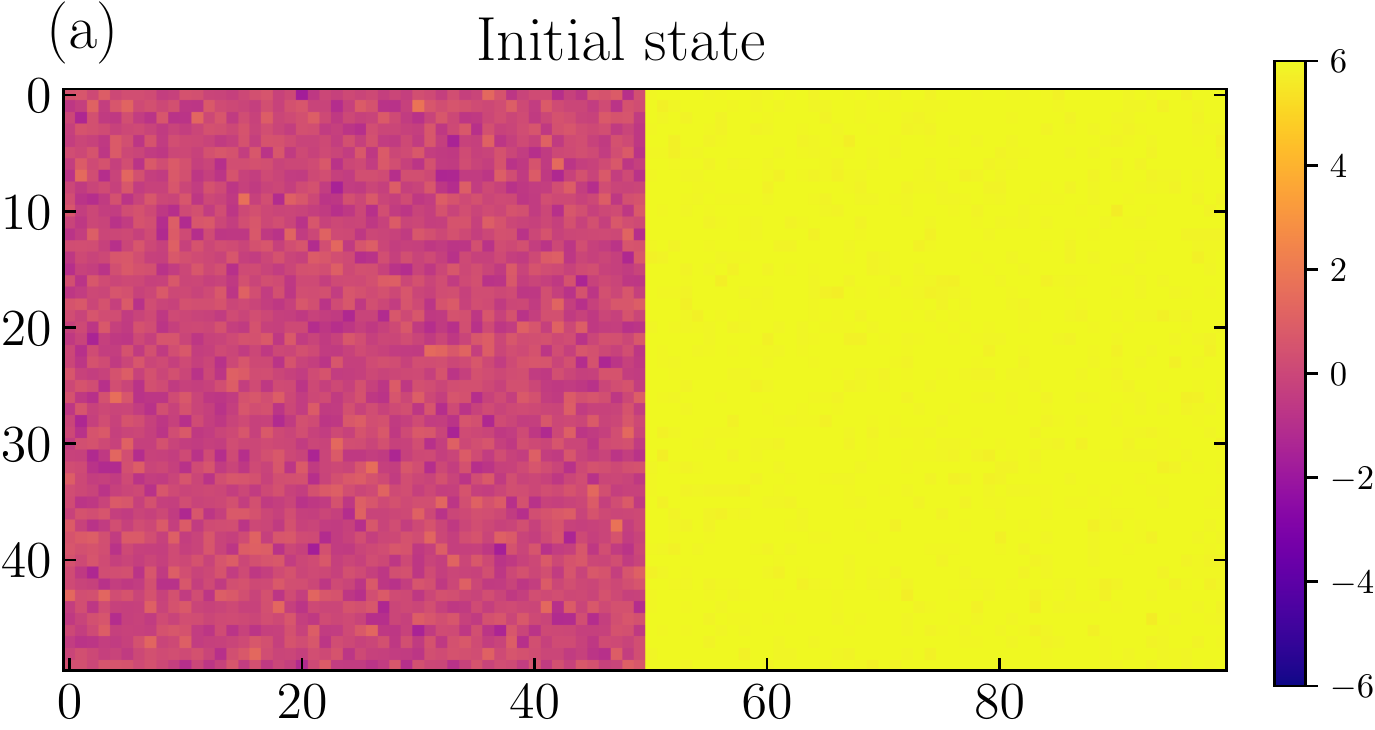}\\
\includegraphics[width=0.7\columnwidth]{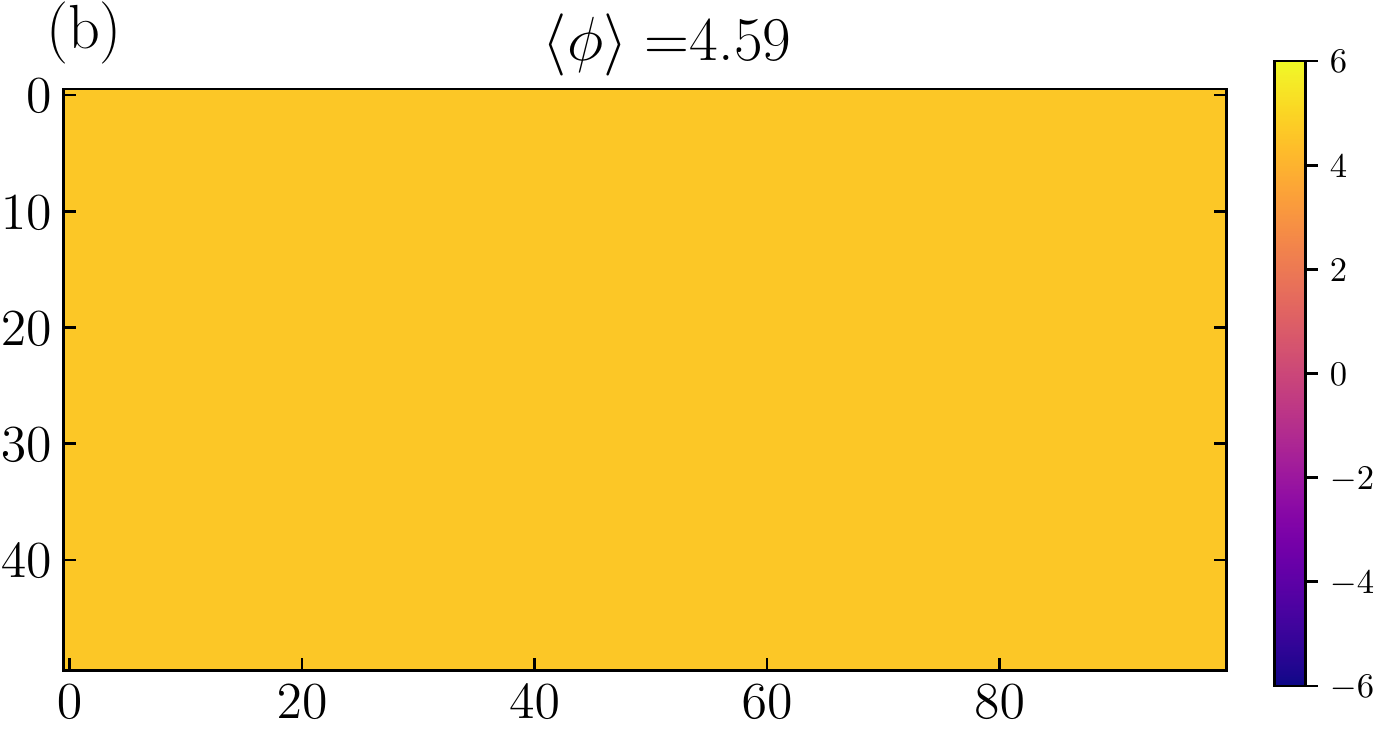}\\
\includegraphics[width=0.7\columnwidth]{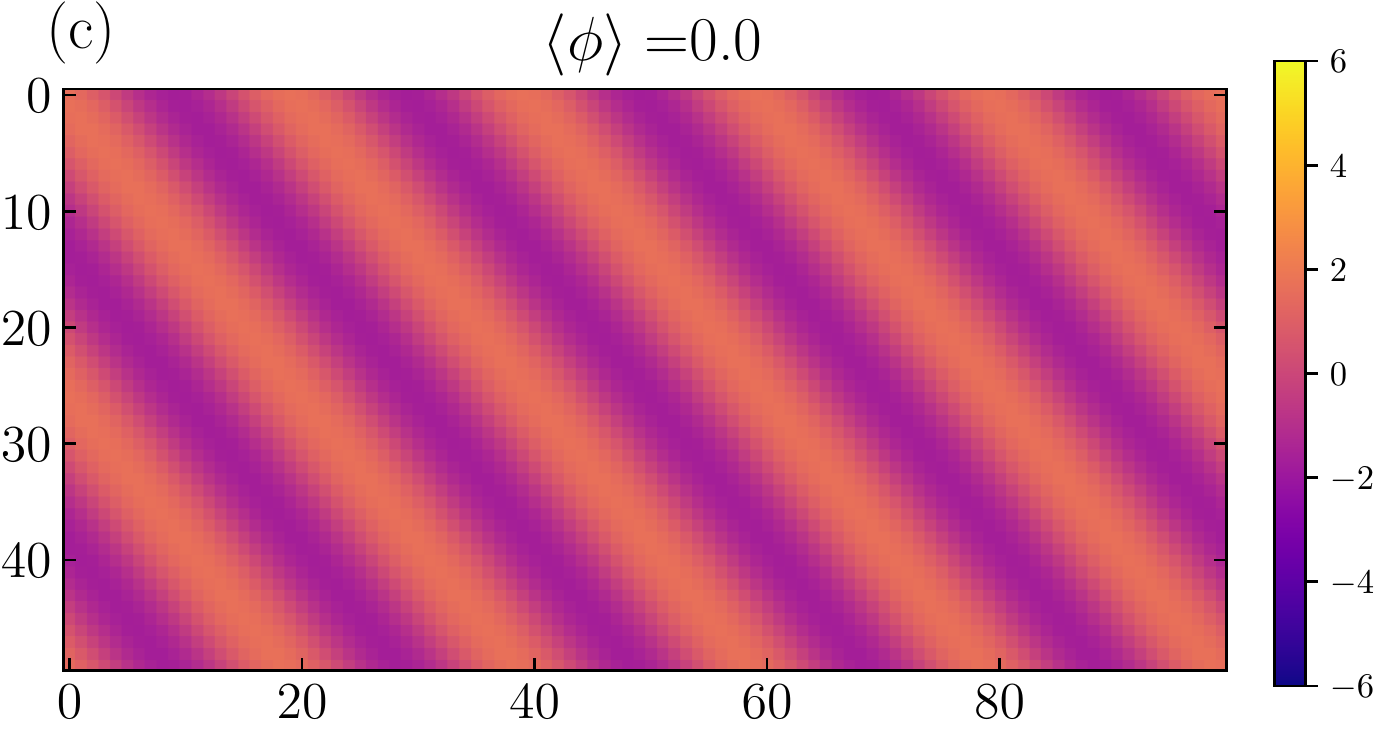}
\caption{Stationary state for the field $\phi$ as given by the PDE evolution Eqs.~\eqref{eq:evolution_rho+_mixed}, \eqref{eq:evolution_rho-_mixed}, \eqref{eq:evolution_phi_naive_mixed}  from a non-homogeneous initial state. (a) Initial conditions for the field $\phi$ with $\langle\phi\rangle=0$ in the left part of the box and $\langle\phi\rangle=6$ in the right part of the box, and $\rho=0.5+(\mathrm{noise})$ everywhere. 
(b) Steady state for $\phi$ when $\omega=0.1$. 
(c) Steady state for $\phi$ when $\omega=0.3$.
Other parameters: $r=0.01$, $B=0.3$, $\rho_0=0.5$, $\phi_0=5$, $\mu=1$, $\eta=0.005$ and $s=0.5$.  }
\label{fig:mixed_abc}
\end{figure}

\subsection{Linear stability analysis}

We now perform a linear stability analysis of equations \eqref{eq:evolution_rho+_mixed}, \eqref{eq:evolution_rho-_mixed} and \eqref{eq:evolution_phi_naive_mixed}, and study the stability for the different solutions of the self-consistent equation $\phi_h=G(\phi_h)$. We restrict ourselves to the case $s=1/2$ which contains already rich physics. With this choice of $s$, we still have an up down symmetry for the spins, thus $\phi_h=0$ is always a homogeneous and stationary solution. We perform LSA around this solution and the results are shown on Fig.~\ref{fig:3figures_mixedActiveEq}(a). The results are interesting when compared to the existence of other solutions of the self-consistent equations (see Fig.~\ref{fig:3figures_mixedActiveEq}(b)). In particular, the homogeneous state develops patterns before other solutions for $\phi_h$ appear (case $\omega=0.45$ for instance). In addition, we observe in th explicit PDEs' solutions that the final state of the system depends on initial conditions: even if the homogeneous state $\phi_h=0$ is not stable, the system may prefer creating patterns instead of having a full ferromagnetic order, which is also stable. To predict, at low cost, as a way of rationalizing our results, the final state of the system in this bi-stability regime, we can exhibit a mean field ``free energy" $F$ whose minima are the possible homogeneous solutions for $\phi$ ($F$ is not a free energy since we are far from equilibrium). For a homogeneous density $\rho_0$, the evolution equation of the homogeneous field $\phi$ simply becomes
\begin{align}
\partial_t\phi=-\frac{\partial F}{\partial \phi},
\end{align}
with
$F(\phi)=(r+B\rho_0)\phi^2/2-\rho_0\ln\left(\omega+2\eta\cosh(B\phi_0\phi)   \right)$
an even function of $\phi$ displayed on Fig.~\ref{fig:3figures_mixedActiveEq}(c) for different values of $\omega$. The global minimum of the function $F$ corresponds indeed to the final state of the magnetization, namely $\langle \phi\rangle=0$ or $\langle \phi \rangle>0$, as observed in the PDE solution (see Fig.~\ref{fig:mixed_abc}).

To sum up this section, we have seen that introducing a small amount of equilibrium in the dynamics of particle flip does not destroy the patterns. Furthermore, depending on its initial state, the system is now able to display either patterns, or  ferromagnetic order, in striking contrast with a full active or equilibrium case where only one option is accessible. 
We now turn to the analysis of energy dissipation in the active system, and more precisely, we address the question of origin and of the location of entropy production.

\section{Entropy production}
\label{sec:Entropy_creation}

Entropy production is a quantity that provides a measure of the degree of irreversibility of the dynamics of the system. In some cases it can simply be connected to the rate of energy dissipated by the system into the environment. We view entropy production as an elegant way to pinpoint the physical ingredients that are responsible for driving the system out of equilibrium. When spatially resolved, in the spirit of \cite{nardini_2017}, entropy production may be used to connect the emerging structures, at a local level, with dissipation. While the nonequilibrium drive has been identified as the key ingredient for the generation of patterns, whether the genuinely nonequilibrium processes operate at the pattern boundaries, or within the bulk of the system, is a question of interest to us.

\begin{figure}
\includegraphics[width=.49\columnwidth]{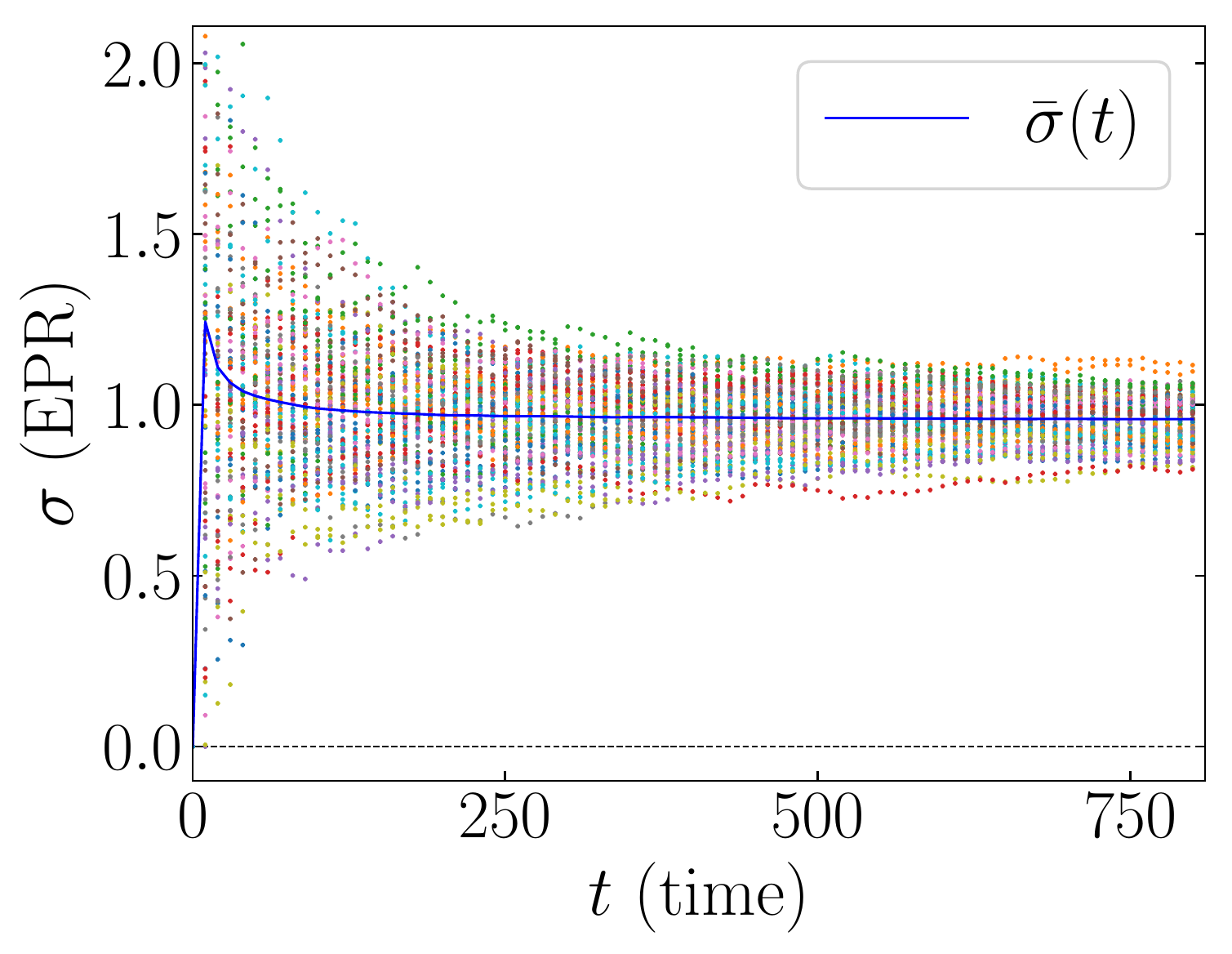}
\includegraphics[width=.49\columnwidth]{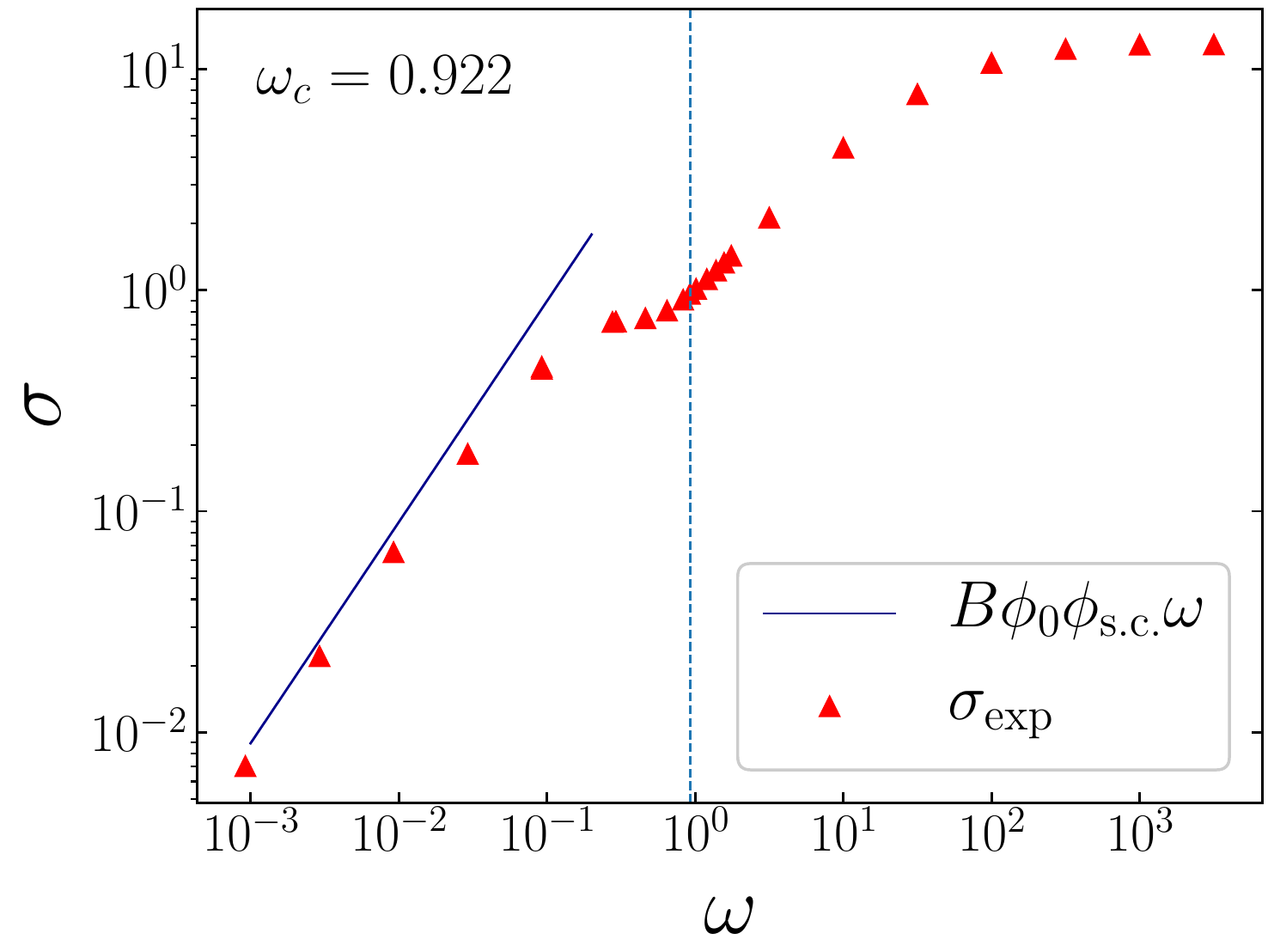}
\caption{\textbf{Left:} Entropy Production Rate of $100$ particles out of $N=6479$ particles in the simulation. Each dot represents the entropy production rate $\sigma_k(t)$ of a particle $k$ at time $t$. Blue line: average entropy production rate $\bar\sigma(t)=\frac{1}{N}\sum_{k=1}^N \sigma_k(t)$. Parameters: $r=0.01$, $B=0.18$, $\rho_0=0.2$, $\phi_0=8$, $\mu=5$, $s=0.5$, $\omega=0.922$. \textbf{Right:} Entropy production rate as a function of the flipping  parameter $\omega$. The dashed blue vertical line indicates the pattern apparition threshold $\omega_c$ predicted by mean field analysis. Parameters: $r=0.01$, $B=0.18$, $\rho_0=0.2$, $\phi_0=8$, $\mu=5$ and $s=0.5$.}
\label{fig:EPR_d20B18eps0}
\end{figure}

To estimate the total entropy production along a trajectory (in the whole phase space), we have to evaluate the probability of a trajectory relative to the probability of the time reversed trajectory~\cite{Lebowitz-Spohn-1999}. This question can be asked for various collections of degrees of freedom; we choose to focus on a single particle of position $X(t)\equiv X_t$ interacting with the Gaussian field $\phi(x,t)\equiv\phi_t(x)$. We restrict our derivation to a one dimensional system to keep the notation simple. The system evolves according to Eqs.~\eqref{eq:field_evol} and \eqref{eq:langevin} with $T=\Gamma=1$, and the spin flips from $+1$ to $-1$ (resp. from $-1$ to $+1$) with finite rate $\gamma$ (resp. $\alpha$). The spin $S_t$ jumps a finite number of times over an interval $[0,t_F]$, and $S_t$ is right continuous. The Hamiltonian $H$ will also be right continuous as a function of time. 
On an interval $[0,t_F]$ we define $t_j=jt_F/N$ and $t_{j+1}-t_j=t_F/N=\Delta t$. The probability of a noise history for the particle is given by 
\begin{align}
P[\xi|\xi_I]=&\exp\left(-\frac{1}{2}\sum_{j=0}^{N-1} \Delta\xi_j^2 \right).
\end{align}
Using the It\=o convention, the probability of a trajectory $\{X(t)\}_{0<t<t_F}$ is equal to the probability of observing the corresponding noise history. We thus have
\begin{align}
P[X|X_I]&=P[\xi|\xi_I]\\
=&\exp\frac{-1}{4\mu}\sum_{j=0}^{N-1} \Delta t \bigg[ \dot{X}_{t_j}+\mu \frac{\partial H}{\partial X_{t_j}}[X_{t_j},\phi_{t_j}(X_{t_j})]\bigg]^2\\
\simeq&\exp\frac{-1}{4\mu}\int_0^{t_F}d\tau\bigg[ \dot{X}_\tau+\mu \frac{\partial H}{\partial X_\tau}[X_\tau,\phi_{\tau}(X_\tau)]\bigg]^2,
\end{align}
where the set of points at which $H$ is discontinuous is of measure zero.
At initial time $t_I=0$, we start with $X(0)=X_I$, $\phi(x,0)=\phi_I(x)$ and $S(0)=S_I$ and the system evolves to a final time $t_F$. We can define the time reversed noise history through $\tilde{\xi}(\tau)=\xi(t_F-\tau)$ and the reversed trajectory is then $\tilde{X}(\tau)=X(t_F-\tau)$ such that $\tilde{X}(0)=X(t_F)\equiv \tilde X_I$. Entropy creation along a path is given by
\begin{align}
&\ln\frac{P[X(\tau)|X_I]}{P[\tilde{X}(\tau)|\tilde{X}_I]}\nonumber\\
&=-\frac{1}{4\mu}\int_0^{t_F}(\dot{X}_\tau+\mu \frac{\partial H}{\partial X_\tau})^2-(\dot{X}_\tau-\mu \frac{\partial H}{\partial X}_\tau)^2 d\tau\\
&=-\int_0^{t_F}\dot{X}_\tau\frac{\partial H}{\partial X_\tau}d\tau.
\end{align}
Similarly, since we have a Langevin equation for the field $\phi$, entropy creation of a field trajectory reads
\begin{align}
\ln\frac{P[\phi(x,\tau)|\phi_I(x)] }{ P[\tilde \phi(x,\tau)|\tilde \phi_I(x)]}
=-\int_0^{t_F} d\tau\int_x\, \partial_\tau\phi(x,\tau)\frac{\delta H}{\delta \phi(x,\tau)}.
\end{align}
We also have entropy creation related to the realization of the sequence of flips. If we start from a down configuration and if we slice time into intervals of duration $\Delta t$ then entropy creation for a flip history writes
\begin{align}
\ln\frac{P[S(\tau)|S_I]}{P[\tilde{S}(\tau)|\tilde S_I]}=\delta_{t_F} \ln\frac{\alpha}{\gamma},
\end{align}
where $\delta_{t_F}=-1$, $0$ or $1$, depending on the initial and final values of the spin.
We want to relate these previous results to the energy difference between the final time and the initial time. Let us start from the energy difference to see what terms appear in the calculation. We have
\begin{align}
\begin{split}
H(t_F)-H(0)=&\sum_{j=0}^{N-1}  H[X_{t_{j+1}},\phi_{t_{j+1}},S_{t_{j+1}}]\\ 
& \quad -H[X_{t_j},\phi_{t_j},S_{t_j}]
\end{split}\\
\begin{split}
=&\sum_{j=0}^{N-1} \Delta t \dot X_{t_j}\frac{\partial H}{\partial X}[X_{t_j},\phi_{t_j},S_{t_{j+1}}]\\
&\quad+\Delta t \int_x\dot \phi_{t_j}\frac{\delta H}{\delta \phi(x)}[X_{t_j},\phi_{t_j},S_{t_{j+1}}]\\
&\quad+H[X_{t_{j}},\phi_{t_{j}},S_{t_{j+1}}]-H[X_{t_{j}},\phi_{t_{j}},S_{t_{j}}]\\
&\quad+o(\Delta t)
\end{split}\\
\begin{split}
\simeq&\int_0^{t_F} \!\!d\tau \bigg[\dot X_\tau\frac{\partial H}{\partial X_\tau}[X_{\tau},\phi_{\tau},S_{\tau}]\\
&\quad +\int_x\dot \phi_{\tau}\frac{\delta H}{\delta \phi(x)}[X_{\tau},\phi_{\tau},S_\tau]\bigg]\\
&+\sum_{t_\alpha} \bigg[H[X_{t_\alpha^+},\phi_{t_\alpha^+},S_{t_\alpha^+}]\\
&\quad-H[X_{t_\alpha^-},\phi_{t_\alpha^-},S_{t_\alpha^-}]\bigg],
\end{split}
\end{align}
and where the spin flips at time $t_\alpha$, with $t_\alpha^+$ and $t_\alpha^-$ denoting the times right after and right before the flip, respectively.
We can now compute the entropy produced along any path:
\begin{align}
&\ln\frac{P[X_\tau,\phi_\tau(x),S_\tau|X_I,\phi_I(x),S_I]}{P[\tilde X_\tau,\tilde\phi_\tau(x),\tilde S_\tau|\tilde X_I, \tilde \phi_I(x),\tilde S_I]}\nonumber\\
\begin{split}
=&-\int_0^t d\tau \left(\dot{X}_\tau\frac{\partial H}{\partial X}+\int_x \partial_\tau \phi(x,\tau) \frac{\delta H}{\delta \phi(x,\tau)}\right)\\
&+\delta_{t_F}\ln\frac{\alpha}{\gamma}
\end{split}\\
\begin{split}
=&-[H(t_F)-H(0)]+\delta_{t_F}\ln\frac{\alpha}{\gamma}\\
&+\sum_{t_\alpha} \bigg[H[X_{t_\alpha^+},\phi_{t_\alpha^+},S_{t_\alpha^+}]-H[X_{t_\alpha^-},\phi_{t_\alpha^-},S_{t_\alpha^-}]\bigg]
\end{split}\\
\begin{split}
=&-[H(t_F)-H(0)]+\delta_{t_F}\ln\frac{\alpha}{\gamma}\\
&+\sum_{t_\alpha} 2B \phi_0 S_{t_\alpha^-}\,\phi_{t_\alpha}(X_{t_\alpha}).
\label{eq:entropy_production}
\end{split}
\end{align}
Dividing this result by $t_F$ and taking the limit $t_F\to \infty$ yields the entropy production rate $\sigma$. We immediately see that the first two terms in Eq.~\eqref{eq:entropy_production} vanish when divided by the total duration $t_F$ as $t_F$ is taken asymptotically large, since they are bounded. In the stationary state, the entropy production thus simplifies into 
\begin{align}
\sigma&=\lim_{t_F\to\infty}\frac{2B \phi_0}{t_F}\sum_{0<t_\alpha<t_F}  S_{t_\alpha^-}\,\phi_{t_\alpha}(X_{t_\alpha})
\label{eq:def_entropy_production_rate}\\
&=\lim_{t_F\to\infty}\frac{2B \phi_0}{t_F}N_{t_F}\langle S_{t_\alpha^-}\,\phi_{t_\alpha}(X_{t_\alpha})\rangle,
\end{align}
with $N_{t_F}$ the number of flips in $[0,t_F]$. For a variable that flips between two states at fixed rates $\alpha$ and $\gamma$, this number is given by
\begin{align}
N_{t_F}=2\frac{\alpha\gamma}{\alpha+\gamma} t_F,
\end{align}
when $t_F\to\infty$, and thus scales like $O(t_F)$. We can further simplify the expression of entropy production:
\begin{align}
\sigma&=4B \phi_0\frac{\alpha\gamma}{\alpha+\gamma} \langle S_{t_\alpha^-}\,\phi_{t_\alpha}(X_{t_\alpha})\rangle\\
&=4 B \phi_0\, \omega\, s (1-s)\langle S_{t_\alpha^-}\,\phi_{t_\alpha}(X_{t_\alpha})\rangle.
\end{align}
The average $\langle S_{t_\alpha^-}\,\phi_{t_\alpha}(X_{t_\alpha})\rangle$ is however more complicated to compute because it depends on the whole dynamics. We are now considering two important limiting cases: (i) the time between two flips is large with respect to the particle--field dynamics, (ii) the time between two flips is small in that respect. 
In (i), we typically witness pattern formation. In this situation, the field at the particle's location has the same sign as the spin before the flip, and thus scales like $O(\phi_0S_{t^-})$. We can further say that the field can equilibrate between flips, thus the field $\phi$ is equal to $\phi_\mathrm{s.c.}$ which satisfies the self-consistent equation $\phi_\mathrm{s.c.}=B\phi_0\rho_0 \tanh(B\phi_0\phi_\mathrm{s.c.})/(r+B\rho_0)$ (see Sec.~\ref{sec:mean-field_equilibrium}) in the bulk of each microphase. In this case the entropy production rate reads:
\begin{align}
\sigma_{\omega\ll\omega_c}^\mathrm{max}&=4 B \phi_0\phi_\mathrm{s.c.}\omega s(1-s).
\label{eq:EPR_lowFlip}
\end{align}
In regime (ii) of fast flipping, patterns disappear, and $S(t)$  and  $\phi_t(X_t)$ are almost uncorrelated. We can actually predict that the entropy production rate saturates when $\omega\to \infty$.  We notice that for $N$ particles we have, by definition, $\psi(x,t)= \sum_{k=1}^N S_k(t)\delta(x-X_k(t))$ and thus for one particle $\psi(x,t)\sim S_t$. We thus approximate $\langle S_{t_\alpha^-}\,\phi_{t_\alpha}(X_{t_\alpha})\rangle$ with its continuum description, namely $\langle \psi_t \phi_t(X_t) \rangle$. From the evolution equations of the fields, we compute $\phi\times\p_t\psi +\psi\times\p_t\phi$ (where $\p_t\psi$ is given in Eq.~\eqref{eq:evolution_psi} and $\p_t\phi$ is given in Eq.~\eqref{eq:evolution_phi}) to reconstruct a time derivative of a correlation, which is $0$ in steady state. In the particular case of symmetric flips $s=1/2$, we have:
\begin{align}
\begin{split}0&=\partial_t\langle \psi\phi\rangle\\
&=\bigg\langle\phi\big(\mu \nabla^2\psi+\mu B\,\bm\nabla\cdot\left[(\psi\phi-\rho\phi_0)\bm\nabla \phi\right]-\omega \,\psi\big) \\
&\quad\quad\quad +\psi\big(\nabla^2 \phi - r\phi- B \rho\phi+ B\phi_0\psi\big)\bigg\rangle
\end{split}\\
&\!\!\!\!\underset{\omega\to\infty}{=}-\omega\langle\psi\phi\rangle+B\phi_0\langle\psi^2\rangle,
\end{align}
which yields
$\langle\psi\phi\rangle\underset{\omega\to\infty}{\to} B\phi_0\langle\psi^2\rangle/\omega$,
from which we infer the produced entropy for $s=1/2$:
\begin{align}
\sigma_{\omega\to\infty}\propto B^2\phi_0^2.
\end{align}
In our numerical experiments, we start to measure entropy production at time $0$ and we define the entropy production rate for a particle $k$ at time $t$ as 
\begin{align}
\sigma_k(t)= \frac{2B \phi_0}{t}\sum_{0<t_{\alpha k}<t}  S^k_{t_{\alpha k}^-}\,\phi_{t_{\alpha k}}(X^k_{t_{\alpha k}}),
\end{align}
where the $t_{\alpha k}$ are the time of flips of particle $k$. In Fig.~\ref{fig:EPR_d20B18eps0} (left), we display the convergence of the entropy production rate towards its stationary state value for $100$ particles (out of $N=6479$ in the simulation), starting from a homogeneous state $\langle\phi\rangle=0$ and all particles spin up. Measuring entropy production for different values of $\omega$, we recover that entropy production rate saturates when $\omega\to \infty$,  and we find the value we predicted in Eq.~\eqref{eq:EPR_lowFlip} for $\omega\to 0$. These results are displayed in Fig.~\ref{fig:EPR_d20B18eps0} (right), where the graph also shows that the critical flipping rate predicted by the LSA matches the transition observed in entropy production.

Finally, another way to extract interesting information from our calculation for entropy production is to define a local entropy production rate or density of entropy production such that $\sigma=\int d^2 r\sigma(\bm r)$. Returning to a two-dimensional system, from Eq.~\eqref{eq:def_entropy_production_rate}, we can identify such an entropy production density for a system with $N$ particles:
\begin{align}
\sigma(\bm r)=\lim_{t \to \infty}\frac{1}{t}\sum_{k=1}^N 2 B\phi_0\sum_{t_{\alpha k}<t}S^k_{t_{\alpha k}^-}\phi_{t_{\alpha k}}(\bm r)\delta(\bm r-\bm X^k_{t_{\alpha k}}),
\end{align}
where the $(t_{\alpha k})_{\alpha\in\mathbb{N}}$ are the instants of flip of particle $k$, and where $\bm X^k_t$ is the position of particle $k$ at time $t$. We are now able to establish a map of the entropy production rate within the stationary state. In our simulations, though we observe diffusion of the whole pattern, the entropy production rate converges over a much smaller time scale, and we thus reach a ``stationary state" before pattern blurring. In Fig.~\ref{fig:entropyMap_d30B0,106}, we see that entropy production is localized within the bulk of the stripes. In other words, dissipation occurs in the bulk and not specifically at the boundaries of patterns. While the existence of patterns is a genuine nonequilibrium effect, one cannot interpret the role of the nonequilibrium drive in terms of a stabilizing effective surface tension at the boundaries of the ordered domains.
\begin{figure}
\includegraphics[width=.99\columnwidth]{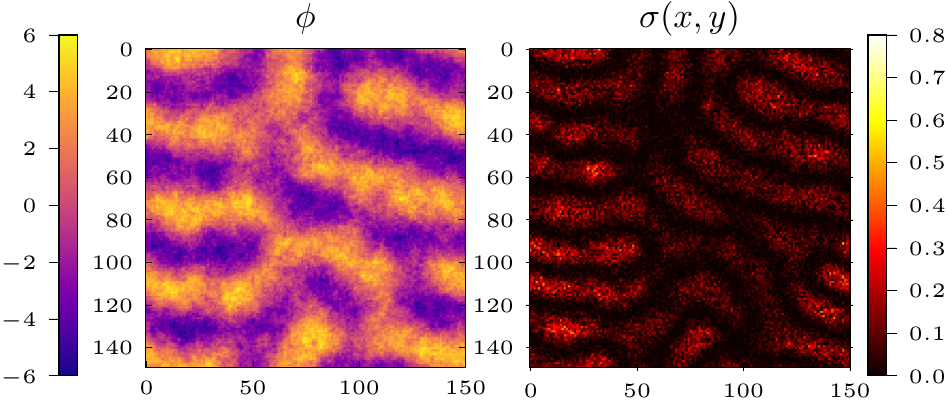}
\caption{Snapshot of the system in pattern forming regime (iii). Left: field $\phi$. Right: entropy production density. Parameters: $r=0.01$, $B=0.106$, $\rho_0=0.3$, $\phi_0=8$, $\mu=5$.}
\label{fig:entropyMap_d30B0,106}
\end{figure}

\section{Conclusion}

Our goal was to explore and predict the emergence of collective phenomena in assemblies of active particles whose interactions are mediated by a fluctuating medium. We have done so on the basis of a minimal model in which particles diffuse while locally constraining the medium deformation. Activity is introduced by means of an internal degree of freedom that controls the interaction with the background field. This internal degree of freedom fluctuates independently of the bath temperature, and thus breaks the equilibrium nature of the dynamics of the whole system.

By means of Monte Carlo simulations and of a mean-field analysis of the dynamical equations, we have shown that this system displays a wealth of pattern formation regimes. When patterns appear, their wavelength is given by the geometric mean of the characteristic correlation length of the underlying elastic field and of the diffusion length of particles between two active flips. This geometric mean property is reminiscent of the typical wavelength emerging in crystal growth and in the Mullins-Sekerka instability~\cite{langer_1977}. This coincidence might \textit{a posteriori} be perceived as little surprising since we have at stake, in both systems, interactions favoring phase separation (a surface tension ingredient) competing with a diffusive process.  In addition, as the number of particles is conserved in the system, patterns can be localized on a small fraction of the system size~\cite{matthews_2000,cox_2003}. 
We have also examined how to interpolate between equilibrium dynamics and active dynamics for the flips since we reasonably expect that the flips might also feature temperature induced fluctuations. This interpolation has shown that the patterns could survive a moderate amount of equilibrium.
Finally, we addressed the question of energy dissipation and entropy production in the active system. We have seen that entropy production vanishes for low flipping rates, as expected, and that it saturates for large flipping rates. We have also seen that entropy is produced within the bulk of the patterns, as opposed to other active systems where it is localized at the phase boundaries~\cite{nardini_2017}.

We are now at a stage where our model should be made more realistic. This may be achieved in a variety of  directions.
The Hamiltonian for the field can be adapted to specific systems we 
want to describe. Typically, we could use a Helfrich Hamiltonian to work on biological membranes. 
The field dynamics may also be changed. If the field now stands for a molecular density, we expect it to evolve according to a conserved dynamics (Cahn-Hilliard, Allen-Cahn). 
To focus on active proteins in the biological membrane, we also believe that hydrodynamic effects should be taken into account. This would certainly imply dealing with non-local equations, with the drag in a two-dimensional liquid layer (the lipid leaflet), and with the three-dimensional bulk liquid, which drives the system to another level of complexity, along with a (probably) richer behavior.

\acknowledgements 
R. Z. would like to thank Fridtjof Brauns for useful discussions.
The authors are indebted to S. Fauve and J. Tailleur for their very useful feedback.



\begin{thebibliography}{38}%
\makeatletter
\providecommand \@ifxundefined [1]{%
 \@ifx{#1\undefined}
}%
\providecommand \@ifnum [1]{%
 \ifnum #1\expandafter \@firstoftwo
 \else \expandafter \@secondoftwo
 \fi
}%
\providecommand \@ifx [1]{%
 \ifx #1\expandafter \@firstoftwo
 \else \expandafter \@secondoftwo
 \fi
}%
\providecommand \natexlab [1]{#1}%
\providecommand \enquote  [1]{``#1''}%
\providecommand \bibnamefont  [1]{#1}%
\providecommand \bibfnamefont [1]{#1}%
\providecommand \citenamefont [1]{#1}%
\providecommand \href@noop [0]{\@secondoftwo}%
\providecommand \href [0]{\begingroup \@sanitize@url \@href}%
\providecommand \@href[1]{\@@startlink{#1}\@@href}%
\providecommand \@@href[1]{\endgroup#1\@@endlink}%
\providecommand \@sanitize@url [0]{\catcode `\\12\catcode `\$12\catcode
  `\&12\catcode `\#12\catcode `\^12\catcode `\_12\catcode `\%12\relax}%
\providecommand \@@startlink[1]{}%
\providecommand \@@endlink[0]{}%
\providecommand \url  [0]{\begingroup\@sanitize@url \@url }%
\providecommand \@url [1]{\endgroup\@href {#1}{\urlprefix }}%
\providecommand \urlprefix  [0]{URL }%
\providecommand \Eprint [0]{\href }%
\providecommand \doibase [0]{http://dx.doi.org/}%
\providecommand \selectlanguage [0]{\@gobble}%
\providecommand \bibinfo  [0]{\@secondoftwo}%
\providecommand \bibfield  [0]{\@secondoftwo}%
\providecommand \translation [1]{[#1]}%
\providecommand \BibitemOpen [0]{}%
\providecommand \bibitemStop [0]{}%
\providecommand \bibitemNoStop [0]{.\EOS\space}%
\providecommand \EOS [0]{\spacefactor3000\relax}%
\providecommand \BibitemShut  [1]{\csname bibitem#1\endcsname}%
\let\auto@bib@innerbib\@empty
\bibitem [{\citenamefont {Fisher}\ and\ \citenamefont
  {de~Gennes}(1978)}]{Fisher78}%
  \BibitemOpen
  \bibfield  {author} {\bibinfo {author} {\bibfnamefont {M.~E.}\ \bibnamefont
  {Fisher}}\ and\ \bibinfo {author} {\bibfnamefont {P.~G.}\ \bibnamefont
  {de~Gennes}},\ }\href@noop {} {\bibfield  {journal} {\bibinfo  {journal} {C.
  R. Acad. Sci. Paris Ser. B}\ }\textbf {\bibinfo {volume} {287}},\ \bibinfo
  {pages} {207} (\bibinfo {year} {1978})}\BibitemShut {NoStop}%
\bibitem [{\citenamefont {Hertlein}\ \emph {et~al.}(2008)\citenamefont
  {Hertlein}, \citenamefont {Helden}, \citenamefont {Gambassi}, \citenamefont
  {Dietrich},\ and\ \citenamefont {Bechinger}}]{Hertlein08}%
  \BibitemOpen
  \bibfield  {author} {\bibinfo {author} {\bibfnamefont {C.}~\bibnamefont
  {Hertlein}}, \bibinfo {author} {\bibfnamefont {L.}~\bibnamefont {Helden}},
  \bibinfo {author} {\bibfnamefont {A.}~\bibnamefont {Gambassi}}, \bibinfo
  {author} {\bibfnamefont {S.}~\bibnamefont {Dietrich}}, \ and\ \bibinfo
  {author} {\bibfnamefont {C.}~\bibnamefont {Bechinger}},\ }\href {\doibase
  10.1038/nature06443} {\bibfield  {journal} {\bibinfo  {journal} {Nature}\
  }\textbf {\bibinfo {volume} {451}},\ \bibinfo {pages} {172} (\bibinfo {year}
  {2008})}\BibitemShut {NoStop}%
\bibitem [{\citenamefont {Ajdari}\ \emph {et~al.}(1991)\citenamefont {Ajdari},
  \citenamefont {Peliti},\ and\ \citenamefont {Prost}}]{Ajdari91}%
  \BibitemOpen
  \bibfield  {author} {\bibinfo {author} {\bibfnamefont {A.}~\bibnamefont
  {Ajdari}}, \bibinfo {author} {\bibfnamefont {L.}~\bibnamefont {Peliti}}, \
  and\ \bibinfo {author} {\bibfnamefont {J.}~\bibnamefont {Prost}},\ }\href
  {\doibase 10.1103/PhysRevLett.66.1481} {\bibfield  {journal} {\bibinfo
  {journal} {Phys. Rev. Lett.}\ }\textbf {\bibinfo {volume} {66}},\ \bibinfo
  {pages} {1481} (\bibinfo {year} {1991})}\BibitemShut {NoStop}%
\bibitem [{\citenamefont {Poulin}\ \emph {et~al.}(1997)\citenamefont {Poulin},
  \citenamefont {Stark}, \citenamefont {Lubensky},\ and\ \citenamefont
  {Weitz}}]{poulin_1997}%
  \BibitemOpen
  \bibfield  {author} {\bibinfo {author} {\bibfnamefont {P.}~\bibnamefont
  {Poulin}}, \bibinfo {author} {\bibfnamefont {H.}~\bibnamefont {Stark}},
  \bibinfo {author} {\bibfnamefont {T.~C.}\ \bibnamefont {Lubensky}}, \ and\
  \bibinfo {author} {\bibfnamefont {D.~A.}\ \bibnamefont {Weitz}},\ }\href
  {\doibase 10.1126/science.275.5307.1770} {\bibfield  {journal} {\bibinfo
  {journal} {Science}\ }\textbf {\bibinfo {volume} {275}},\ \bibinfo {pages}
  {1770} (\bibinfo {year} {1997})}\BibitemShut {NoStop}%
\bibitem [{\citenamefont {Nicolson}(1949)}]{Nicolson49}%
  \BibitemOpen
  \bibfield  {author} {\bibinfo {author} {\bibfnamefont {M.~M.}\ \bibnamefont
  {Nicolson}},\ }\href@noop {} {\bibfield  {journal} {\bibinfo  {journal}
  {Proc. Cambridge Philos. Soc.}\ }\textbf {\bibinfo {volume} {45}},\ \bibinfo
  {pages} {288} (\bibinfo {year} {1949})}\BibitemShut {NoStop}%
\bibitem [{\citenamefont {Hu}\ and\ \citenamefont {Bush}(2005)}]{Hu05Nat}%
  \BibitemOpen
  \bibfield  {author} {\bibinfo {author} {\bibfnamefont {D.~L.}\ \bibnamefont
  {Hu}}\ and\ \bibinfo {author} {\bibfnamefont {J.~W.~M.}\ \bibnamefont
  {Bush}},\ }\href {\doibase 10.1038/nature03995} {\bibfield  {journal}
  {\bibinfo  {journal} {Nature}\ }\textbf {\bibinfo {volume} {437}},\ \bibinfo
  {pages} {733} (\bibinfo {year} {2005})}\BibitemShut {NoStop}%
\bibitem [{\citenamefont {Goulian}\ \emph {et~al.}(1993)\citenamefont
  {Goulian}, \citenamefont {Bruinsma},\ and\ \citenamefont
  {Pincus}}]{Goulian93EPL}%
  \BibitemOpen
  \bibfield  {author} {\bibinfo {author} {\bibfnamefont {M.}~\bibnamefont
  {Goulian}}, \bibinfo {author} {\bibfnamefont {R.}~\bibnamefont {Bruinsma}}, \
  and\ \bibinfo {author} {\bibfnamefont {P.}~\bibnamefont {Pincus}},\ }\href
  {http://stacks.iop.org/0295-5075/22/i=2/a=012} {\bibfield  {journal}
  {\bibinfo  {journal} {EPL}\ }\textbf {\bibinfo {volume} {22}},\ \bibinfo
  {pages} {145} (\bibinfo {year} {1993})}\BibitemShut {NoStop}%
\bibitem [{\citenamefont {Dan}\ \emph {et~al.}(1993)\citenamefont {Dan},
  \citenamefont {Pincus},\ and\ \citenamefont {Safran}}]{Dan93Lang}%
  \BibitemOpen
  \bibfield  {author} {\bibinfo {author} {\bibfnamefont {N.}~\bibnamefont
  {Dan}}, \bibinfo {author} {\bibfnamefont {P.}~\bibnamefont {Pincus}}, \ and\
  \bibinfo {author} {\bibfnamefont {S.~A.}\ \bibnamefont {Safran}},\ }\href
  {\doibase 10.1021/la00035a005} {\bibfield  {journal} {\bibinfo  {journal}
  {Langmuir}\ }\textbf {\bibinfo {volume} {9}},\ \bibinfo {pages} {2768}
  (\bibinfo {year} {1993})}\BibitemShut {NoStop}%
\bibitem [{\citenamefont {Dommersnes}\ and\ \citenamefont
  {Fournier}(1999)}]{Dommersnes99EPJB}%
  \BibitemOpen
  \bibfield  {author} {\bibinfo {author} {\bibfnamefont {P.}~\bibnamefont
  {Dommersnes}}\ and\ \bibinfo {author} {\bibfnamefont {J.-B.}\ \bibnamefont
  {Fournier}},\ }\href {\doibase 10.1007/s100510050968} {\bibfield  {journal}
  {\bibinfo  {journal} {Eur. Phys. J. B}\ }\textbf {\bibinfo {volume} {12}},\
  \bibinfo {pages} {9} (\bibinfo {year} {1999})}\BibitemShut {NoStop}%
\bibitem [{\citenamefont {Bitbol}\ \emph {et~al.}(2012)\citenamefont {Bitbol},
  \citenamefont {Constantin},\ and\ \citenamefont {Fournier}}]{BitbolPlos12}%
  \BibitemOpen
  \bibfield  {author} {\bibinfo {author} {\bibfnamefont {A.-F.}\ \bibnamefont
  {Bitbol}}, \bibinfo {author} {\bibfnamefont {D.}~\bibnamefont {Constantin}},
  \ and\ \bibinfo {author} {\bibfnamefont {J.-B.}\ \bibnamefont {Fournier}},\
  }\href {\doibase 10.1371/journal.pone.0048306} {\bibfield  {journal}
  {\bibinfo  {journal} {PLoS ONE}\ }\textbf {\bibinfo {volume} {7}},\ \bibinfo
  {pages} {e48306} (\bibinfo {year} {2012})}\BibitemShut {NoStop}%
\bibitem [{\citenamefont {Van Der~Wel}\ \emph {et~al.}(2016)\citenamefont {Van
  Der~Wel}, \citenamefont {Vahid}, \citenamefont {{\v{S}}ari{\'c}},
  \citenamefont {Idema}, \citenamefont {Heinrich},\ and\ \citenamefont
  {Kraft}}]{vanderWel16}%
  \BibitemOpen
  \bibfield  {author} {\bibinfo {author} {\bibfnamefont {C.}~\bibnamefont {Van
  Der~Wel}}, \bibinfo {author} {\bibfnamefont {A.}~\bibnamefont {Vahid}},
  \bibinfo {author} {\bibfnamefont {A.}~\bibnamefont {{\v{S}}ari{\'c}}},
  \bibinfo {author} {\bibfnamefont {T.}~\bibnamefont {Idema}}, \bibinfo
  {author} {\bibfnamefont {D.}~\bibnamefont {Heinrich}}, \ and\ \bibinfo
  {author} {\bibfnamefont {D.~J.}\ \bibnamefont {Kraft}},\ }\href {\doibase
  10.1038/srep32825} {\bibfield  {journal} {\bibinfo  {journal} {Scientific
  Reports}\ }\textbf {\bibinfo {volume} {6}} (\bibinfo {year} {2016}),\
  10.1038/srep32825}\BibitemShut {NoStop}%
\bibitem [{\citenamefont {Brown}\ \emph {et~al.}(2000)\citenamefont {Brown},
  \citenamefont {Smith},\ and\ \citenamefont {Rennie}}]{brown_2000}%
  \BibitemOpen
  \bibfield  {author} {\bibinfo {author} {\bibfnamefont {A.~B.~D.}\
  \bibnamefont {Brown}}, \bibinfo {author} {\bibfnamefont {C.~G.}\ \bibnamefont
  {Smith}}, \ and\ \bibinfo {author} {\bibfnamefont {A.~R.}\ \bibnamefont
  {Rennie}},\ }\href {\doibase 10.1103/PhysRevE.62.951} {\bibfield  {journal}
  {\bibinfo  {journal} {Phys. Rev. E}\ }\textbf {\bibinfo {volume} {62}},\
  \bibinfo {pages} {951} (\bibinfo {year} {2000})}\BibitemShut {NoStop}%
\bibitem [{\citenamefont {Noguchi}\ and\ \citenamefont
  {Fournier}(2017)}]{noguchi_2017}%
  \BibitemOpen
  \bibfield  {author} {\bibinfo {author} {\bibfnamefont {H.}~\bibnamefont
  {Noguchi}}\ and\ \bibinfo {author} {\bibfnamefont {J.-B.}\ \bibnamefont
  {Fournier}},\ }\href {\doibase 10.1039/C7SM00305F} {\bibfield  {journal}
  {\bibinfo  {journal} {Soft Matter}\ }\textbf {\bibinfo {volume} {13}},\
  \bibinfo {pages} {4099} (\bibinfo {year} {2017})}\BibitemShut {NoStop}%
\bibitem [{\citenamefont {Miller}\ and\ \citenamefont
  {Bassler}(2001)}]{miller_2001}%
  \BibitemOpen
  \bibfield  {author} {\bibinfo {author} {\bibfnamefont {M.~B.}\ \bibnamefont
  {Miller}}\ and\ \bibinfo {author} {\bibfnamefont {B.~L.}\ \bibnamefont
  {Bassler}},\ }\href {\doibase 10.1146/annurev.micro.55.1.165} {\bibfield
  {journal} {\bibinfo  {journal} {Annual Review of Microbiology}\ }\textbf
  {\bibinfo {volume} {55}},\ \bibinfo {pages} {165} (\bibinfo {year} {2001})},\
  \bibinfo {note} {pMID: 11544353},\ \Eprint
  {http://arxiv.org/abs/https://doi.org/10.1146/annurev.micro.55.1.165}
  {https://doi.org/10.1146/annurev.micro.55.1.165} \BibitemShut {NoStop}%
\bibitem [{\citenamefont {Stenhammar}\ \emph {et~al.}(2017)\citenamefont
  {Stenhammar}, \citenamefont {Nardini}, \citenamefont {Nash}, \citenamefont
  {Marenduzzo},\ and\ \citenamefont {Morozov}}]{stenhammar_2017}%
  \BibitemOpen
  \bibfield  {author} {\bibinfo {author} {\bibfnamefont {J.}~\bibnamefont
  {Stenhammar}}, \bibinfo {author} {\bibfnamefont {C.}~\bibnamefont {Nardini}},
  \bibinfo {author} {\bibfnamefont {R.~W.}\ \bibnamefont {Nash}}, \bibinfo
  {author} {\bibfnamefont {D.}~\bibnamefont {Marenduzzo}}, \ and\ \bibinfo
  {author} {\bibfnamefont {A.}~\bibnamefont {Morozov}},\ }\href {\doibase
  10.1103/PhysRevLett.119.028005} {\bibfield  {journal} {\bibinfo  {journal}
  {Phys. Rev. Lett.}\ }\textbf {\bibinfo {volume} {119}},\ \bibinfo {pages}
  {028005} (\bibinfo {year} {2017})}\BibitemShut {NoStop}%
\bibitem [{\citenamefont {Fribourg}\ \emph {et~al.}(2014)\citenamefont
  {Fribourg}, \citenamefont {Chami}, \citenamefont {Sorzano}, \citenamefont
  {Gubellini}, \citenamefont {Marabini}, \citenamefont {Marco}, \citenamefont
  {Jault},\ and\ \citenamefont {Lévy}}]{fribourg_2014}%
  \BibitemOpen
  \bibfield  {author} {\bibinfo {author} {\bibfnamefont {P.~F.}\ \bibnamefont
  {Fribourg}}, \bibinfo {author} {\bibfnamefont {M.}~\bibnamefont {Chami}},
  \bibinfo {author} {\bibfnamefont {C.~O.~S.}\ \bibnamefont {Sorzano}},
  \bibinfo {author} {\bibfnamefont {F.}~\bibnamefont {Gubellini}}, \bibinfo
  {author} {\bibfnamefont {R.}~\bibnamefont {Marabini}}, \bibinfo {author}
  {\bibfnamefont {S.}~\bibnamefont {Marco}}, \bibinfo {author} {\bibfnamefont
  {J.-M.}\ \bibnamefont {Jault}}, \ and\ \bibinfo {author} {\bibfnamefont
  {D.}~\bibnamefont {Lévy}},\ }\href {\doibase 10.1016/j.jmb.2014.03.002}
  {\bibfield  {journal} {\bibinfo  {journal} {Journal of Molecular Biology}\
  }\textbf {\bibinfo {volume} {426}},\ \bibinfo {pages} {2059} (\bibinfo {year}
  {2014})}\BibitemShut {NoStop}%
\bibitem [{\citenamefont {Sumino}\ \emph {et~al.}(2014)\citenamefont {Sumino},
  \citenamefont {Yamamoto}, \citenamefont {Iwamoto}, \citenamefont {Dewa},\
  and\ \citenamefont {Oiki}}]{sumino_2014}%
  \BibitemOpen
  \bibfield  {author} {\bibinfo {author} {\bibfnamefont {A.}~\bibnamefont
  {Sumino}}, \bibinfo {author} {\bibfnamefont {D.}~\bibnamefont {Yamamoto}},
  \bibinfo {author} {\bibfnamefont {M.}~\bibnamefont {Iwamoto}}, \bibinfo
  {author} {\bibfnamefont {T.}~\bibnamefont {Dewa}}, \ and\ \bibinfo {author}
  {\bibfnamefont {S.}~\bibnamefont {Oiki}},\ }\href {\doibase
  10.1021/jz402491t} {\bibfield  {journal} {\bibinfo  {journal} {The Journal of
  Physical Chemistry Letters}\ }\textbf {\bibinfo {volume} {5}},\ \bibinfo
  {pages} {578} (\bibinfo {year} {2014})}\BibitemShut {NoStop}%
\bibitem [{\citenamefont {Evans}\ \emph {et~al.}(2013)\citenamefont {Evans},
  \citenamefont {Sun}, \citenamefont {Senyuk}, \citenamefont {Keller},
  \citenamefont {Pergamenshchik}, \citenamefont {Lee},\ and\ \citenamefont
  {Smalyukh}}]{evans_2013}%
  \BibitemOpen
  \bibfield  {author} {\bibinfo {author} {\bibfnamefont {J.~S.}\ \bibnamefont
  {Evans}}, \bibinfo {author} {\bibfnamefont {Y.}~\bibnamefont {Sun}}, \bibinfo
  {author} {\bibfnamefont {B.}~\bibnamefont {Senyuk}}, \bibinfo {author}
  {\bibfnamefont {P.}~\bibnamefont {Keller}}, \bibinfo {author} {\bibfnamefont
  {V.~M.}\ \bibnamefont {Pergamenshchik}}, \bibinfo {author} {\bibfnamefont
  {T.}~\bibnamefont {Lee}}, \ and\ \bibinfo {author} {\bibfnamefont {I.~I.}\
  \bibnamefont {Smalyukh}},\ }\href {\doibase 10.1103/PhysRevLett.110.187802}
  {\bibfield  {journal} {\bibinfo  {journal} {Physical Review Letters}\
  }\textbf {\bibinfo {volume} {110}} (\bibinfo {year} {2013}),\
  10.1103/PhysRevLett.110.187802}\BibitemShut {NoStop}%
\bibitem [{\citenamefont {Criante}\ \emph {et~al.}(2013)\citenamefont
  {Criante}, \citenamefont {Bracalente}, \citenamefont {Lucchetti},
  \citenamefont {Simoni},\ and\ \citenamefont {Brasselet}}]{criante_2013}%
  \BibitemOpen
  \bibfield  {author} {\bibinfo {author} {\bibfnamefont {L.}~\bibnamefont
  {Criante}}, \bibinfo {author} {\bibfnamefont {F.}~\bibnamefont {Bracalente}},
  \bibinfo {author} {\bibfnamefont {L.}~\bibnamefont {Lucchetti}}, \bibinfo
  {author} {\bibfnamefont {F.}~\bibnamefont {Simoni}}, \ and\ \bibinfo {author}
  {\bibfnamefont {E.}~\bibnamefont {Brasselet}},\ }\href {\doibase
  10.1039/c3sm50273b} {\bibfield  {journal} {\bibinfo  {journal} {Soft Matter}\
  }\textbf {\bibinfo {volume} {9}},\ \bibinfo {pages} {5459} (\bibinfo {year}
  {2013})}\BibitemShut {NoStop}%
\bibitem [{\citenamefont {Grawitter}\ and\ \citenamefont
  {Stark}(2018)}]{grawitter_2018}%
  \BibitemOpen
  \bibfield  {author} {\bibinfo {author} {\bibfnamefont {J.}~\bibnamefont
  {Grawitter}}\ and\ \bibinfo {author} {\bibfnamefont {H.}~\bibnamefont
  {Stark}},\ }\href {\doibase 10.1039/C7SM02101A} {\bibfield  {journal}
  {\bibinfo  {journal} {Soft Matter}\ }\textbf {\bibinfo {volume} {14}},\
  \bibinfo {pages} {1856} (\bibinfo {year} {2018})}\BibitemShut {NoStop}%
\bibitem [{\citenamefont {Chen}(2004)}]{chen_2004}%
  \BibitemOpen
  \bibfield  {author} {\bibinfo {author} {\bibfnamefont {H.-Y.}\ \bibnamefont
  {Chen}},\ }\href {\doibase 10.1103/PhysRevLett.92.168101} {\bibfield
  {journal} {\bibinfo  {journal} {Physical Review Letters}\ }\textbf {\bibinfo
  {volume} {92}} (\bibinfo {year} {2004}),\
  10.1103/PhysRevLett.92.168101}\BibitemShut {NoStop}%
\bibitem [{\citenamefont {Reigada}\ \emph
  {et~al.}(2005{\natexlab{a}})\citenamefont {Reigada}, \citenamefont {Buceta},\
  and\ \citenamefont {Lindenberg}}]{reigada_2005_1}%
  \BibitemOpen
  \bibfield  {author} {\bibinfo {author} {\bibfnamefont {R.}~\bibnamefont
  {Reigada}}, \bibinfo {author} {\bibfnamefont {J.}~\bibnamefont {Buceta}}, \
  and\ \bibinfo {author} {\bibfnamefont {K.}~\bibnamefont {Lindenberg}},\
  }\href {\doibase 10.1103/PhysRevE.71.051906} {\bibfield  {journal} {\bibinfo
  {journal} {Physical Review E}\ }\textbf {\bibinfo {volume} {71}} (\bibinfo
  {year} {2005}{\natexlab{a}}),\ 10.1103/PhysRevE.71.051906}\BibitemShut
  {NoStop}%
\bibitem [{\citenamefont {Reigada}\ \emph
  {et~al.}(2005{\natexlab{b}})\citenamefont {Reigada}, \citenamefont {Buceta},\
  and\ \citenamefont {Lindenberg}}]{reigada_2005_2}%
  \BibitemOpen
  \bibfield  {author} {\bibinfo {author} {\bibfnamefont {R.}~\bibnamefont
  {Reigada}}, \bibinfo {author} {\bibfnamefont {J.}~\bibnamefont {Buceta}}, \
  and\ \bibinfo {author} {\bibfnamefont {K.}~\bibnamefont {Lindenberg}},\
  }\href {\doibase 10.1103/PhysRevE.72.051921} {\bibfield  {journal} {\bibinfo
  {journal} {Physical Review E}\ }\textbf {\bibinfo {volume} {72}} (\bibinfo
  {year} {2005}{\natexlab{b}}),\ 10.1103/PhysRevE.72.051921}\BibitemShut
  {NoStop}%
\bibitem [{\citenamefont
  {Riecke}(1992{\natexlab{a}})}]{riecke1992ginzburg_landau}%
  \BibitemOpen
  \bibfield  {author} {\bibinfo {author} {\bibfnamefont {H.}~\bibnamefont
  {Riecke}},\ }\href {\doibase https://doi.org/10.1016/0167-2789(92)90169-N}
  {\bibfield  {journal} {\bibinfo  {journal} {Physica D: Nonlinear Phenomena}\
  }\textbf {\bibinfo {volume} {61}},\ \bibinfo {pages} {253 } (\bibinfo {year}
  {1992}{\natexlab{a}})}\BibitemShut {NoStop}%
\bibitem [{\citenamefont {Matthews}\ and\ \citenamefont
  {Cox}(2000)}]{matthews_2000}%
  \BibitemOpen
  \bibfield  {author} {\bibinfo {author} {\bibfnamefont {P.~C.}\ \bibnamefont
  {Matthews}}\ and\ \bibinfo {author} {\bibfnamefont {S.~M.}\ \bibnamefont
  {Cox}},\ }\href {\doibase 10.1088/0951-7715/13/4/317} {\bibfield  {journal}
  {\bibinfo  {journal} {Nonlinearity}\ }\textbf {\bibinfo {volume} {13}},\
  \bibinfo {pages} {1293} (\bibinfo {year} {2000})},\ \bibinfo {note} {arXiv:
  nlin/0006002}\BibitemShut {NoStop}%
\bibitem [{\citenamefont {Cox}\ and\ \citenamefont
  {Matthews}(2003)}]{cox_2003}%
  \BibitemOpen
  \bibfield  {author} {\bibinfo {author} {\bibfnamefont {S.}~\bibnamefont
  {Cox}}\ and\ \bibinfo {author} {\bibfnamefont {P.}~\bibnamefont {Matthews}},\
  }\href {\doibase 10.1016/S0167-2789(02)00733-9} {\bibfield  {journal}
  {\bibinfo  {journal} {Physica D: Nonlinear Phenomena}\ }\textbf {\bibinfo
  {volume} {175}},\ \bibinfo {pages} {196} (\bibinfo {year}
  {2003})}\BibitemShut {NoStop}%
\bibitem [{\citenamefont {Winterbottom}(2006)}]{winterbottom_2005}%
  \BibitemOpen
  \bibfield  {author} {\bibinfo {author} {\bibfnamefont {D.~M.}\ \bibnamefont
  {Winterbottom}},\ }\href {http://eprints.nottingham.ac.uk/10180/} {\enquote
  {\bibinfo {title} {Pattern formation with a conservation law},}\ } (\bibinfo
  {year} {2006})\BibitemShut {NoStop}%
\bibitem [{\citenamefont {Helfrich}(1973)}]{helfrich_1973}%
  \BibitemOpen
  \bibfield  {author} {\bibinfo {author} {\bibfnamefont {W.}~\bibnamefont
  {Helfrich}},\ }\href@noop {} {\bibfield  {journal} {\bibinfo  {journal} {Z.
  NaturForsch.}\ }\textbf {\bibinfo {volume} {C 28}},\ \bibinfo {pages} {693}
  (\bibinfo {year} {1973})}\BibitemShut {NoStop}%
\bibitem [{\citenamefont {West}\ \emph {et~al.}(2009)\citenamefont {West},
  \citenamefont {Brown},\ and\ \citenamefont {Schmid}}]{west_2009}%
  \BibitemOpen
  \bibfield  {author} {\bibinfo {author} {\bibfnamefont {B.}~\bibnamefont
  {West}}, \bibinfo {author} {\bibfnamefont {F.~L.}\ \bibnamefont {Brown}}, \
  and\ \bibinfo {author} {\bibfnamefont {F.}~\bibnamefont {Schmid}},\ }\href
  {\doibase 10.1529/biophysj.108.138677} {\bibfield  {journal} {\bibinfo
  {journal} {Biophysical Journal}\ }\textbf {\bibinfo {volume} {96}},\ \bibinfo
  {pages} {101} (\bibinfo {year} {2009})}\BibitemShut {NoStop}%
\bibitem [{\citenamefont {Krauth}(2006)}]{krauth_2006}%
  \BibitemOpen
  \bibfield  {author} {\bibinfo {author} {\bibfnamefont {W.}~\bibnamefont
  {Krauth}},\ }\href@noop {} {\emph {\bibinfo {title} {Statistical Mechanics
  Algorithms and Computations}}}\ (\bibinfo  {publisher} {Oxford University
  Press},\ \bibinfo {address} {Oxford},\ \bibinfo {year} {2006})\BibitemShut
  {NoStop}%
\bibitem [{\citenamefont {Dean}(1996)}]{dean1996langevin}%
  \BibitemOpen
  \bibfield  {author} {\bibinfo {author} {\bibfnamefont {D.~S.}\ \bibnamefont
  {Dean}},\ }\href@noop {} {\bibfield  {journal} {\bibinfo  {journal} {Journal
  of Physics A: Mathematical and General}\ }\textbf {\bibinfo {volume} {29}},\
  \bibinfo {pages} {L613} (\bibinfo {year} {1996})}\BibitemShut {NoStop}%
\bibitem [{\citenamefont {Cross}\ and\ \citenamefont
  {Hohenberg}(1993)}]{cross_93}%
  \BibitemOpen
  \bibfield  {author} {\bibinfo {author} {\bibfnamefont {M.~C.}\ \bibnamefont
  {Cross}}\ and\ \bibinfo {author} {\bibfnamefont {P.~C.}\ \bibnamefont
  {Hohenberg}},\ }\href {\doibase 10.1103/RevModPhys.65.851} {\bibfield
  {journal} {\bibinfo  {journal} {Reviews of Modern Physics}\ }\textbf
  {\bibinfo {volume} {65}},\ \bibinfo {pages} {851} (\bibinfo {year}
  {1993})}\BibitemShut {NoStop}%
\bibitem [{\citenamefont {Swift}\ and\ \citenamefont
  {Hohenberg}(1977)}]{swift_hohenberg_77}%
  \BibitemOpen
  \bibfield  {author} {\bibinfo {author} {\bibfnamefont {J.}~\bibnamefont
  {Swift}}\ and\ \bibinfo {author} {\bibfnamefont {P.~C.}\ \bibnamefont
  {Hohenberg}},\ }\href {\doibase 10.1103/PhysRevA.15.319} {\bibfield
  {journal} {\bibinfo  {journal} {Physical Review A}\ }\textbf {\bibinfo
  {volume} {15}},\ \bibinfo {pages} {319} (\bibinfo {year} {1977})}\BibitemShut
  {NoStop}%
\bibitem [{\citenamefont {Coullet}\ and\ \citenamefont
  {Fauve}(1985)}]{coullet_1985}%
  \BibitemOpen
  \bibfield  {author} {\bibinfo {author} {\bibfnamefont {P.}~\bibnamefont
  {Coullet}}\ and\ \bibinfo {author} {\bibfnamefont {S.}~\bibnamefont
  {Fauve}},\ }\href {\doibase 10.1103/PhysRevLett.55.2857} {\bibfield
  {journal} {\bibinfo  {journal} {Physical Review Letters}\ }\textbf {\bibinfo
  {volume} {55}},\ \bibinfo {pages} {2857} (\bibinfo {year}
  {1985})}\BibitemShut {NoStop}%
\bibitem [{\citenamefont
  {Riecke}(1992{\natexlab{b}})}]{riecke_self-trapping_1992}%
  \BibitemOpen
  \bibfield  {author} {\bibinfo {author} {\bibfnamefont {H.}~\bibnamefont
  {Riecke}},\ }\href {\doibase 10.1103/PhysRevLett.68.301} {\bibfield
  {journal} {\bibinfo  {journal} {Physical Review Letters}\ }\textbf {\bibinfo
  {volume} {68}},\ \bibinfo {pages} {301} (\bibinfo {year}
  {1992}{\natexlab{b}})}\BibitemShut {NoStop}%
\bibitem [{\citenamefont {Nardini}\ \emph {et~al.}(2017)\citenamefont
  {Nardini}, \citenamefont {Fodor}, \citenamefont {Tjhung}, \citenamefont {van
  Wijland}, \citenamefont {Tailleur},\ and\ \citenamefont
  {Cates}}]{nardini_2017}%
  \BibitemOpen
  \bibfield  {author} {\bibinfo {author} {\bibfnamefont {C.}~\bibnamefont
  {Nardini}}, \bibinfo {author} {\bibfnamefont {E.}~\bibnamefont {Fodor}},
  \bibinfo {author} {\bibfnamefont {E.}~\bibnamefont {Tjhung}}, \bibinfo
  {author} {\bibfnamefont {F.}~\bibnamefont {van Wijland}}, \bibinfo {author}
  {\bibfnamefont {J.}~\bibnamefont {Tailleur}}, \ and\ \bibinfo {author}
  {\bibfnamefont {M.~E.}\ \bibnamefont {Cates}},\ }\href {\doibase
  10.1103/PhysRevX.7.021007} {\bibfield  {journal} {\bibinfo  {journal} {Phys.
  Rev. X}\ }\textbf {\bibinfo {volume} {7}},\ \bibinfo {pages} {021007}
  (\bibinfo {year} {2017})}\BibitemShut {NoStop}%
\bibitem [{\citenamefont {Lebowitz}\ and\ \citenamefont
  {Spohn}(1999)}]{Lebowitz-Spohn-1999}%
  \BibitemOpen
  \bibfield  {author} {\bibinfo {author} {\bibfnamefont {J.~L.}\ \bibnamefont
  {Lebowitz}}\ and\ \bibinfo {author} {\bibfnamefont {H.}~\bibnamefont
  {Spohn}},\ }\href {\doibase 10.1023/A:1004589714161} {\bibfield  {journal}
  {\bibinfo  {journal} {Journal of Statistical Physics}\ }\textbf {\bibinfo
  {volume} {95}},\ \bibinfo {pages} {333} (\bibinfo {year} {1999})}\BibitemShut
  {NoStop}%
\bibitem [{\citenamefont {Langer}\ and\ \citenamefont
  {Müller-Krumbhaar}(1977)}]{langer_1977}%
  \BibitemOpen
  \bibfield  {author} {\bibinfo {author} {\bibfnamefont {J.}~\bibnamefont
  {Langer}}\ and\ \bibinfo {author} {\bibfnamefont {J.}~\bibnamefont
  {Müller-Krumbhaar}},\ }\href {\doibase 10.1016/0022-0248(77)90171-3}
  {\bibfield  {journal} {\bibinfo  {journal} {Journal of Crystal Growth}\
  }\textbf {\bibinfo {volume} {42}},\ \bibinfo {pages} {11} (\bibinfo {year}
  {1977})}\BibitemShut {NoStop}%
\end{thebibliography}


%

\end{document}